\documentclass[a4paper,11pt]{article}
\pdfoutput=1

\usepackage{jheppub}
\usepackage{mathrsfs}
\usepackage{verbatim}
\usepackage{booktabs}
\usepackage[tight]{subfigure}

\usepackage[latin1]{inputenc} 
\usepackage{latexsym} 
\usepackage{graphicx} 
\usepackage{amsfonts}
\usepackage{amsmath} 
\usepackage{amssymb} 
\usepackage{parskip}
\usepackage{array}
\usepackage{color}

\newcommand{\be}{\begin{equation}}
\newcommand{\ee}{\end{equation}}
\newcommand{\bea}{\begin{eqnarray}}
\newcommand{\eea}{\end{eqnarray}}

\def\simgt{\stackrel{>}{{}_\sim}}
\newcommand{\gsim}{\lower.7ex\hbox{$\;\stackrel{\textstyle>}{\sim}\;$}}
\newcommand{\lsim}{\lower.7ex\hbox{$\;\stackrel{\textstyle<}{\sim}\;$}}

\newcommand{\cl}{\text{CL }}

\newcommand{\BR}{BR}

\newcommand{\brbsgamma}{\BR(\overline{B}\rightarrow X_s\gamma)}

\newcommand\brbsmumu{\BR(\overline{B}_s\to\mu^+\mu^-)}

\newcommand\RBtaunu{\frac{\BR(B_u \to \tau \nu)}{\BR(B_u \to \tau \nu)_{SM}}}
\newcommand\DeltaO{\Delta_{0-}}
\newcommand\RBDtaunuBDenu{\frac{\BR(B \to D \tau \nu)}{\BR(B \to D e \nu)}}
\newcommand\Rl{R_{l23}}

\newcommand\Dstaunu{\BR(D_s \to \tau \nu)}
\newcommand\Dsmunu{\BR(D_s\to \mu \nu)} 
\newcommand\Dmunu{\BR(D \to \mu \nu)}

\newcommand{\mhl}{m_h}

\newcommand{\params}{{\mathbf \Theta}}

\newcommand{\like}{{\mathcal L}}

\newcommand{\sigmaSI}{\sigma^\text{SI}_{\chi N}}
\newcommand{\sigmaSD}{\sigma^\text{SD}_{\chi N}}
\newcommand{\neut}{{\tilde{\chi}^0_1}}

\newcommand{\OhLSP}{\Omega_\text{LSP} h^2}
\newcommand{\siW}{\sigma_{\text{PLANCK}}}
\newcommand{\muW}{\mu_\text{PLANCK}}

\title{LHC and dark matter phenomenology of the NUGHM}
\author[a]{Maria Eugenia Cabrera,}
\author[b]{Alberto Casas,}
\author[c]{Roberto Ruiz de Austri}
\author[a]{and Gianfranco Bertone}

\affiliation[a]{Institute of Theoretical Physics, GRAPPA, \\
                    University of Amsterdam, \\
                   Amsterdam, The Netherlands}
\affiliation[b]{Instituto de F\'isica Te\'orica, IFT-UAM/CSIC, \\
        U.A.M., Cantoblanco, \\
        28049 Madrid, Spain} 
\affiliation[c]{Instituto de F\'isica Corpuscular, IFIC-UV/CSIC, \\
        Valencia, Spain}
\abstract{We present a Bayesian analysis of the NUGHM, a supersymmetric scenario with non-universal gaugino masses and Higgs masses, including all the relevant experimental observables and dark matter constraints. The main merit of the NUGHM is that it essentially includes all the possibilities for dark matter (DM) candidates within the MSSM, since the neutralino and chargino spectrum -and composition- are as free as they can be in the general MSSM.
We identify the most probable regions in the NUHGM parameter space, and study the associated phenomenology at the LHC and the prospects for DM direct detection. Requiring that the neutralino makes all of the DM in the Universe, we identify two preferred regions around $m_{\chi_1^0}= 1\ {\rm TeV},\; 3\ {\rm TeV}$, which correspond to the (almost) pure Higgsino and wino case. There exist other marginal regions (e.g. Higgs-funnel), but with much less statistical weight. The prospects for detection at the LHC in this case are quite pessimistic, but future direct detection experiments like LUX and XENON1T, will be able to probe this scenario. In contrast, when allowing other DM components, the prospects for detection at the LHC become more encouraging -- the most promising signals being, beside the production of gluinos and squarks, the production of the heavier chargino and neutralino states, which lead to WZ and same-sign WW final states -- and direct detection remains a complementary, and even more powerful, way to probe the scenario.}
 
\preprint{IFT-UAM/CSIC-13-128, IFIC/13-92}

\begin{document}
\maketitle
\flushbottom

%\documentclass[12pt]{article}
%
%\textwidth=16cm
%
%\textheight=23cm 
%\topmargin -1cm 
%\oddsidemargin -0.cm 
%\evensidemargin -0.cm 
%\hoffset -0 cm
%\footskip 24pt
%
%\usepackage[latin1]{inputenc} 
%\usepackage{latexsym} 
%\usepackage{graphicx} 
%\usepackage{amsfonts}
%\usepackage{amsmath} 
%\usepackage{amssymb} 
%%\usepackage{mathrsfs}
%\usepackage{parskip}
%\usepackage{array}
%
%\title{Non Universal gaugino + non universal Higgsino masses MSSM } 
%\author{Maria Eugenia Cabrera}
%\date{}
%
%\begin{document}
%
%\maketitle

%%%%%%%%%%%%%%%%%%%%%%%%%%%%%%%%%%%%%%%%%%%%%%%%%%%%%%%%%%%%%%%%%%
\section{Introduction}
%%%%%%%%%%%%%%%%%%%%%%%%%%%%%%%%%%%%%%%%%%%%%%%%%%%%%%%%%%%%%%%%%%

Despite the impressive performance of the LHC and the discovery of the Higgs boson, we do not have yet any hints of physics beyond the Standard Model (BSM). Still, there is a reasonable hope to find such new physics in the next LHC runs, especially if the famous `hierarchy problem' is actually a sound case for BSM. In this sense, supersymmetric (SUSY) scenarios \cite{Martin:1997ns}, in particular the minimal supersymmetric Standard Model (MSSM), continue to be one of the best-motivated candidates for BSM. In addition there is a reasonable hope to detect dark matter (DM) in the present and future experiments, especially if the DM particles are weakly interacting massive particles (WIMPs), as would be the natural case in a supersymmetric scenario\cite{Jungman:1995df,Bergstrom:2000pn,Munoz:2003gx,Bertone:2004pz}.

The MSSM is already constrained by current data, since a $\sim 125-126$ GeV Higgs generically requires rather large SUSY masses \cite{Ellis:1990nz, Haber:1990aw,Casas:1994us}, which is in tension with the naturalness of the electroweak breaking at the correct scale. However, there are many  acceptable regions of the MSSM parameter space (even with relatively light supersymmetric particles) which are still to be probed. Recall in this sense that, strictly speaking, the Higgs mass only puts limits on the stop masses. Moreover, some supersymmetric particles, e.g. charginos and neutralinos (electroweakinos), are not as constrained by data as squarks and gluinos. On the other hand, part of the difficulty to put robust and handy constraints on SUSY comes from the fact that the MSSM has $\sim 100$ independent parameters, mainly soft terms related to the unknown mechanism of SUSY breaking and its transmission to the observable sector; and it is not straightforward to translate LHC data into intelligible limits on such complex parameter-space.

A usual strategy is to present the LHC data as constraints in a simplified version of the MSSM, typically the CMSSM, i.e. demanding universal scalar masses, gaugino masses and trilinear scalar couplings at a high scale, $M_X$. A slightly extended version of this model is the non-universal Higgs mass model (NUHM), where the Higgs soft-masses are allowed to be different than the rest of the scalar masses at $M_X$. With the present data, it turns out that a big portion of the previously-acceptable CMSSM and NUHM parameter-space becomes excluded \cite{Akula:2012kk,Buchmueller:2012hv,Strege:2012bt,Fowlie:2012im,Cabrera:2012vu,Boehm:2013qva}. The reason is that in these models the stop and slepton (chargino and neutralino) masses are strongly correlated to the squark (gluino) masses, so that the scenario typically requires {\em all} the supersymmetric particles to be heavy, most probably beyond the LHC reach. In addition, these models have great difficulties to have a neutralino as dark matter (DM) particle, consistent with all DM constraints. However, the CMSSM and NUHM are probably over-simplified models, as they are based on too-constraining initial conditions. Note e.g. that the universality of gaugino masses is not motivated by any phenomenological or theoretical fact (except by an hypothetical GUT theory below the scale of SUSY-breaking transmission).

Another strategy that has gained relevance is the use of so-called ``simplified models". A simplified model can be described by a small number of masses and cross-secctions, which are directly related to collider-physics observables. The idea is to mimic the collider signatures of a physical scenario (as the MSSM) with a dominant simplified model (or a reduced set of them) in each region of the parameter space. This potentially makes more efficient the exploration of complex models. However, the great intricacy of the generic MSSM would demand an enormous proliferation of simplified models in order to cover the parameter-space.

In this paper we will follow a strategy which is potentially very powerful to optimize SUSY searches. First, we consider a quite generic MSSM model, namely one with non-universal gaugino masses and Higgs masses (NUGHM). As discussed in sect. 2, though not completely general, this scenario is well-motivated by a number of theoretical and phenomenological facts and goes far beyond the CMSSM and NUHM in complexity and phenomenological richness. In addition, this scenario does capture the most natural DM candidates of the MSSM far better than the CMSSM. Then we perform a Bayesian analysis to identify the preferred regions in the associated parameter-space. Finally, we examine the typical (i.e. most likely) phenomenology emerging from this scenario. For this task we will focus on the regions that are potentially accessible by experiment. This analysis will allow to identify the most representative simplified models, i.e. the processes that more faithfully describe the phenomenology in the most relevant regions of the parameter-space.

On the other hand, along the paper it will become clear the great importance of DM searches as a complementary, and often more powerful, way to probe the supersymmetric parameter-space in the future.

In sect.~2 we define the NUGHM and discuss its main merits. In sect.~3 we explain the characteristics of our Bayesian analysis of the NUGHM, and the experimental and observational data used for the computation of the likelihood. This section includes a discussion of the naturalness issue and the choice of priors. These matters have been treated in an improved way with respect to previous literature. In sect.~4 we present the results. In subsect.~4.1 (single-component CDM), we consider the possibility that the neutralino makes all the DM in the Universe. Subsect.~4.2 is devoted to special regions with little statistical weight but with physical interest (like the Higgs-funnel region). In subsect.~4.3 (multi-component CDM) we relax the condition on DM, allowing that the supersymmetric contribution is equal or less than the actual DM abundance. In sect.4.4 (``Low Energy" NUGHM) we focus on the region of the NUGHM parameter space which is potentially accesible to LHC. We determine what are the most characteristic and visible signals at LHC and compare with DM detection prospects. The complementarity of both experimental strategies will become very clear. Finally, in sect.~5 we present our conclusions.

%%%%%%%%%%%%%%%%%%%%%%%%%%%%%%%%%%%%%%%%%%%%%%%%%%%%%%%%%%%%%%%%%%
\section{The NUGHM model}
%%%%%%%%%%%%%%%%%%%%%%%%%%%%%%%%%%%%%%%%%%%%%%%%%%%%%%%%%%%%%%%%%%

There are two main directions along which the simplest CMSSM scenario can be extended. First, the scalar masses and trilinear couplings do not need to be universal (as it is assumed in the CMSSM). Constraints from flavour- and CP-violating processes certainly require (very) accurate universality for sfermions of the same type and for the two first generations. But otherwise, there is a broad scope for non-degeneracy of sfermion masses. The second direction is that the initial gaugino masses do not need to be universal, as assumed in the CMSSM. Actually, there is no theoretical or phenomenological reason for such simplification, apart from a possible GUT scenario at the scale of the transmission of SUSY breaking to the observable sector (or below it). Here we will focus on this second direction. More precisely, we will consider  throughout the paper an extension of the CMSSM that
allows non-universal gaugino masses and non-universal Higgs masses, which will be denoted as NUGHM in what follows.
Thus the NUGHM has an 8-dimension
parameter space, defined by
\begin{displaymath}
\{s, M_1,\ M_2,\ M_3,\ A_0, \ m_H,\ m_0, B, \mu \};
\end{displaymath}
where $M_3$, $M_2$ and $M_1$ are the gaugino masses of $SU(3)\times SU(2)\times U(1)_Y$; $A_0, m_0$ are the (universal) trilinear scalar coupling and sfermion mass (except for the Higgs sector); $m_H$ is the universal mass soft-term for the two Higgs doublets; and $\mu$ is the Higgs mass term in the superpotential. The first seven parameters are soft-terms, while $\mu$ is a parameter in the superpotential (but probably with a similar origin). Finally, $s$ stands for the SM-like `nuisance' parameters (gauge and Yukawa couplings). All the parameters are defined at a high scale, $M_X$.

The main merit of the NUGHM is that it essentially includes all the
possibilities for DM candidates within the MSSM, since the neutralino spectrum
and composition (as well as the chargino ones, which are relevant for
co-annihilation processes) are as free as they can be in the general MSSM
framework; and they are not correlated to the gluino mass, which is severely
constrained by LHC. On the other hand, even if the sfermion masses are heavy
--as the experimental Higgs mass, the present bounds on squarks and
constraints on flavor-violation processes may suggest-- there are reasons to
expect fermionic supersymmetric states to be around the TeV range. Namely,
this is required in order to keep the succesful gauge unification that occurs
in the MSSM; and it is also required for DM issues, as mentioned above. In
addition the presence of light charginos and neutralinos is probably the most
robust consequence of ``Natural SUSY" scenarios, i.e. those with
as-small-as-possible fine-tuning
\cite{Barbieri:1987fn,Casas:2003jx,Papucci:2011wy,Hardy:2013ywa}.

Thus, the production of charginos and neutralinos ("electroweakinos") at the LHC may be one of the best motivated avenues to detect SUSY at the LHC. In this sense, the NUGHM captures the rich phenomenology associated to these states in the general MSSM.

Finally, although we are assuming a universal sfermion mass, we allow for different Higgs mass terms. One reason for that is that the Higgs sector in the MSSM (or in the ordinary SM) is clearly different from the other matter states, which come in three families. Thus it is reasonable to expect that the transmission of SUSY breaking to this sector may also be different. Second, in the electroweak (EW) breaking process, the degeneracy of the Higgs masses and the rest of the sfermion masses leads to unnecessary correlations between e.g. the values of $\mu$ and $\tan\beta$ (which play a crucial role in phenomenology and DM issues), and the rest of the spectrum. 

%%%%%%%%%%%%%%%%%%%%%%%%%%%%%%%%%%%%%%%%%%%%%%%%%%%%%%%%%%%%%%%%%%
\section{The analysis}
%%%%%%%%%%%%%%%%%%%%%%%%%%%%%%%%%%%%%%%%%%%%%%%%%%%%%%%%%%%%%%%%%%

\subsection{Bayesian statistics}

The goal of the Bayesian approach is to generate a forecast, i.e. a map of the relative probability of the different regions of the parameter space of the model under consideration (NUGHM in our case), using all the available (theoretical and experimental) information. This is the so-called {\em posterior} or probability density function (pdf), $p(\theta_i|{\rm data})$, where `data' stands for all the experimental information and $\theta_i$ represent the various parameters of the model. The posterior is given by the Bayes' Theorem
\bea
\label{Bayes}
p(\theta_i|{\rm data})\ =\ p({\rm data}|\theta_i)\ p(\theta_i)\ \frac{1}{p({\rm data})}\ ,
\eea
where $p({\rm data}|\theta_i)$ is the likelihood (sometimes denoted by ${\cal L}$), i.e. the probability density of observing the given data if nature has chosen to be at the $\{\theta_i\}$ point of the parameter space (this is the quantity used in frequentist approaches); $p(\theta_i)$ is the prior, i.e. the ``theoretical" probability density that we assign a priori to the point in the parameter space; and, finally, $p({\rm data})$ is a normalization factor which plays no role unless one wishes to compare different classes of models.
One can say that in eq.~(\ref{Bayes}) the first factor (the likelihood) is objective, while the second (the prior) contains our prejudices about 
how the probability is distributed a priori in the parameter space, given all our previous knowledge about the model. Certainly, the prior piece is the most disgusting one, but it is inescapable if one attempts to obtain a forecast of the model.

Ignoring the prior factor is not necessarily the most reasonable or ``free of prejudices" attitude. Such procedure amounts to an implicit choice for the prior, namely 
a {\it flat} prior in the parameters. However, one needs some theoretical basis to establish, at least, the parameters whose prior can be reasonably taken as flat.

On the other hand, a choice for the allowed ranges of the various parameters is necessary in order to make statistical statements. (Indeed, the ranges of the parameters can be considered as part of the prior, since they are equivalent to steps in the prior functions.) Often one is interested in showing the probability density of one (or several) of the initial parameters, say $\theta_i,\ i=1,...,N_1$, but not in the others, $\theta_i,\ i=N_1+1,...,N$.
Then, one has to {\em marginalize} the latter, i.e. integrate in the parameter space:
\bea
\label{marg}
p(\theta_i,\ i=1,...,N_1|{\rm data})\ = \int d\theta_{N_1+1},...,d\theta_N\ p(\theta_i,\ i=1,...,N|{\rm data})\ 
 .
\eea
This procedure is very useful and common to make predictions about the values of particularly interesting parameters. It is also useful to get ride of the nuisance parameters (gauge couplings and physical masses of observed particles). Now, in order to perform the marginalization, we need an input for the prior functions (which, besides the likelihood determines the posterior in eq.~(\ref{Bayes})) {\em and} for the range of allowed values of the parameters, which determines the range of the definite integration (\ref{marg}). A choice for these ingredients is therefore inescapable when one performs Bayesian LHC forecasts. We will come back to this point in the next subsections.

%%%%%%%%%%%%%%%%%%%%%%%%%%%%%%%%%%%%%%%%%%%%%%%%%%%%%%%%%%%%%%%%%%
\subsection{Naturalness and Bayesian statistics}
%%%%%%%%%%%%%%%%%%%%%%%%%%%%%%%%%%%%%%%%%%%%%%%%%%%%%%%%%%%%%%%%%%

It is a common assumption that the parameters of the MSSM should not be too far from the experimental EW scale in order to avoid unnatural fine-tunings to obtain the correct size of the EW breaking. On the other hand, since the naturalness arguments are at bottom statistical arguments, one might expect that an effective penalization of fine-tunings arises automatically from the Bayesian analysis, with no need of introducing 
``naturalness priors"  or restricting the soft terms to the low-energy scale. It was shown in ref. \cite{Cabrera:2008tj} that this is indeed the case. 

The key is, instead of solving $\mu$ in terms of $M_Z$ and the other supersymmetric parameters using the minimization conditions, treat $M_Z^{\rm exp}$ as experimental data on a similar footing with the others, entering the total likelihood, ${\cal L}$. Approximating the $M_Z$ likelihood as a Dirac delta,
\bea
\label{likelihood}
p({\rm data}|s, M_1, M_2, M_3, A_0, m_0, m_H, B, \mu)\ \simeq\ \delta(M_Z-M_Z^{\rm exp})\ {\cal L}_{\rm rest}\ ,
\eea
where ${\cal L}_{\rm rest}$ is the likelihood associated to all the physical observables, except $M_Z$, one can marginalize the $\mu-$parameter
\bea
\label{marg_mu}
%{}\hspace{-1cm}
&&p(s, M_1, M_2, M_3, A_0, m_0, m_H, B| \ {\rm data} ) = \int d\mu\ p(s, M_1, M_2, M_3, A_0, m_0, m_H, B, \mu | 
{\rm data} )
\nonumber\\ 
&&\hspace{3cm}\simeq\ {\cal L}_{\rm rest}  \left|\frac{d\mu}{d M_Z}\right|_{\mu_Z}
p(s, M_1, M_2, M_3, A_0, m_0, m_H, B, \mu_Z)\ ,
\eea
where we have used eq.~(\ref{Bayes}). Here $\mu_Z$ is the value of $\mu$ that reproduces $M_Z^{\rm exp}$ for the given values of $\{s, m_0, m_{1/2}, A, B\}$, and $p(s, M_1, M_2, M_3, A_0, m_0, m_H, B, \mu)$ is the prior in the initial parameters (still undefined). Note that the Jacobian factor, $\frac{d\mu}{d M_Z}$, can be written as $\frac{2\mu}{M_Z}\ \frac{1}{c_\mu}$, where 
$c_\mu = \left|\frac{\partial \ln M_Z^2}{\partial \ln \mu}\right|$ is the
conventional Barbieri-Giudice measure \cite{Ellis:1986yg, Barbieri:1987fn} of
the degree of fine-tuning. Thus, the above Jacobian factor incorporates the  fine-tuning penalization, with no ad hoc assumptions.
An important consequence is that the high-scale region of the parameter space, say the region of soft terms $\gsim\ {\cal O} (10)\ {\rm TeV}$, becomes statistically insignificant. This allows to consider wide ranges for the soft parameters (up to the very $M_X$). In consequence, the results of our analysis are essentially independent on the upper limits of the MSSM parameters, an important advantage over other Bayesian approaches.

In practice, beside the $\mu-$parameter, one also trades the fermionic Yukawa couplings by the fermion masses (particularly the top one) and the $B-$parameter by $\tan\beta$. Hence, calling $J$ the Jacobian of the $\{\mu,y_t,B\}\ \rightarrow\  \{M_Z,m_t,\tan\beta\}$ transformation, the {\em effective prior} in the new variables reads
\bea
\label{eff_prior}
p_{\rm eff}(g_i, m_t, m_0, m_{1/2}, A, \tan\beta)\  \equiv\ 
J|_{\mu=\mu_Z}\  p(g_i, y_t, m_0, m_{1/2}, A, B, \mu=\mu_Z) \ .
\eea

In this work we have computed $J$ numerically. An analytical and quite accurate expression of $J$ can be found in refs.\cite{Cabrera:2008tj,Cabrera:2010dh}, namely
\bea
\label{J}
J\propto\ \frac{\tan^2\beta -1}{\tan\beta(1+\tan^2\beta)} \frac{B_{\rm low}}{\mu} \ ,
%\frac{1}{\sin \beta},
\eea
where the ``low'' subscript indicates that the quantity is evaluated at low scale.

An important point to stress is that the Jacobian (\ref{J}) is a model-independent factor, valid for any MSSM, and in particular for the NUGHM, which must be multiplied by whatever prior is chosen for the initial parameters. It {\em cannot} be ignored. In addition, it contains the above-discussed penalization of fine-tuned regions. Note in this sense that, besides the penalization for large $\mu$, the Jacobian factor 
contains a penalization of large $\tan\beta$, reflecting the smaller statistical weight of this possibility. The implicit fine-tuning associated to a large $\tan\beta$ was already noted in refs.~\cite{Hall:1993gn, Nelson:1993vc}, where it was estimated to be of order $1/\tan\beta$, in agreement with eq.(\ref{J}). 

%%%%%%%%%%%%%%%%%%%%%%%%%%%%%%%%%%%%%%%%%%%%%%%%%%%%%%%%%%%%%%%%%%
\subsection{Priors}
%%%%%%%%%%%%%%%%%%%%%%%%%%%%%%%%%%%%%%%%%%%%%%%%%%%%%%%%%%%%%%%%%%

The choice of the prior for the initial parameters is an unavoidable decision in order to construct the probability distribution function in the parameter-space, as is clear from the Bayes theorem, eq.(\ref{Bayes}). Admittedly, this is a rather subjective issue. A reasonable prior must reflect our knowledge about the parameters, before consideration of the experimental data (to be included in the likelihood piece). Concerning the prior-dependence of the results, a conservative attitude is to use two different, though still reasonable, priors, and compare the results. 

A standard choice is to adopt a logarithmic prior, i.e. to assume that, in principle, the typical order of magnitude of the soft terms can be anything (below $M_X$) with equal probability. This is certainly quite reasonable, since it amounts to consider all the possible magnitudes of the SUSY breaking in the observable sector on the same footing (as occurs e.g. in conventional SUSY breaking by gaugino condensation in a hidden sector). On the other hand, this idea can be realized in two different fashions, which we describe below. Along the paper we will present the results obtained from both fashions, as a measure of the prior-dependence.

{\em Standard Log Prior (S-log)}

One simply assumes that each independent parameter, $\theta_i$, has an independent logarithmic prior, $p(\theta_i)\propto 1/\theta_i$, so that the total prior is the product of the individual priors; in our case
\begin{eqnarray}
  \label{eq:logPrior}
    p(M_1, M_2, M_3, A_0, m_0, m_H, B, \mu)\ \propto\ \frac{1}{\left|M_1\ M_2\ M_3\  A_0\ m_0\  m_H\  B\  \mu\right|} \ .
\end{eqnarray}
The use of this kind of prior is common in Bayesian analyses of the MSSM. However, it has some drawbacks. First, it presents divergences when the parameters take very small values. To avoid that, just for the purpose of prior evaluation, whenever one parameter is smaller than 10 GeV, we take it equal to 10 GeV in eq.(\ref{eq:logPrior}). A more disturbing fact is that very large values of some initial parameters can be compensated in eq.(\ref{eq:logPrior}) by very small values of another ones. E.g. the S-log prior for $A_0= M_1=1$ TeV is the same as for $A_0=10$ GeV, $M_1=100$ TeV, something which is not very realistic if all the soft terms have a common origin (the SUSY breaking mechanism). 
Consequently, the S-log prior can artificially favor large (or even huge) splittings between the initial parameters.

This kind of problems are avoided with the following improved log prior.

{\em Improved Log Prior (I-log)}

Since the soft-breaking terms share a common origin it is logical to assume that their sizes are also similar, say $M_S\sim F/\Lambda$, where $F$ is the SUSY breaking scale, which corresponds to the dominant VEV among the auxiliary fields in the SUSY breaking sector (it can be an $F-$term or a $D-$term) and $\Lambda$ is the messenger scale, associated to the interactions that transmit the breaking to the observable sector. Of course, there are several contributions to a particular soft term, which depend on the details of the superpotential, the K\"ahler potential and the gauge kinetic function of the complete theory (see e.g. ref. \cite{Kaplunovsky:1993rd}). So, it is reasonable to assume
  that a particular soft term can get any value (with essentially flat probability) of the order of $M_S$ or smaller. The $\mu-$parameter is not a soft term, but a parameter of the superpotential. However, it is desirable that its size is related (e.g. through the Giudice-Masiero mechanism \cite{Giudice:1988yz}) to the SUSY breaking scale. Otherwise, one has to face the so-called $\mu-$problem, i.e. why should be the size of $\mu$ similar to the soft terms', as is required for a correct electroweak breaking. Thus, concerning the prior, we can consider $\mu$ on a similar footing with the other soft terms.
  
The next step is to marginalize the typical scale of the soft terms, $M_S$, using a logarithmic prior for it. 
This leaves a prior which depends just on the initial $\{M_1, M_2, M_3, A_0, m_0, m_H, B, \mu\}$ parameters. This procedure was described and carried out in section 2.4 of ref.\cite{Cabrera:2009dm}, which we follow here. The result, for our set of independent parameters, is remarkably simple,
\begin{eqnarray}
  \label{eq:CCRlogPrior}
  \hspace{-0.4cm}p(M_1, M_2, M_3, A_0, m_0, m_H, B, \mu)\propto\frac{1}{\max\{|M_1|, |M_2|, |M_3|,  |A_0|, m_0, m_H, |B|, |\mu|\}^8}\ .
\end{eqnarray}
It is worth-noticing that, in the absence of a likelihood, if one marginalizes all the parameters but one, say $M_3$, the I-log prior --eq.~(\ref{eq:CCRlogPrior})-- and the S-log one --eq.~(\ref{eq:logPrior})-- produce the same individual logarithmic prior for $M_3$, namely $p(M_3)\propto 1/|M_3|$. However, the expression for the I-log prior shows a non-trivial correlation between the parameters. For the I-log prior it does not pay to increase a parameter at the expense of decreasing another; what matters is the typical size of the parameters. This feature, besides being conceptually appealing, does avoid the above-mentioned drawbacks of the S-log prior. First, since some initial parameters are necessarily different from zero (e.g. $|\mu|\simgt 100$ GeV to keep charginos above the LEP limit), the I-log prior is never singular, so one does not need to impose minimum values on the parameters for the purpose of the prior evaluation. This makes the I-prior simpler and prevents spurious dependences on those minimum values required by the S-log prior. Second, by using I-log priors we do not find bizarre situations where abnormally large values of some initial parameters are compensated by the smallness of others. This can be illustrated by the example used above: unlike the S-log prior, the I-log prior for $A_0= M_1=1$ TeV is much larger than for $A_0=10$ GeV, $M_1=100$ TeV. Consequently, the I-log prior does not favor huge splittings between parameters, which is satisfactory since all the soft parameters have a common physical origin. This also makes the scanning more efficient and stable since it gets rid of bizarre throats in the parameter space where the S-log prior becomes large thanks to the unusual smallness of some parameter(s) even if others get large.

Thus, in our opinion, the results using the I-log prior are more reliable than those using the S-log one; but we will use both in order to compare the results and study the prior-dependence.

We finish this subsection commenting on the ranges for the independent parameters. We recall here that, 
aside from the naturalness argument, there is no reason to choose an ${\cal O}$(TeV) upper
limit for the SUSY parameters. Since, as discussed in subsect. 3.2, the Bayesian analysis automatically takes care of the
fine-tuning penalization, one should not put further limits on the parameters. Thus we have allowed the SUSY
parameters to vary from zero to $10^6$ GeV, though, as already dicussed, the size of the upper limit is irrelevant in practice.

%%%%%%%%%%%%%%%%%%%%%%%%%%%%%%%%%%%%%%%%%%%%%%%%%%%%%%%%%%%%%%%%%%
\subsection{The data and the numerical algorithm}
%%%%%%%%%%%%%%%%%%%%%%%%%%%%%%%%%%%%%%%%%%%%%%%%%%%%%%%%%%%%%%%%%%

\begin{table*}
\begin{center}
\begin{tabular}{l l l l }
\hline
\hline
\multicolumn{4}{c}{SM nuisance parameters} \\
\hline
 & Gaussian prior  & Range scanned & ref. \\
\hline
$M_t$ [GeV] & $173.2 \pm 0.9$  & (167.0, 178.2) &  \cite{moriond2013} \\
$m_b(m_b)^{\bar{MS}}$ [GeV] & $4.20\pm 0.07$ & (3.92, 4.48) &  \cite{pdg07}\\
$[\alpha_{em}(M_Z)^{\bar{MS}}]^{-1}$ & $127.955 \pm 0.030$ & (127.835, 128.075) &  \cite{pdg07}\\
$\alpha_s(M_Z)^{\bar{MS}}$ & $0.1176 \pm  0.0020$ &  (0.1096, 0.1256) &  \cite{Hagiwara:2006jt}\\
\hline
\end{tabular}
\end{center}
\caption{\fontsize{9}{9}\selectfont Nuisance parameters adopted in the scan of the NUGHM parameter space, 
indicating the mean and standard deviation 
adopted for the Gaussian prior on each of them, as well as the range covered in the scan.}\label{tab:nuis_params} 
\end{table*}

The uncertainties on measurements in some of the Standard Model parameters have been shown to have an 
important impact in inferences of SUSY models~\cite{Roszkowski:2007fd}. 
Of particular importance are the top and bottom masses and the electromagnetic and strong coupling constants. 
To account for this we have considered them as nuisance parameters in the analysis and 
using the central values and uncertainties as given in Table \ref{tab:nuis_params}.

The likelihood function is composed of several different parts, corresponding to the different experimental constraints that are 
applied in our analysis:
\begin{equation}
\ln \like = \ln \like_\text{LHC} + \ln \like_\text{PLANCK}  + \ln \like_\text{EW}+ \ln \like_\text{B(D)} + \ln \like_\text{Xe100}.      
\end{equation}

The LHC likelihood implements the most recent experimental constraint from the CMS and ATLAS collaborations on the mass of the lightest 
Higgs boson which after combination is $m_h = 125.66 \pm 0.41$ GeV \cite{moriond2013} 
\footnote{Combined ATLAS and CMS results using method described in the statistics review of \cite{pdg12}.}. 
We use a Gaussian likelihood and add in quadrature 
a theoretical error of 2 GeV to the experimental error.
We also include the new LHCb constraint on $\brbsmumu = (3.2^{+1.5}_{-1.2}) \times 10^{-9}$, derived from a combined analysis 
of 1 fb$^{-1}$ data at $\sqrt{s} = 7$ TeV collision energy and 1.1 fb$^{-1}$ data at $\sqrt{s} = 8$ TeV collision energy \cite{Aaij:2012nna}. 
This constraint is implemented  as a Gaussian distribution with a conservative experimental error of $\sigma = 1.5 \times 10^{-9}$, 
and a $10\%$ theoretical error. 

The constraint from the DM relic abundance is included as a Gaussian in $\ln \like_\text{PLANCK}$. We use the recent 
PLANCK value $\Omega_\chi h^2 = 0.1196 \pm 0.0031$~\cite{Ade:2013zuv} and add a fixed 10\% theoretical uncertainty in quadrature. 

$\ln \like_\text{EW}$ implements precision tests of the electroweak sector. The electroweak precision observables
$M_W$ and $\sin^2\theta_{eff}$ are included with a Gaussian likelihood. 

Relevant constraints from $B$ and $D$ physics are included in $\ln \like_\text{B(D)}$ as a Gaussian likelihood. The full 
list of $B$ and $D$ physics observables included in our analysis is shown in table~\ref{tab:exp_constraints}.

For constraints from direct DM search experiments we use the recent results from XENON100
with 225 live days of data collected  between February 2011 and March 2012 with 34 kg fiducial volume~\cite{Aprile:2012nq}. 
We build the likelihood function $\ln \like_\text{Xe100}$, following ~\cite{Strege:2012bt}, as a Poisson distribution for observing N
 recoil events when $N_s(\params)$ signal plus $N_b$ background events are expected. The expected number of events from the 
background-only hypothesis in the XENON100 run is $N_b$ = 1.0 $\pm$ 0.2, while  the collaboration reported N=2 events 
observed in the pre-defined signal region. 
We use the latest values for the fiducial mass and exposure time of the detector, and include the reduction of the lower energy 
threshold for the analysis to 3 photoelectron events 
 and an update to the response to 122 keV gamma-rays to $2.28$ PE/keVee, obtained from new calibration measurements, in 
accordance with the values reported in Ref.~\cite{Aprile:2012nq}. We make the simplifying assumption of an 
energy-independent acceptance of data quality cuts, and adjust the acceptance-corrected exposure to accurately reproduce 
the exclusion limit in the $(m_{\neut},\sigmaSI)$ plane reported in Ref.~\cite{Aprile:2012nq} in the mass range of 
interest.
For the calculation of the number of expected signal recoil events we fix the astrophysical parameters that describe the density and 
velocity distribution  of DM particles at the commonly adopted benchmark values: local CDM 
density $\rho_{\odot,{\rm CDM}}=0.4\,$GeV\,cm$^{-3}$, 
circular velocity $v_0=235$ km s$^{-1}$ and escape velocity $v_{esc}=550$ 
km s$^{-1}$ (see, e.g., \cite{Pato:2010zk} and references therein for a 
recent discussion of the astrophysical uncertainties on these 
quantities). 
For the contribution of the light quarks to the nucleon form
factors for the spin-independent WIMP-nucleon cross section we have adopted the values $f_{Tu}=0.02698$, $f_{Td}=0.03906$ and 
$f_{Ts}=0.36$ \cite{Ellis:2008hf} derived experimentally from measurements of the pion-nucleon sigma term \footnote{
Recently, in Ref. \cite{deAustri:2013saa}, it has been quantified the impact of the current uncertainty in the determination of the nucleon matrix elements coming 
from an experimental or a lattice QCD approach on direct dark matter detection in the CMSSM.}.

\begin{table*}
\begin{center}
\begin{tabular}{|l | l l l | l|}
\hline
\hline
Observable & Mean value & \multicolumn{2}{c|}{Uncertainties} & Ref. \\
 &   $\mu$      & ${\sigma}$ (exper.)  & $\tau$ (theor.) & \\\hline
$M_W$ [GeV] & 80.399 & 0.023 & 0.015 & \cite{lepwwg} \\
$\sin^2\theta_{eff}$ & 0.23153 & 0.00016 & 0.00015 & \cite{lepwwg} \\
$\brbsgamma \times 10^4$ & 3.55 & 0.26 & 0.30 & \cite{hfag}\\
$R_{\Delta M_{B_s}}$ & 1.04 & 0.11 & - & \cite{Aaij:2011qx} \\
$\RBtaunu$   &  1.63  & 0.54  & - & \cite{hfag}  \\
$\DeltaO  \times 10^{2}$   &  3.1 & 2.3  & - & \cite{Aubert:2008af}  \\
$\RBDtaunuBDenu \times 10^{2}$ & 41.6 & 12.8 & 3.5  & \cite{Aubert:2007dsa}  \\
$\Rl$ & 0.999 & 0.007 & -  &  \cite{Antonelli:2008jg}  \\
$\Dstaunu \times 10^{2}$ & 5.38 & 0.32 & 0.2  & \cite{hfag}  \\
$\Dsmunu  \times 10^{3}$ & 5.81 & 0.43 & 0.2  & \cite{hfag}  \\
$\Dmunu \times 10^{4}$  & 3.82  & 0.33 & 0.2  & \cite{hfag} \\
$\Omega_\chi h^2$ & 0.1196 & 0.0031 & 0.012 & \cite{Ade:2013zuv} \\
$\mhl$ [GeV] & 125.66  & 0.41  & 2.0 & \cite{moriond2013} \\
$\brbsmumu$ &  $3.2 \times 10^{-9}$ & $1.5  \times 10^{-9}$ & 10\% & \cite{Aaij:2012nna}\\
\hline\hline
   &  Limit (95\%~$\cl$)  & \multicolumn{2}{r|}{$\tau$ (theor.)} & Ref. \\ \hline
Sparticle masses  &  \multicolumn{3}{c|}{As in Table~4 of
  Ref.~\cite{deAustri:2006pe}.}  & \\
$m_\chi - \sigmaSI$ & \multicolumn{3}{l|}{XENON100 2012 limits ($224.6 \times 34$ kg days)} & \cite{Aprile:2012nq} \\
\hline
\end{tabular}
\end{center}
\caption{\fontsize{9}{9} \selectfont Summary of the observables used
for the computation of the likelihood function
For each quantity we use a
likelihood function with mean $\mu$ and standard deviation $s =
\sqrt{\sigma^2+ \tau^2}$, where $\sigma$ is the experimental
uncertainty and $\tau$ represents our estimate of the theoretical
uncertainty. Lower part: Observables for which only limits currently
exist. The explicit form of the likelihood function is given in
ref.~\cite{deAustri:2006pe}, including in particular a smearing out of
experimental errors and limits to include an appropriate theoretical
uncertainty in the observables.
\label{tab:exp_constraints}}
\end{table*}

For the exploration of the posterior pdf we have used the
\texttt{SuperBayeS-v2.0} package \cite{SuperBayes} which is interfaced with
SoftSUSY 3.2.7 \cite{SoftSusy} as SUSY spectrum calculator, MicrOMEGAs 2.4
\cite{MicrOMEGAs} to compute the abundance of dark matter, DarkSUSY 5.0.5
\cite{DarkSusy} for the computation of $\sigmaSI$ and $\sigmaSD$, SuperIso 3.0
\cite{SuperIso} to compute B(D) physics observables, and SusyBSG 1.5 for the
determination of $\brbsgamma$ \cite{SusyBSG}. For wino-like and
  higgsino-like LSP the Sommerfeld enhancement of the primordial neutralino
  annihilation has been computed following the lines of
  refs.~\cite{Slatyer:2009vg,Cassel:2009wt,Iengo:2009ni,Visinelli:2010vg},
  with the help of DarkSE \cite{Hryczuk:2010zi,Hryczuk:2011tq}, a package for
  DarkSusy. We created a grid in the $M_2-\mu$ plane and performed
  interpolations to correct the relic density. The \texttt{SuperBayeS-v2.0}
package uses the publicly available MultiNest v2.18
\cite{Feroz:2007kg,Feroz:2008xx} nested sampling algorithm to explore the
NUGHM model parameter space. MultiNest has been developed in such a way as to
be an extremely efficient sampler even for likelihood functions defined over a
parameter space of large dimensionality with a very complex structure as it is
the case of the NUGHM. The main purpose of the Multinest is the computation of
the Bayesian evidence and its uncertainty but it produces posterior inferences
as a by--product.  Besides, it is also able to reliably evaluate the profile
likelihood, given appropriate MultiNest settings, as demonstrated
in~\cite{Feroz:2011bj}.

Finally, for the marginalization procedure, we have used the above discussed ranges for the priors. Besides, we have considered $2 < \tan \beta < 62$. The lower limit is in fact irrelevant, as such small $\tan\beta$ value requires extremely large values of stop masses, which are drastically disfavoured by the fine-tuning penalization and the prior. The higher limit corresponds to the requirement of perturbativity of the bottom/tau Yukawa couplings.

%%%%%%%%%%%%%%%%%%%%%%%%%%%%%%%%%%%%%%%%%%%%%%%%%%%%%%%%%%%%%%%%%%
\section{Results}
%%%%%%%%%%%%%%%%%%%%%%%%%%%%%%%%%%%%%%%%%%%%%%%%%%%%%%%%%%%%%%%%%%

In this section we present the results of the Bayesian analysis of the NUGHM and examine the associated phenomenology in three different scenarios. In the first one (``Single-Component CDM") we require that the production of supersymmetric cold dark matter (CDM) is consistent with the actual DM abundance. In the second subsection we study especial regions of the parameter space which, though phenomenological viable, do not appear in the pdf plots due to their low statistical weight. In the third one (``Multi-Component CDM") we relax the assumption on DM by requiring only that the supersymmetric CDM production is equal or less than the observed abundance. Finally, in the last  scenario (`Low-Energy SUSY") we re-do the analysis with the extra requirement that SUSY is potentially detectable at LHC. This gives a forecast of the most likely ways in which SUSY may show up in the case it is really there (and consistent with the generic NUGHM scenario).

%%%%%%%%%%%%%%%%%%%%%%%%%%%%%%%%%%%%%%%%%%%%%%%%%%%%%%%%%%%%%%%%%%
\subsection{Single-Component CDM}
%%%%%%%%%%%%%%%%%%%%%%%%%%%%%%%%%%%%%%%%%%%%%%%%%%%%%%%%%%%%%%%%%%

\begin{figure}[ht]
\centering 
\includegraphics[width=0.415\linewidth]{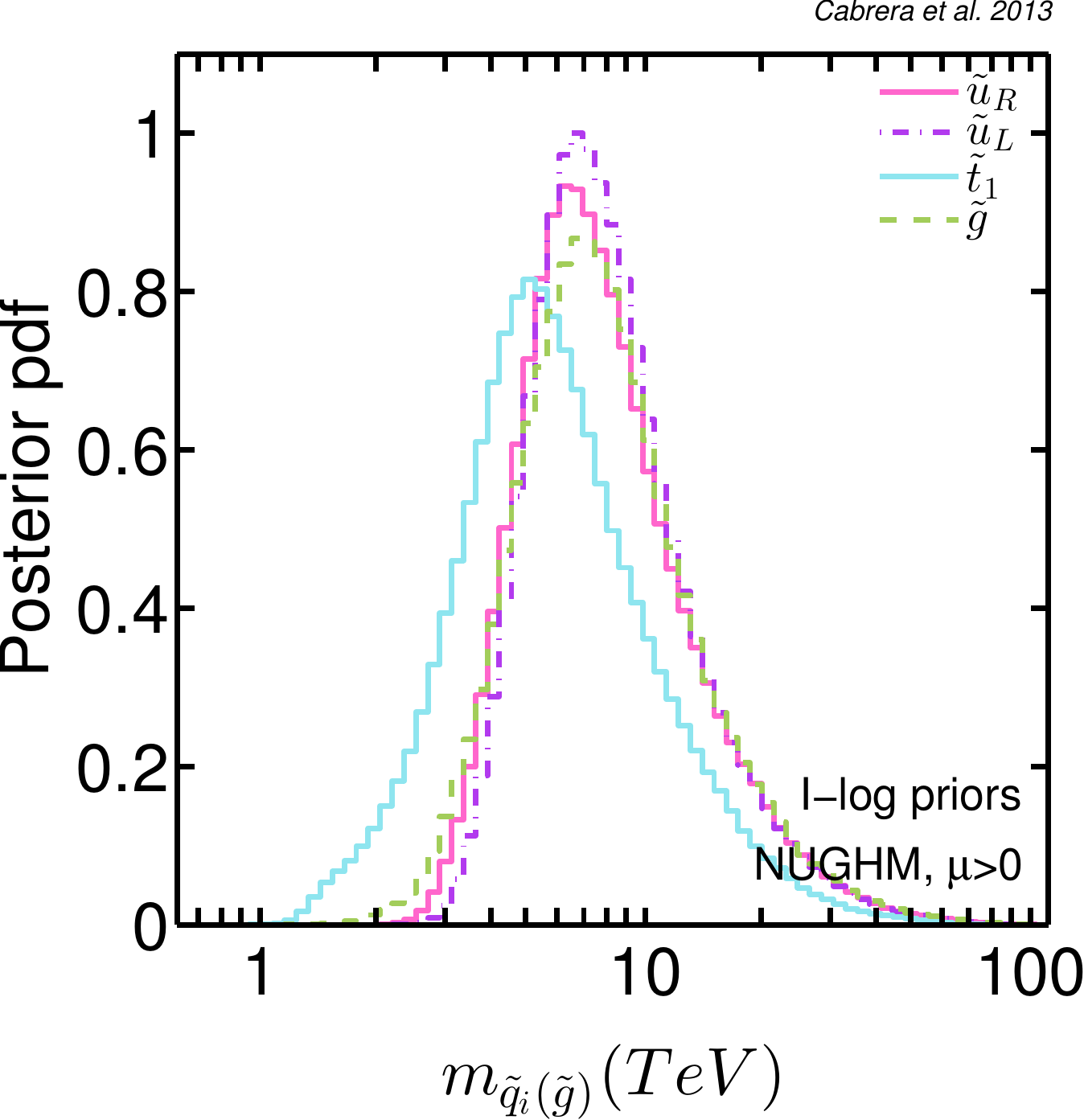}\hspace{1.0cm}
\includegraphics[width=0.4\linewidth]{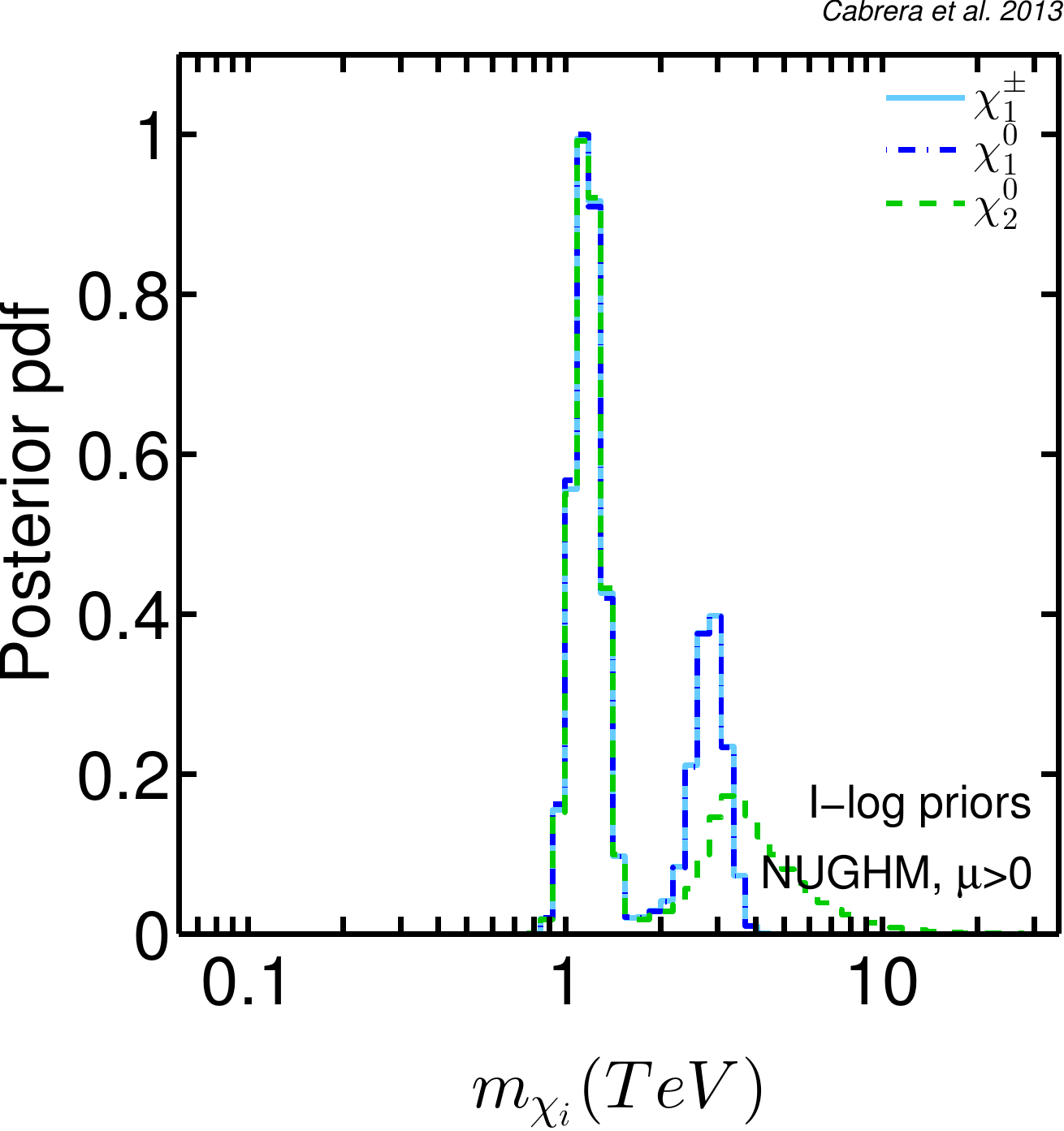}\\
\vspace{0.5cm} 
\includegraphics[width=0.415\linewidth]{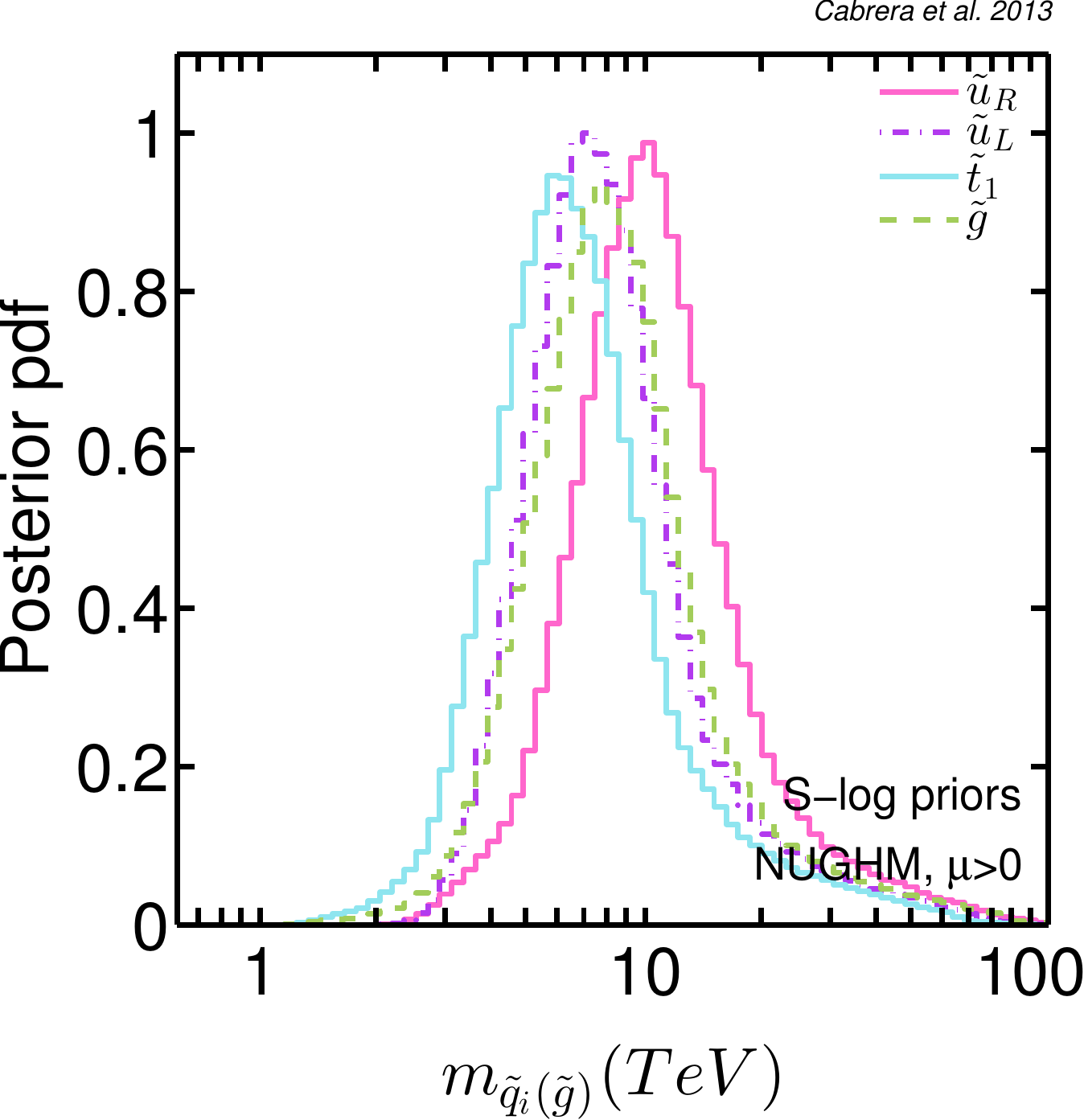}\hspace{1.0cm}
\includegraphics[width=0.4\linewidth]{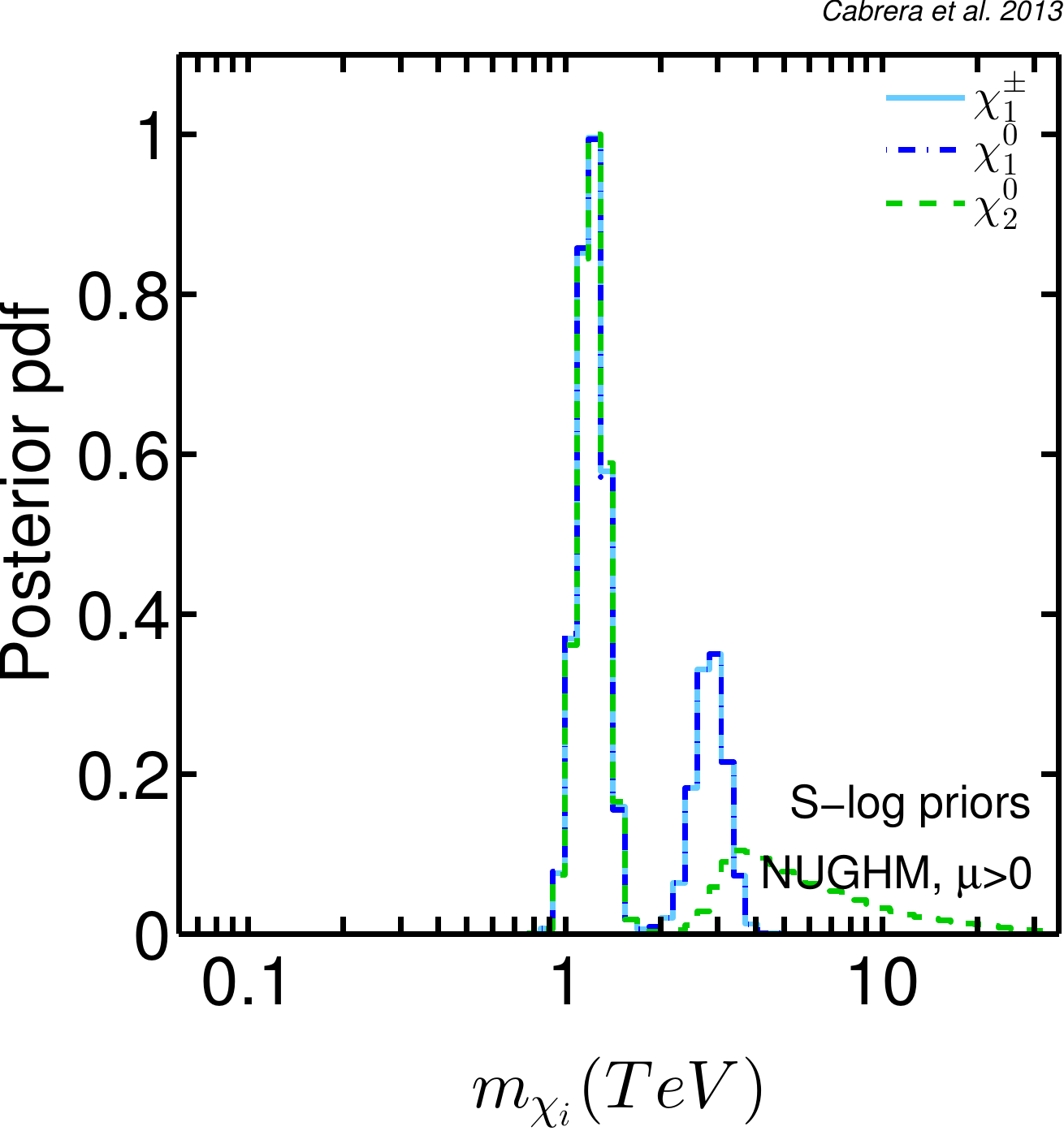}
\caption{1D marginalized posterior probability distribution of masses of supersymmetric particles for single-component CDM. Upper (lower) plots correspond to I-log (S-log) priors}
\label{fig:single1D}
\end{figure}

Let us start with the first scenario, where all the DM has a supersymmetric origin.
Fig. \ref{fig:single1D} shows the 1D pdfs (i.e. probability distribution functions or {\em posteriors}) for different physical masses using I-log and S-log priors (upper and lower plots respectively). Each curve has been obtained after the appropriate marginalization in the parameter-space. Right panels show the pdfs for
$\chi_1^0$ , $\chi_2^0$ and $\chi_1^\pm$ masses. Clearly, two sharp peaks around 1 TeV and 3 TeV are selected.
The first peak corresponds to a situation where $\chi_1^0$ is
Higgsino-like. In this case, the lightest chargino and neutralino states are
essentially determined by the value of $\mu$, which is close to 1 TeV (for the
appearance of this peak in other contexts see refs.\cite{Roszkowski:2009sm,Strege:2012bt,Cabrera:2012vu}); and thus the three states, $\chi_1^0$ , $\chi_2^0$ and $\chi_1^\pm$, are quasi-degenerate. The second peak corresponds to a wino-like $\chi_1^0$. Then, the lightest chargino, $\chi_1^\pm$, is quasi-degenerate with it and their masses are mainly determined by $M_2$. However, in this case the second neutralino, $\chi_2^0$, is not forced to be quasi-degenerate, which is reflected in its pdf. 

Before explaining the appearance of these two sharp peaks, it is worth to comment on the various possibilities to fit the DM abundance in the MSSM.
There used to be four regions in the CMSSM parameter space able lead to acceptable production of DM: bulk region, $A-$funnel region, stau (or stop) co-annihilation region and focus point region. The bulk region, which requires light sleptons, has been essentially ruled out (both in the CMSSM and the NUGHM) by the LHC limits on supersymmetric masses and the large value of the Higgs mass, which requires rather heavy stops. The $A-$funnel region occurs when the mass of the $A-$pseudoscalar is close to twice the mass of the lightest neutralino, which enables the resonant annihilation of neutralinos. This in turn requires large $\tan\beta$ (to decrease the mass of the pseudoscalar). The $A-$funnel region still survives, both in the CMSSM and the NUGHM, however it does not show up in the plots due to its small statistical weight. This comes from the fact that it requires a delicate tuning of parameters so that the $A-$mass has the appropriate value. 
The stau (stop) co-annihilation region occurs when the stau (stop) is close to lightest neutralino. Though squarks must be quite heavy (from LHC direct limits and the value of the Higgs mass), the sleptons could be much lighter, even assuming universality of sfermion masses (as happens in the CMSSM and NUGHM). The reason is that the squarks can be heavy because they get a large contribution from the gluino mass along their RG running. The corresponding contribution (from the wino and bino mass) to sleptons is much less important. In addition, in the NUGHM the wino and bino masses are not related to the gluino mass. Consequently, it is still possible to have staus light enough to enable sufficient coannihilation. Nevertheless, as for the $A-$funnel, this region does not show up either in the plots for the same reason: it has little statistical weight since it implies a fine-tuning of the parameters. However, since these regions are really there, they show up when one displays the pdfs at $99.9\%$ c.l., as it will be shown in subsect.~4.2. It is worth mentioning, that beside the $A-$funnel there exits the possibility of an standard-Higgs funnel, when the lightest neutralino is mostly bino (with a mixture of Higgsino) and its mass is close to $m_h/2$. 
This region is not viable in the CMSSM since it requires light bino mass, which in the CMSSM is correlated to the gluino mass, severely constrained by LHC data. However it is viable in the NUGHM, where the bino and gluino masses are not correlated. This region is really there but, again, its statistical weight is small and it is not visible in the plots. As we will see in subsect.~4.2, this region becomes visible when we zoom the probability distribution in the appropriate area of the parameter space. To summarize, the $A-$funnel, stau-coannihilation and the new Higgs-funnel regions are interesting but quite marginal from the point of view of their statistical weight. Their phenomenology will be examined in subsect.~4.2.

Let us now turn to the focus-point region, where the annihilation takes place
thanks to the non-vanishing Higgsino component of the lightest neutralino. It
turns out that in the mixed case, i.e. when $\chi_1^0$ is a mixture of bino
and Higgsino, this possibility has been ruled out by Xenon100 and
  LUX. However, the focus-point region where $\chi_1^0$ is an almost pure
Higgsino survives.

In addition, in the NUGHM there appears a new possibility, namely that
$\chi_1^0$ is an almost pure wino, which allows for an efficient annihilation
as well. In that case both the primordial and present-day annihilation
  of DM is further enhanced by the Sommerfeld effect, see
  refs.~\cite{Slatyer:2009vg,Cassel:2009wt,Iengo:2009ni,Visinelli:2010vg,Cirelli:2007xd,Cirelli:2008jk}. Actually,
  the bounds from indirect-detection searches give quite stringent bounds;
  e.g. for NFW and Einasto distributions the thermal {\em pure} wino case
  could be excluded by
  H.E.S.S. \cite{Cohen:2013ama,Fan:2013faa,Hryczuk:2014hpa}. In our case, the 
  results for the wino-like situation are consistent with the latter. The only
  noticeable difference is that now the mass splitting between $\chi_1^0$ and
  $\chi_1^\pm$, which plays a crucial role for the size of the enhancement,
  gets modified by the possible small Higgsino component of $\chi_1^0$ and the
  additional radiative corrections. This results in a range of variation of
  about one order of magnitude in the possible annihilation cross-section,
  $\langle \sigma v\rangle$ for a given $M_{\chi_1^0}$. The 95\% CL allowed
  wino-like region corresponds to values of $\langle \sigma v\rangle$ in the
  $10^{-22.3}-10^{-24.8}$ ($cm^3 s^{-1}$) range. These cross-sections are very
  close to the present limits on indirect-detection or even above them,
  depending on the assumed DM profile. On the other hand, these conclusions
  have to be taken with a grain of salt, due to the large uncertainty about
  the DM profile in the inner galaxy and the computation of the Sommerfeld
  effect \cite{Hryczuk:2011vi}.

 Notice that
wino-like DM is not viable in the CMSSM, since due to gaugino universality at
high-scale, the bino is always lighter than the wino. The reason for the
survival of these two regions to the XENON100 limits is that the dominant
diagram for the spin-independent cross section section occurs via
Higgs-interchange in $t$-channel. Then the relevant vertex of the neutralino
is Higgsino-Higgs-Wino(Bino). Consequently, the purer the $\chi_1^0$ state,
the smaller spin-independent cross section, which can thus become consistent
with XENON100 limits.  Note that for a pure Higgsino or wino $\chi_1^0$, the
relic abundance depends on one single parameter, the mass of the neutralino
\cite{ArkaniHamed:2006mb}, which explains the two sharp peaks in the
$\chi_1^0$ instead of wide distributions\footnote{The co-annihilation
  processes with the quasi-degenerate $\chi_1^\pm$ chargino and (for the case
  of pure Higgsino) the quasi-degenerate $\chi_2^0$ neutralino play also a
  relevant role in the determination of the DM abundance; but the mass of
  these quasi-degenerate states is determined by the same parameter
  responsible for the $\chi_1^0$ mass (either $\mu$ or $M_2$ for the
  pure-Higgsino or -wino cases)}. Notice that these results are quite prior
independent, as they are mainly driven by experimental data.

The left panels of Fig. \ref{fig:single1D} show the posterior pdfs for the
squarks and gluino masses. Here is some prior-dependence. Note that 
the splitting
between left and right squarks is larger for S-log priors than for I-log priors. This is mainly due to RG running effects. From $M_X$ down to low scale, the mass of $\tilde{u}_L$ grows with $M_1$, $M_2$ and $M_3$, whereas
the one of $\tilde{u}_R$ grows with $M_1$ (with a different factor from that of $\tilde{u}_L$) and $M_3$. This means that large values
of $M_2$ and $M_1$ favor the splitting between left and right
squarks (though in opposite directions). Of course, increasing $M_2$ or $M_1$ much, amounts to a prior-price, which is much higher for I-log priors whenever one of these parameters gets so large that it becomes the largest soft parameter. As commented in sect.~3, S-log priors are much more tolerant to very large values of some initial parameters, as they can be compensated by the smallness of others. Consequently, for I-log priors the running of the squarks is dominated by $M_3$ and, hence, all of them have similar masses, except the stops and the left-sbottom due to the contribution of the top-Yukawa coupling. They are also similar to the gluino mass.  
The fact that the peak for Higgsino-like $\chi_1^0$ is higher than that for wino-like $\chi_1^0$ comes from the fact the prior penalizes higher supersymmetric masses, see subsects.  3.2, 3.3.% \red{NOTE: A SENTENCE HAS BEEN DELETED}

%The slightly higher peak for Higgsino-like $\chi_1^0$ than for wino-like $\chi_1^0$ in S-log priors with respect to I-log priors has a similar origin: since for S-log priors $M_1$ or/and $M_2$ can be larger than for I-log ones, this gives a volume effect when they are marginalized.

\begin{figure}[ht]
\centering 
\includegraphics[width=0.3\linewidth]{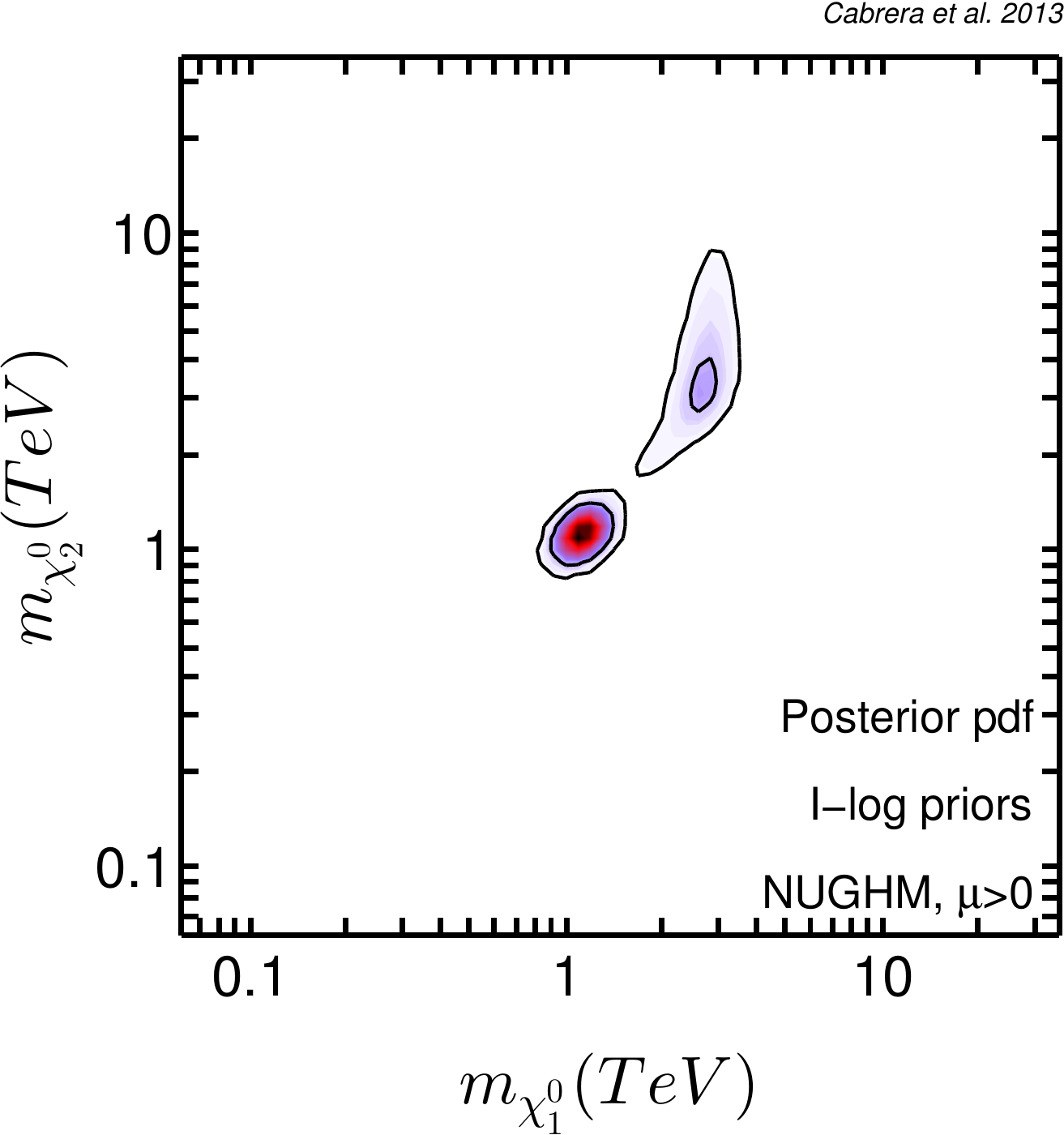}\hspace{0.5cm}
\includegraphics[width=0.3\linewidth]{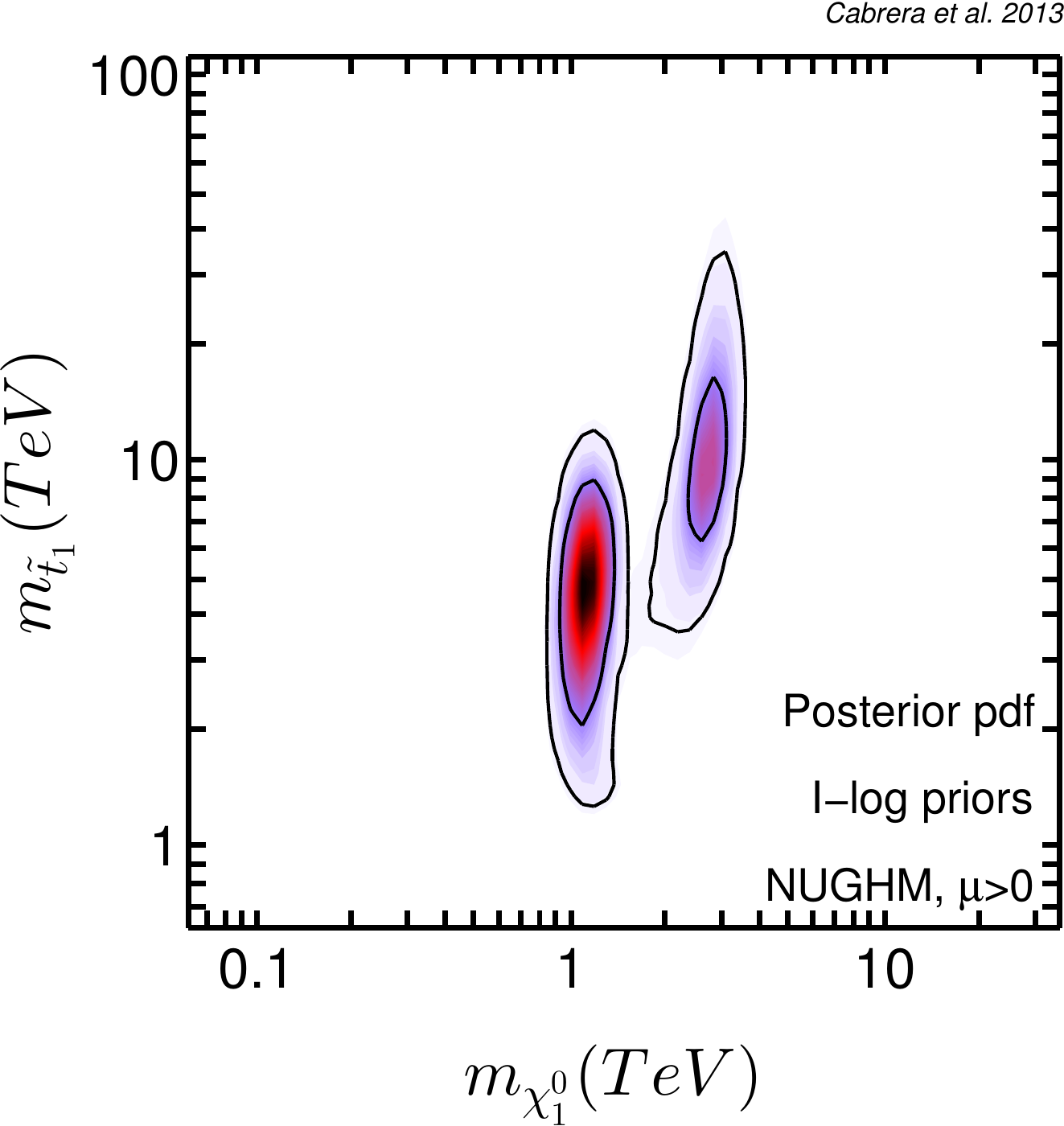}\hspace{0.5cm}
\includegraphics[width=0.3\linewidth]{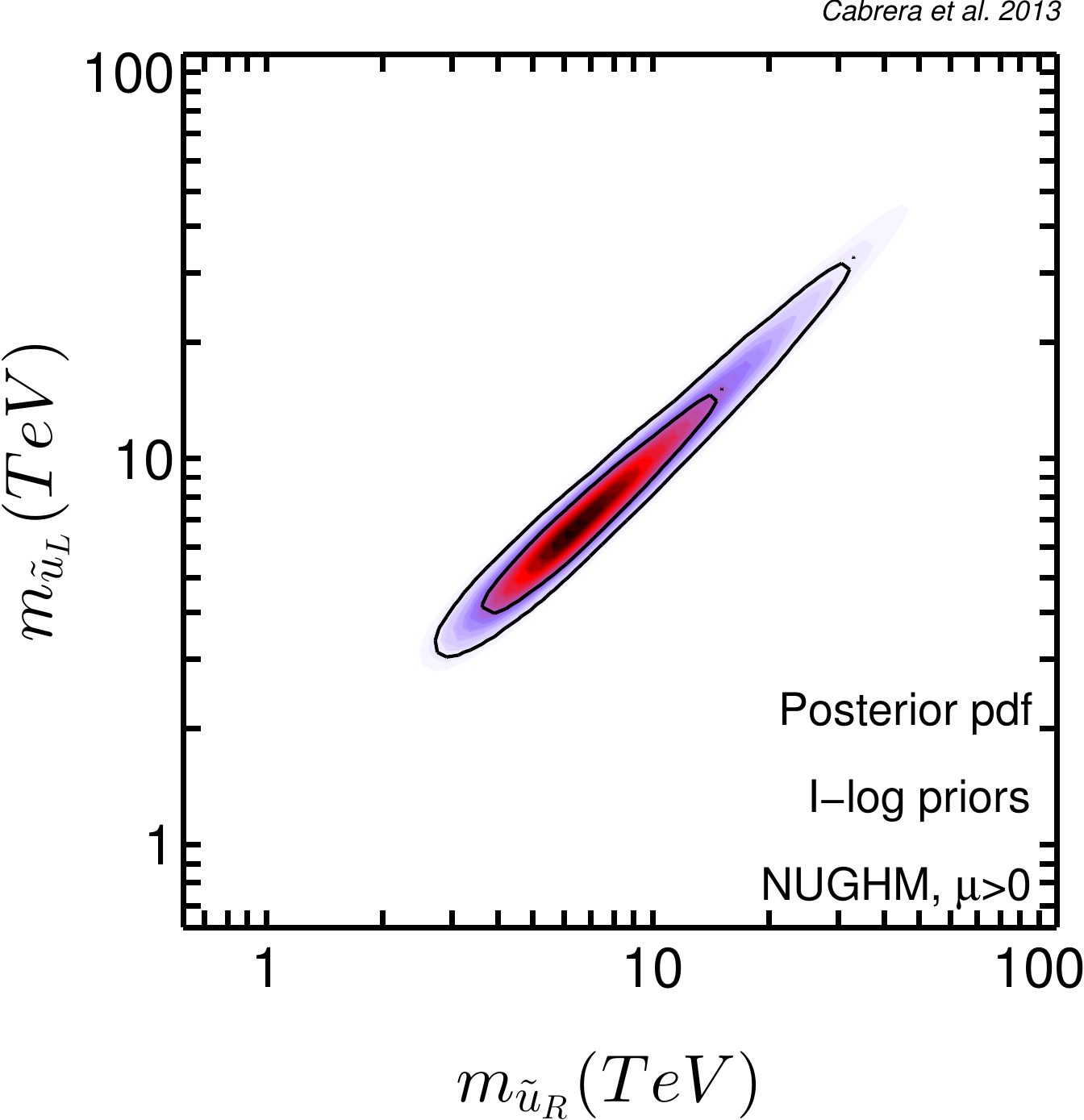}\\
\vspace{0.3cm}
\includegraphics[width=0.3\linewidth]{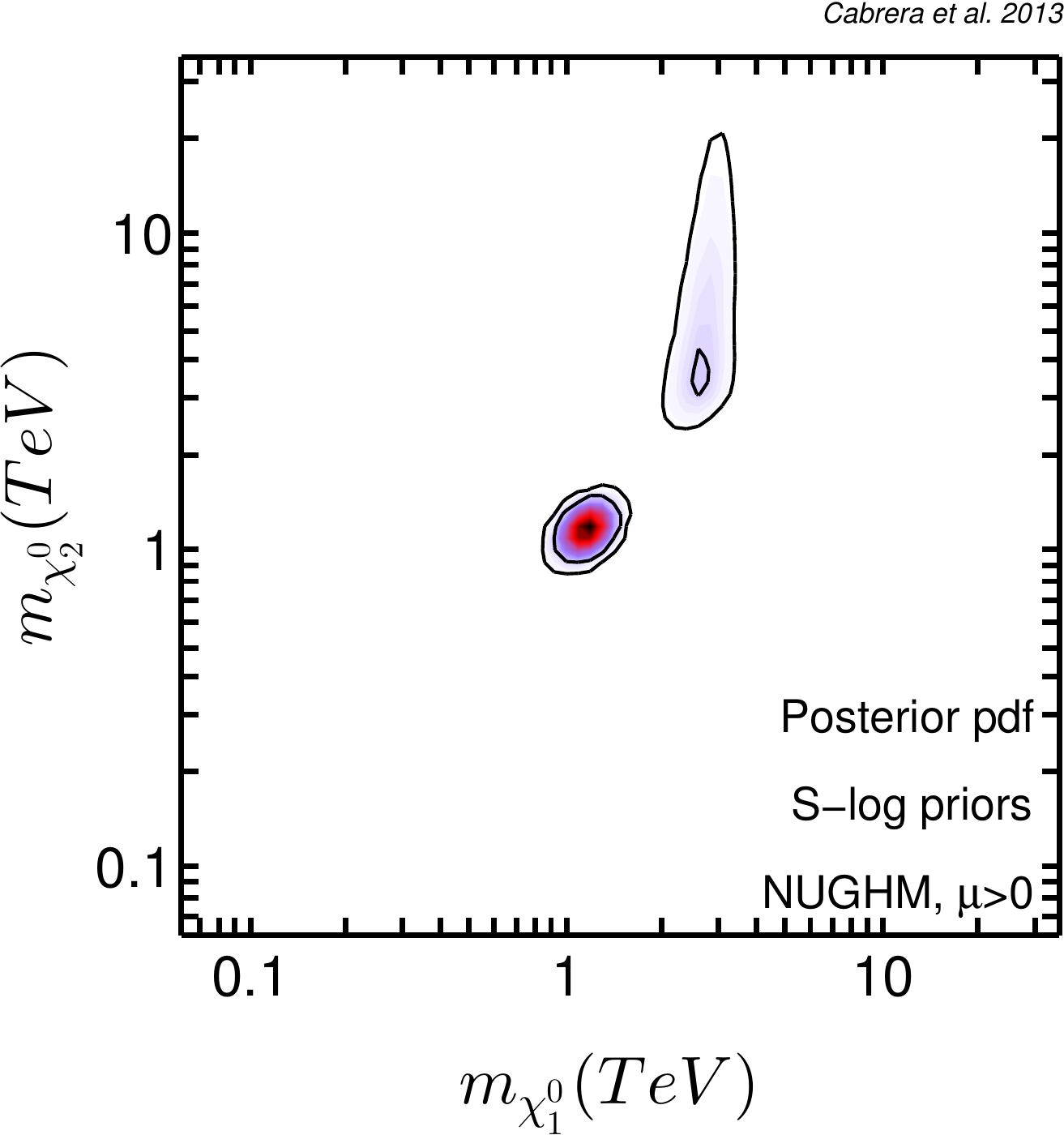}\hspace{0.5cm}
\includegraphics[width=0.3\linewidth]{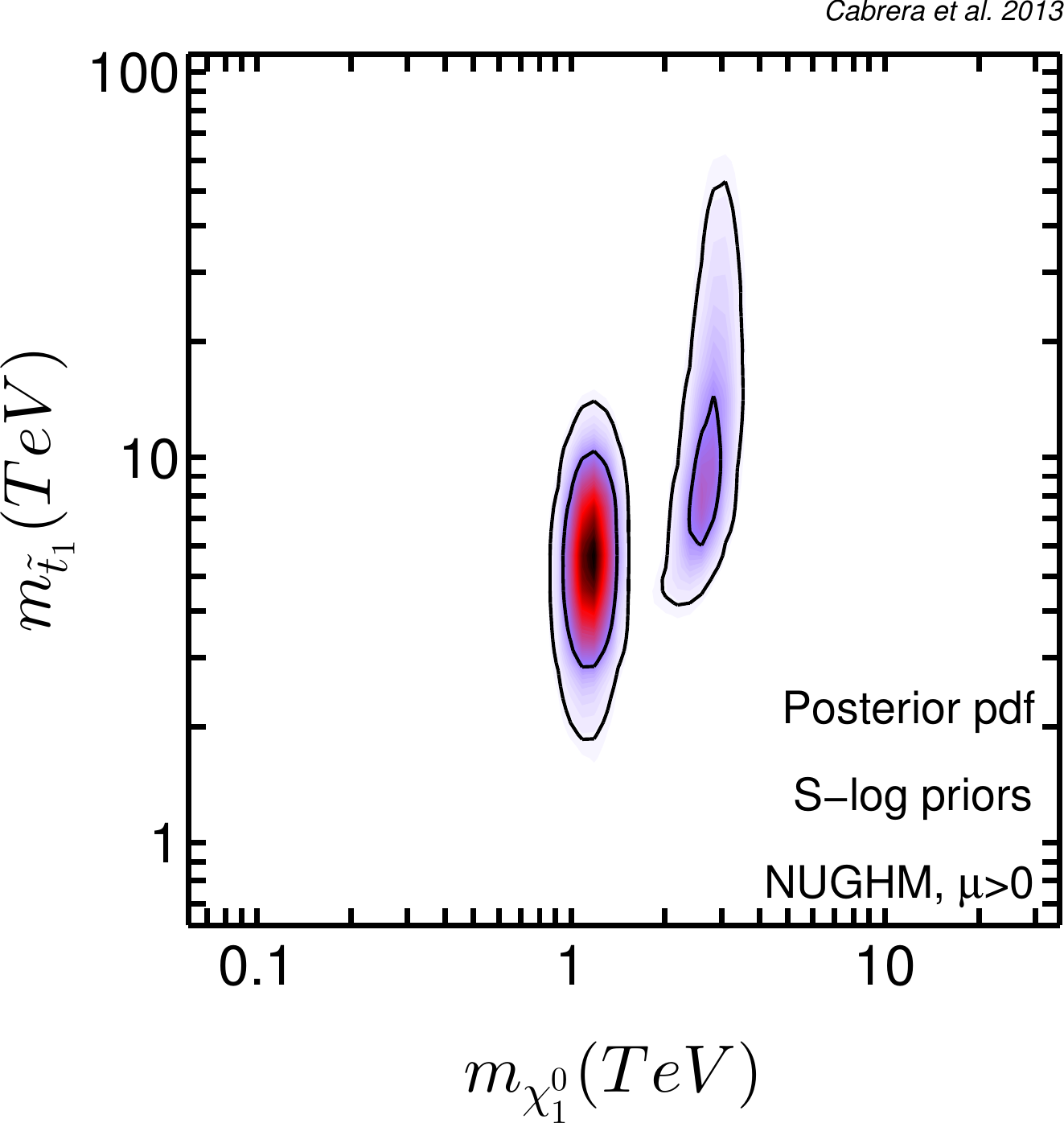}\hspace{0.5cm}
\includegraphics[width=0.3\linewidth]{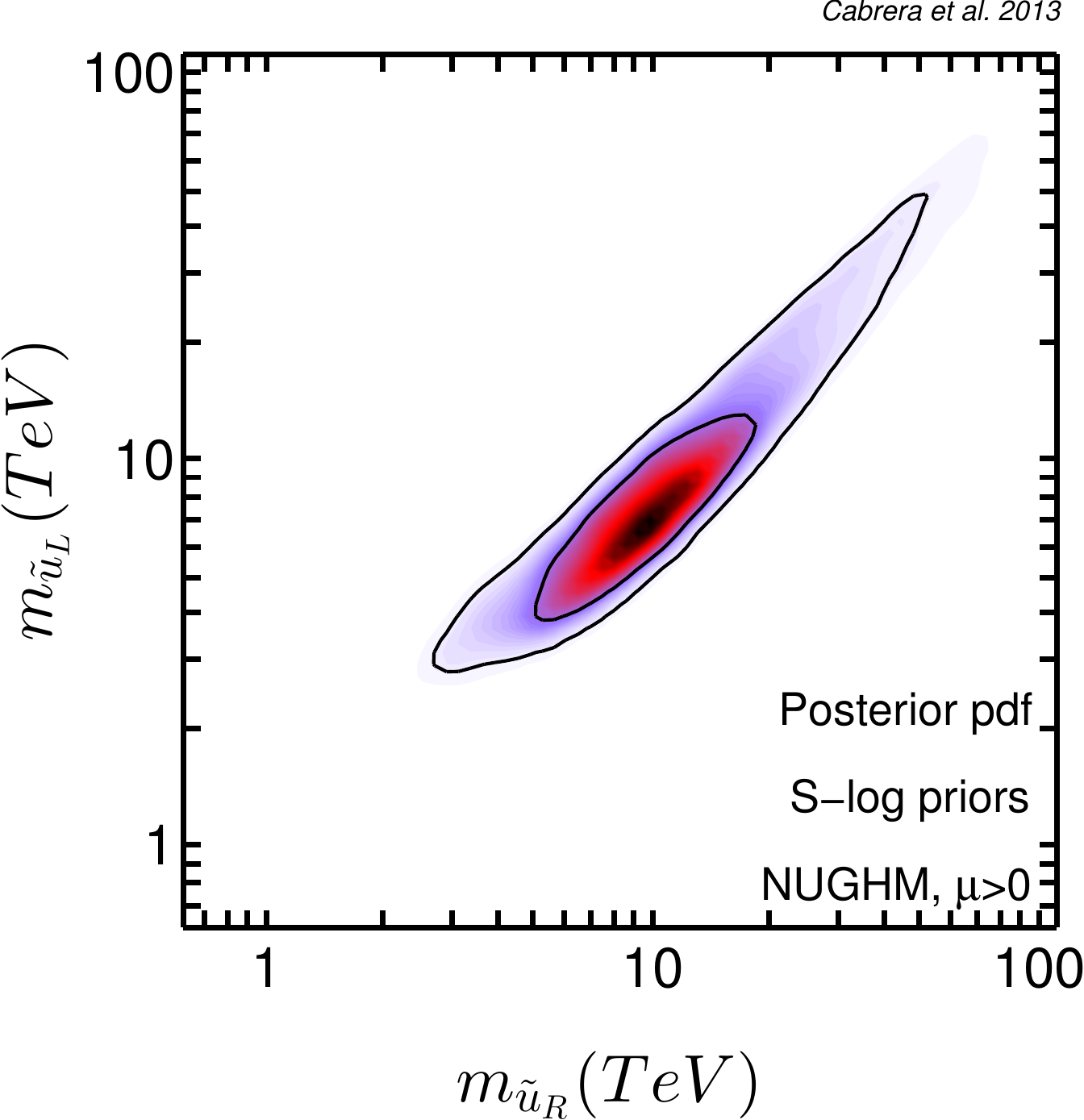}\\
\vspace{0.4cm}
\includegraphics[width=0.35\linewidth]{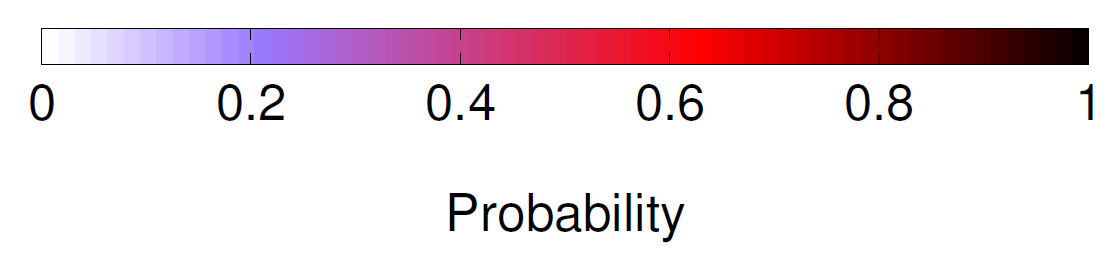}
\caption{2D marginalized posterior probability for single-component CDM on different planes defined by couples of supersymmetric masses. Upper (lower) panels correspond to I-log (S-log) priors. The color code for the probability density (normalized to 1), and 68\% and 95\% CL contours are shown.}
\label{fig:single2D}
\end{figure}

Fig. \ref{fig:single2D} shows the 2D posterior pdf
for $m_{\chi_1^0}-m_{\chi_2^0}$ (left panels),
$m_{\chi_1^0}-m_{\tilde{t}_1}$ (center panels) and $m_{\tilde{u}_L}-m_{\tilde{u}_R}$ (right panels) for S-log (upper panels) and I-log (lower panels) priors. The favored sharp region around $m_{\chi_1^0}\simeq 1$ TeV corresponds to the case where $\chi_1^0$  is Higgsino-like, thus its quasi-degeneracy with $\chi_2^0$. As discussed above and it is clear from the plots, this quasi-degeneracy is broken for the other sharp favoured region, around $m_{\chi_1^0}\simeq 3$ TeV, which corresponds to the case where $\chi_1^0$ is wino-like. The center plots are useful to see the preferred
ranges of $m_{\tilde{t}_1}$ for the Higgsino-like and the wino-like cases. Likewise, the right plots show the preferred ranges 
for $m_{\tilde{u}_L}$ and $m_{\tilde{u}_R}$. In all cases, the lower limits of the preferred ranges are at $1-2$ TeV.

\begin{figure}[ht]
\centering 
\includegraphics[width=0.4\linewidth]{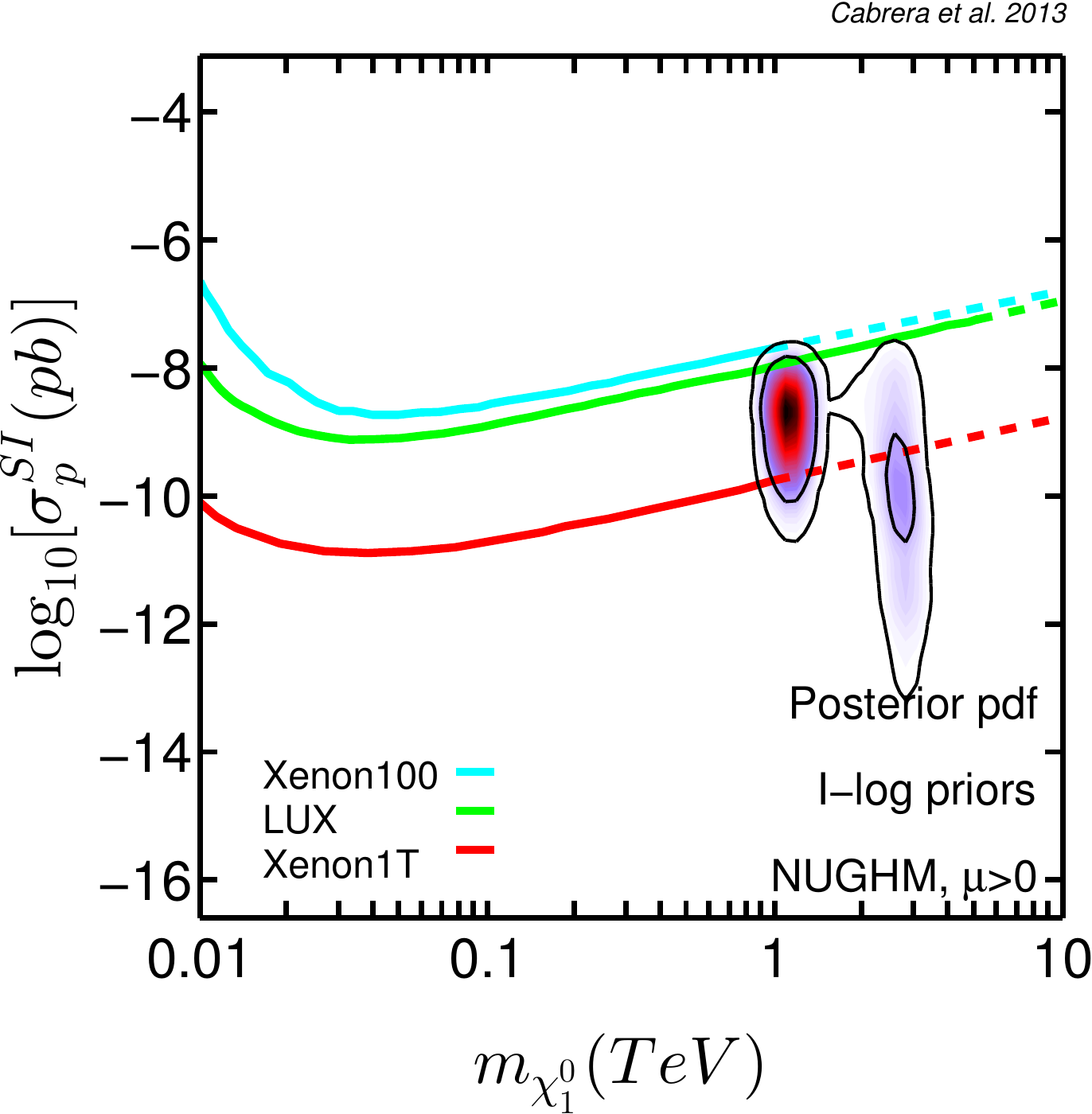}\hspace{1.0cm}
\includegraphics[width=0.4\linewidth]{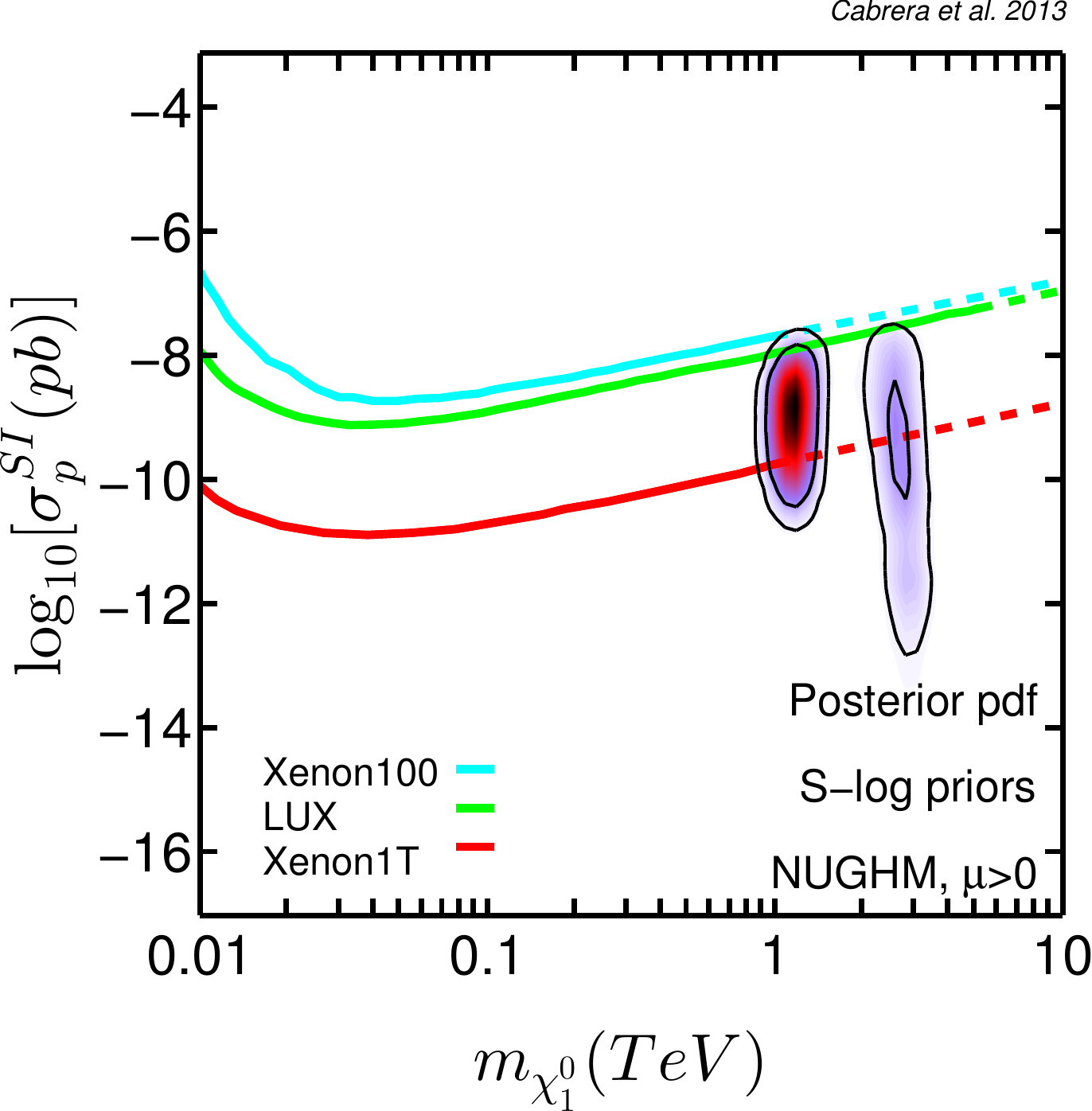}
\caption{2D marginalized posterior probability distribution in the $m_{\chi_1^0}-\sigma^{SI}$ plane for single-component CDM. The left (right) panel corresponds to I-log (S-log) priors. The
  contours enclose respective 68\% and 95\% joint regions. Present and future observational limits are shown. The color code is as in Fig.~\ref{fig:single2D}.}
\label{fig:singleCDM2D}
\end{figure}

Figure \ref{fig:singleCDM2D} shows the pdf in the $m_{\chi_1^0}-\sigma^{SI}$ plane, where $\sigma^{SI}$ is the Spin--Independent cross section for DM direct detection. As already mentioned, the dominant diagram for direct detection occurs via Higgs-interchange in $t$-channel, where the neutralino coupling to the Higgs is a Higgsino-Higgs-wino(bino) vertex. So, for a given $m_{\chi_1^0}$, the purer the ${\chi}_1^0$ state the smaller $\sigma^{SI}$. Beside the XENON100 limit used in the analysis, we have included in the figure the recent limit obtained by LUX \cite{Akerib:2013tjd} (which almost does not probe further the scenario) and the future XENON1T limit \cite{XENON1T} . Remarkably, the latter will probe a substantial fraction of the viable parameter space: 71.6\% (77.4\%) of the total probability for I-log (S-log) priors. Hence, DM searches become not only complementary to LHC searches of SUSY, but even stronger if the dark matter has supersymmetric origin.

In summary, under the assumption of single-component DM, the paramater space of the NUGHM has two well-defined preferred regions around $m_{\chi_1^0}\simeq 1$ TeV and $m_{\chi_1^0}\simeq 3$ TeV, which correspond to Higgsino-like and wino-like $\chi_1^0$ respectively. The colored sector is typically above 3 TeV and the prospects for detection at LHC are quite pessimistic. In contrast, future XENON1T and similar experiments will be able to probe most of the parameter space of the model.

%%%%%%%%%%%%%%%%%%%%%%%%%%%%%%%%%%%%%%%%%%%%%%%%%%%%%%%%%%%%%%%%%%
\subsection{The Higgs-funnel, the A-funnel and the stau-co-annihilation regions\label{sec:HFAFC}}
%%%%%%%%%%%%%%%%%%%%%%%%%%%%%%%%%%%%%%%%%%%%%%%%%%%%%%%%%%%%%%%%%%

As commented above, besides the regions examined in the previous subsection, there exist other viable regions in the NUGHM parameter-space, compatible with the DM constraints, though they have much less statistical weight. In particular, there are Higgs-funnel and $A-$funnel regions, where the annihilation of the LSPs occurs via resonant production of the SM-like Higgs or the supersymmetric $A-$pseudoscalar respectively; and the neutralino-stau co-annihilation region. We consider them below in order. 

%%%%%%%%%%%%%%%%%%%%%%%%%%%%%%%%%%%%%%%%%%%%%%%%%%%%%%%%%%%%%%%%%%
\subsubsection{The Higgs-funnel}
%%%%%%%%%%%%%%%%%%%%%%%%%%%%%%%%%%%%%%%%%%%%%%%%%%%%%%%%%%%%%%%%%%

It is worth-noticing that the possibility of the Higgs-funnel implies that the LSP is mostly bino (if it is wino or Higgsino the lightest chargino would be much lighter than the LEP bound). Consequently, this instance is not  viable in the CMSSM, since it  would imply a far too-light gluino, excluded by LHC. However, once one allows gaugino non-universality, the bino and gluino masses are not linked anymore and the Higgs-funnel becomes viable. Nevertheless, its statistical weight is so small that it is not visible in the pdf plots of the previous subsection. In order to see it we have to zoom into the relevant region, the one with $m_{{\chi}_1^0}$ not far from $m_h/2$. 

\begin{figure}[ht]
  \centering
  \includegraphics[width=0.415\linewidth]{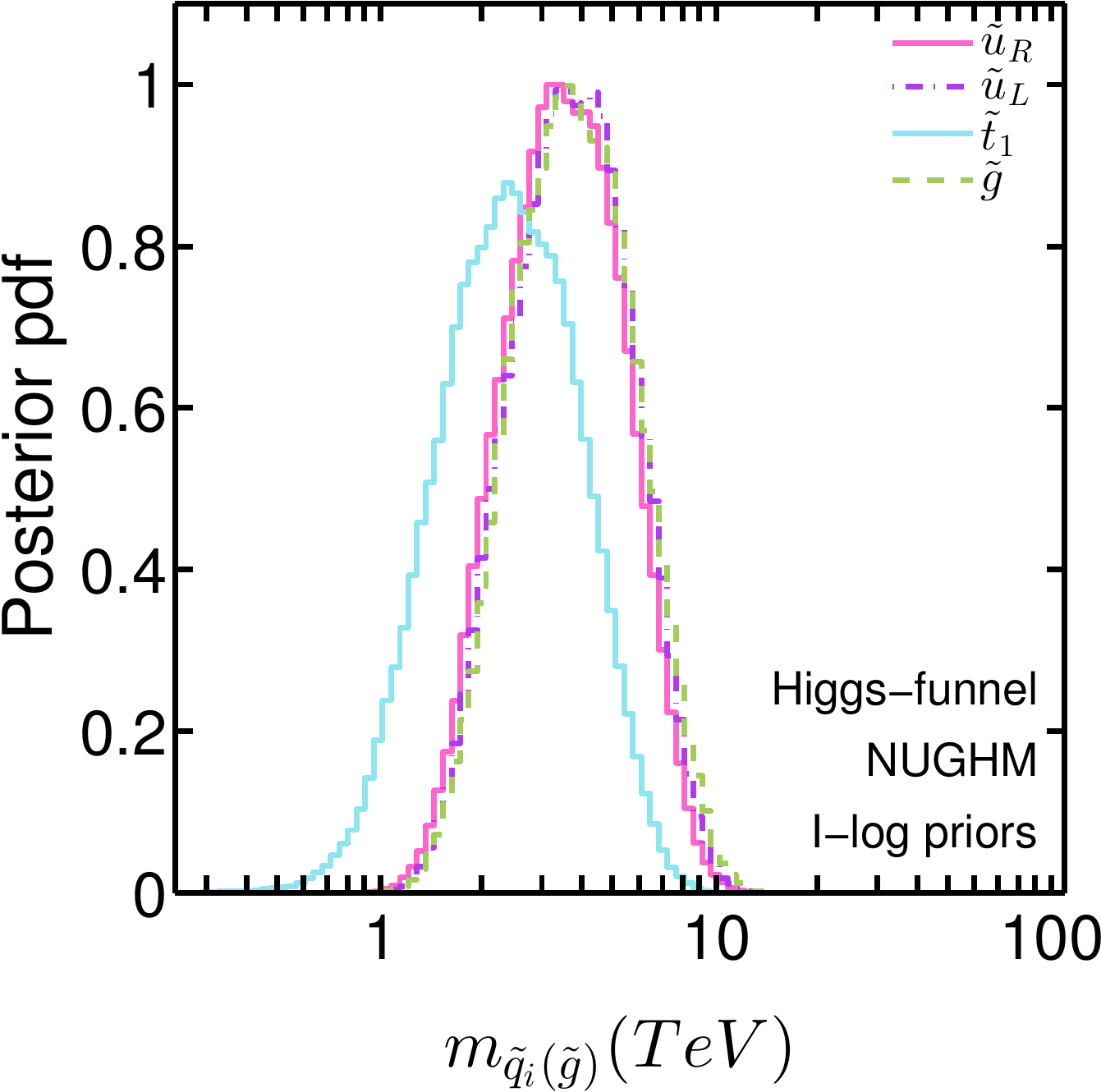}\hspace{1.0cm}
  \includegraphics[width=0.4\linewidth]{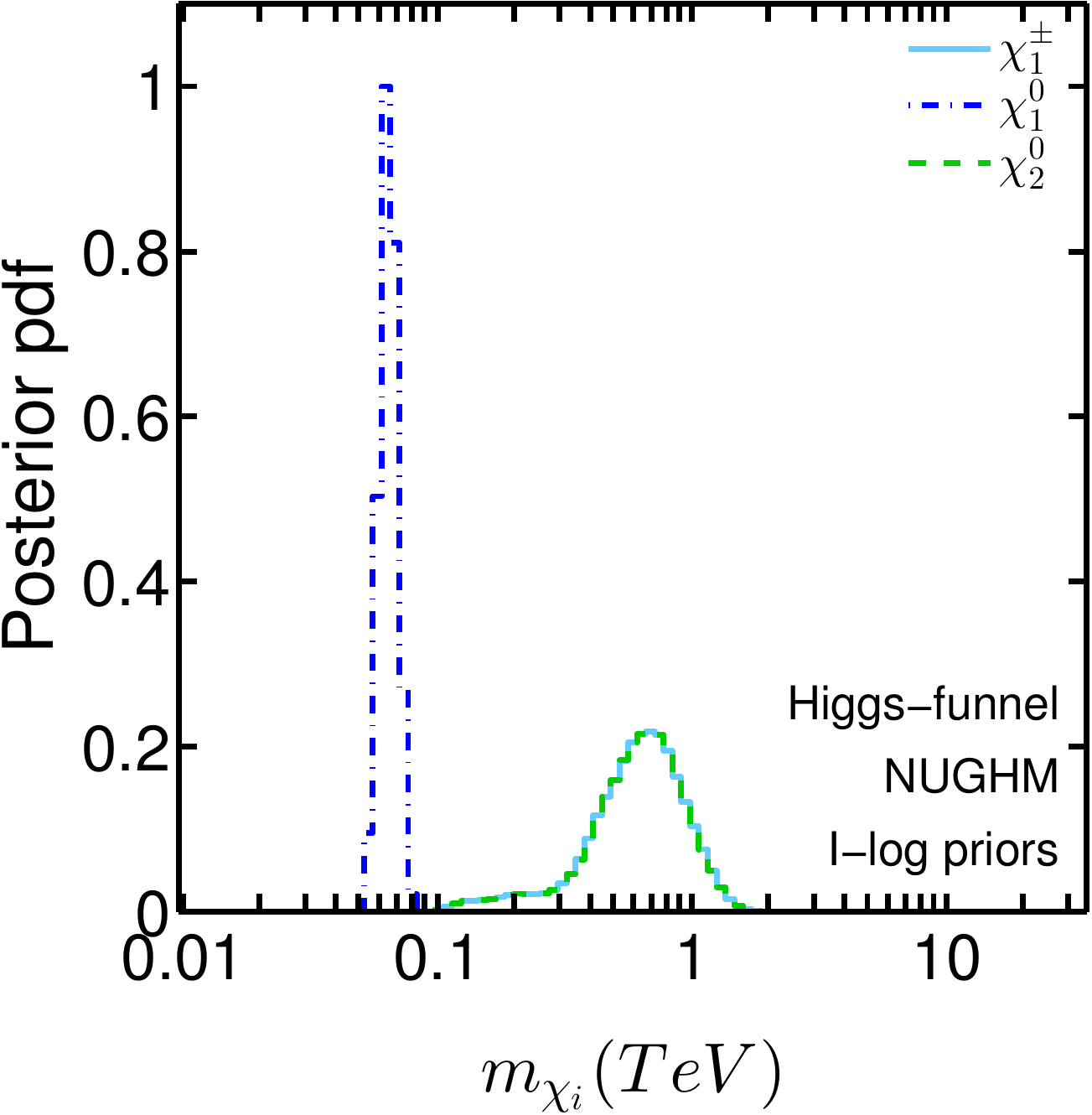}\\
  \vspace{0.5cm} 
  \includegraphics[width=0.415\linewidth]{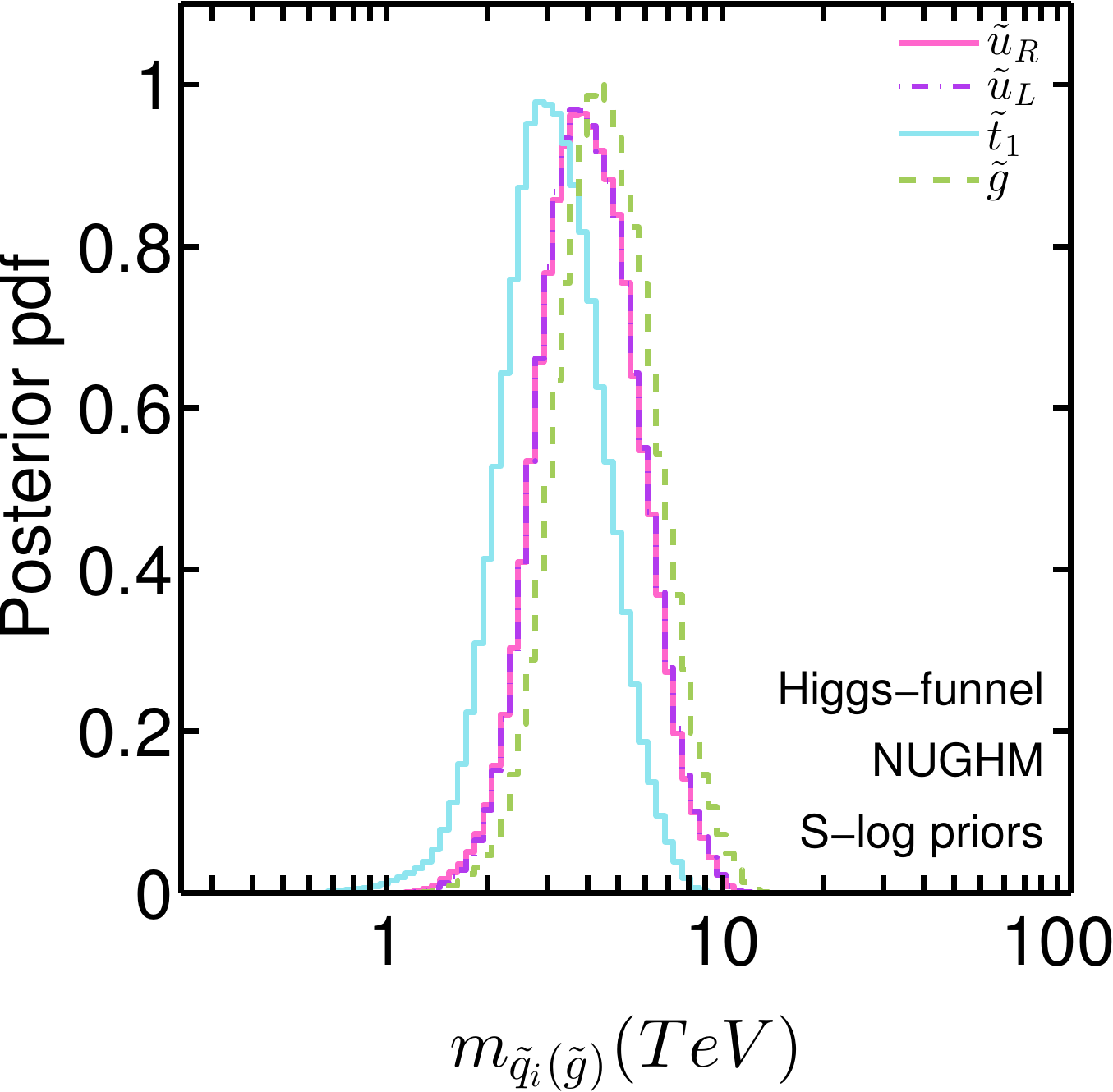}\hspace{1.0cm}
  \includegraphics[width=0.4\linewidth]{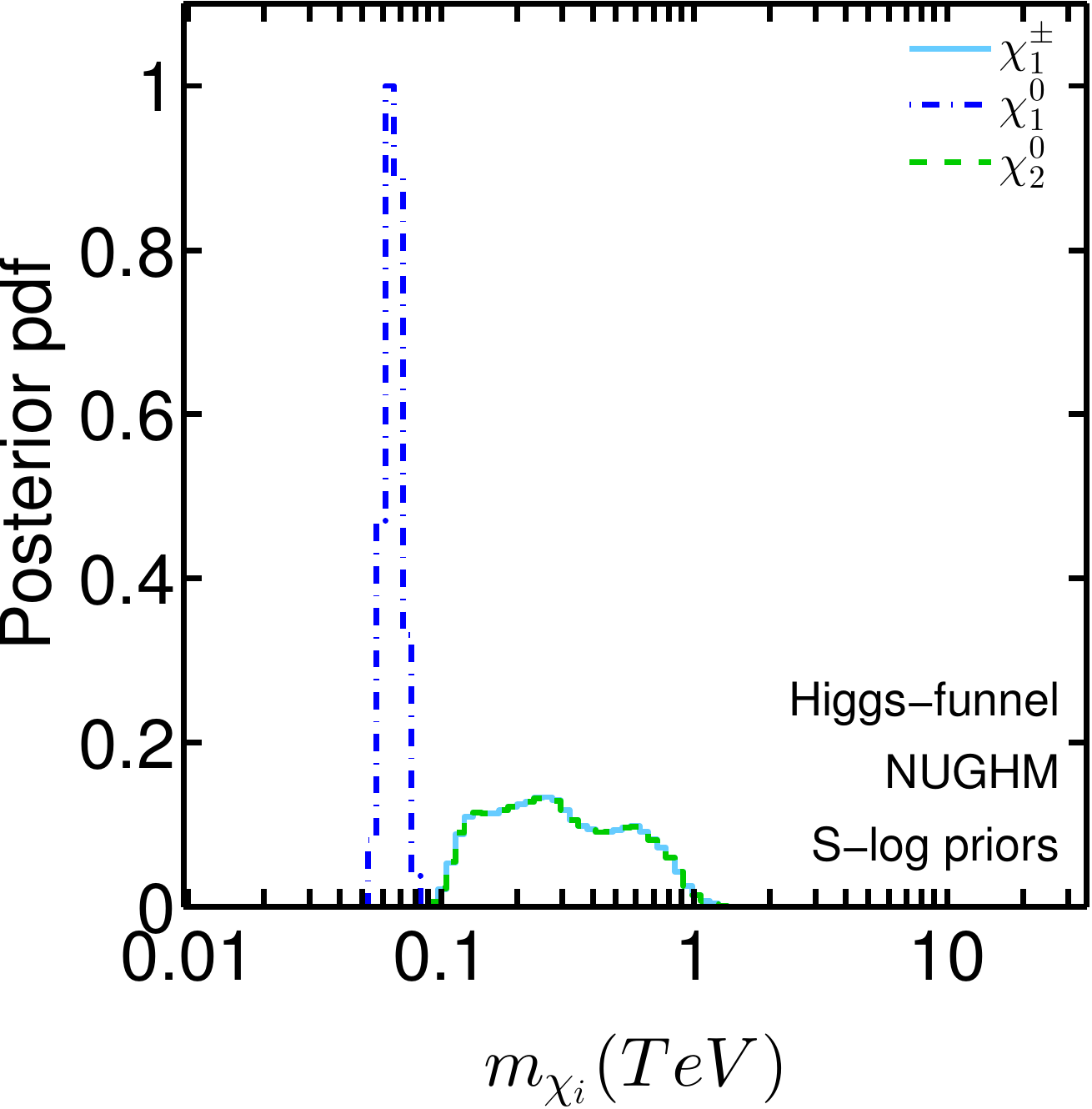}
  \caption{1D marginalized posterior probability distribution of different supersymmetric masses in the Higgs-funnnel region. Upper (lower) panels correspond to I-log (S-log) priors.}
  \label{fig:low_M1_Susy1D}
\end{figure}

Fig.~\ref{fig:low_M1_Susy1D} shows the pdfs of the various supersymmetric masses, for $|m_{{\chi}_1^0}-m_h/2|\leq 5$ GeV. One can clearly see the bump in the pdf of $m_{{\chi}_1^0}$, and the fact that $\chi_2^0$ and $\chi_1^\pm$ are quasi-degenerate, as they are either winos or Higgsinos. 

\begin{figure}[ht]
\centering 
\includegraphics[width=0.4\linewidth]{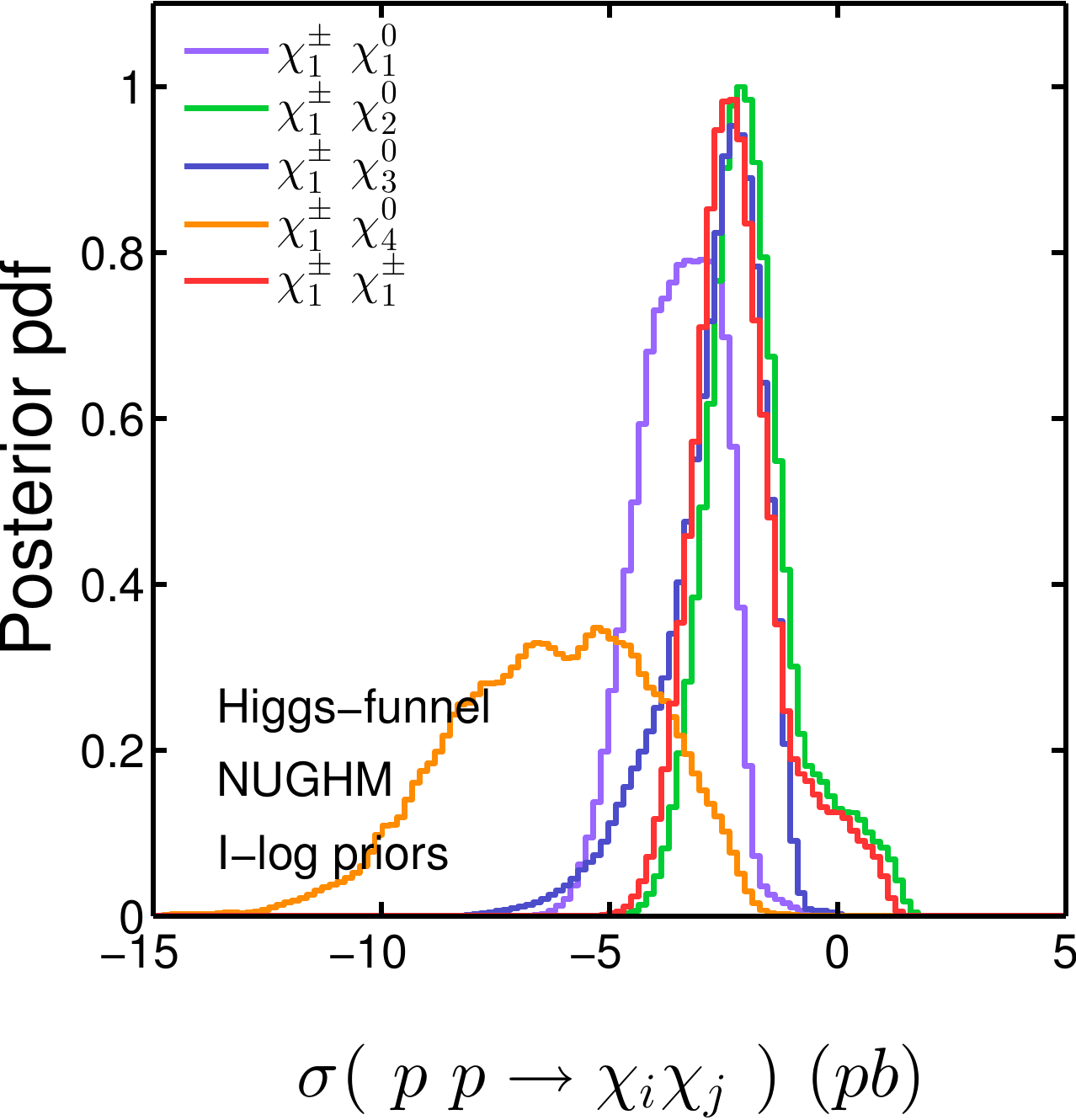}\hspace{0.5cm} 
\includegraphics[width=0.4\linewidth]{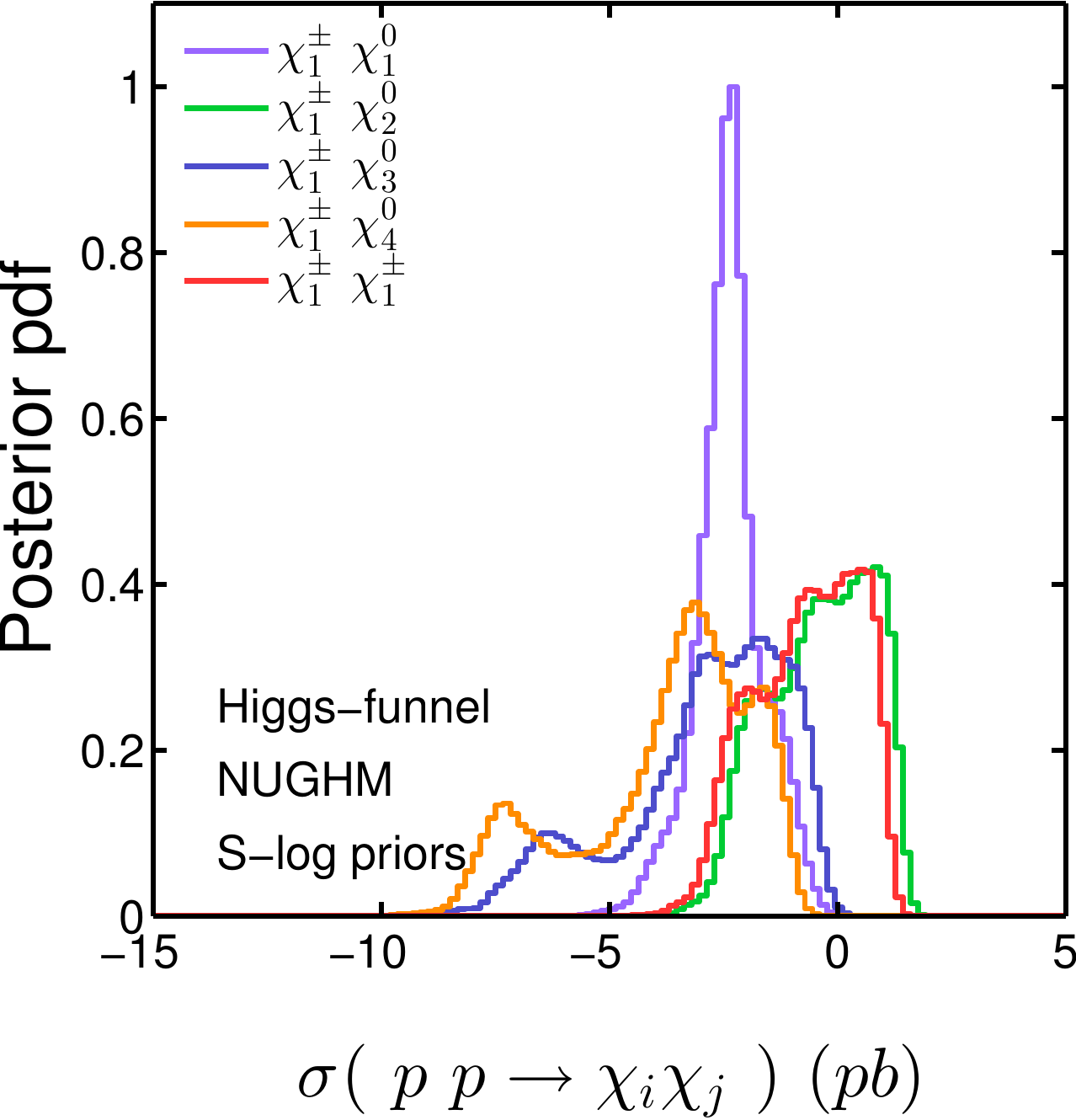}
\caption{1D posterior probability distribution of production cross-section of different electroweakino pairs at $\sqrt{s}=14$ TeV for
  the Higgs-funnel region.}
\label{fig:low_M1_sigma1}
%\label{fig:}
\end{figure}

\begin{figure}[h]
\centering 
\includegraphics[width=0.4\linewidth]{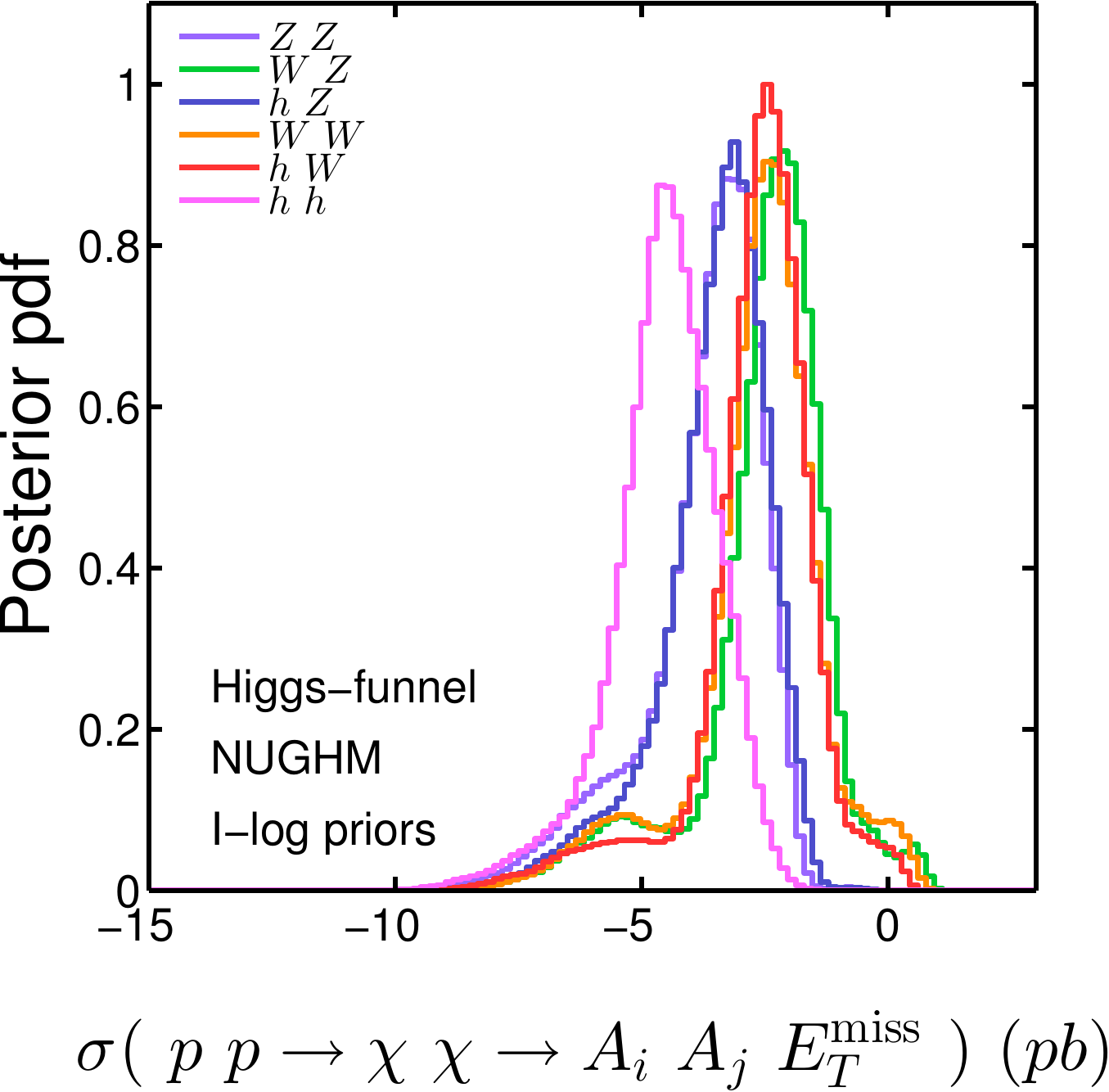}\hspace{0.5cm} 
\includegraphics[width=0.4\linewidth]{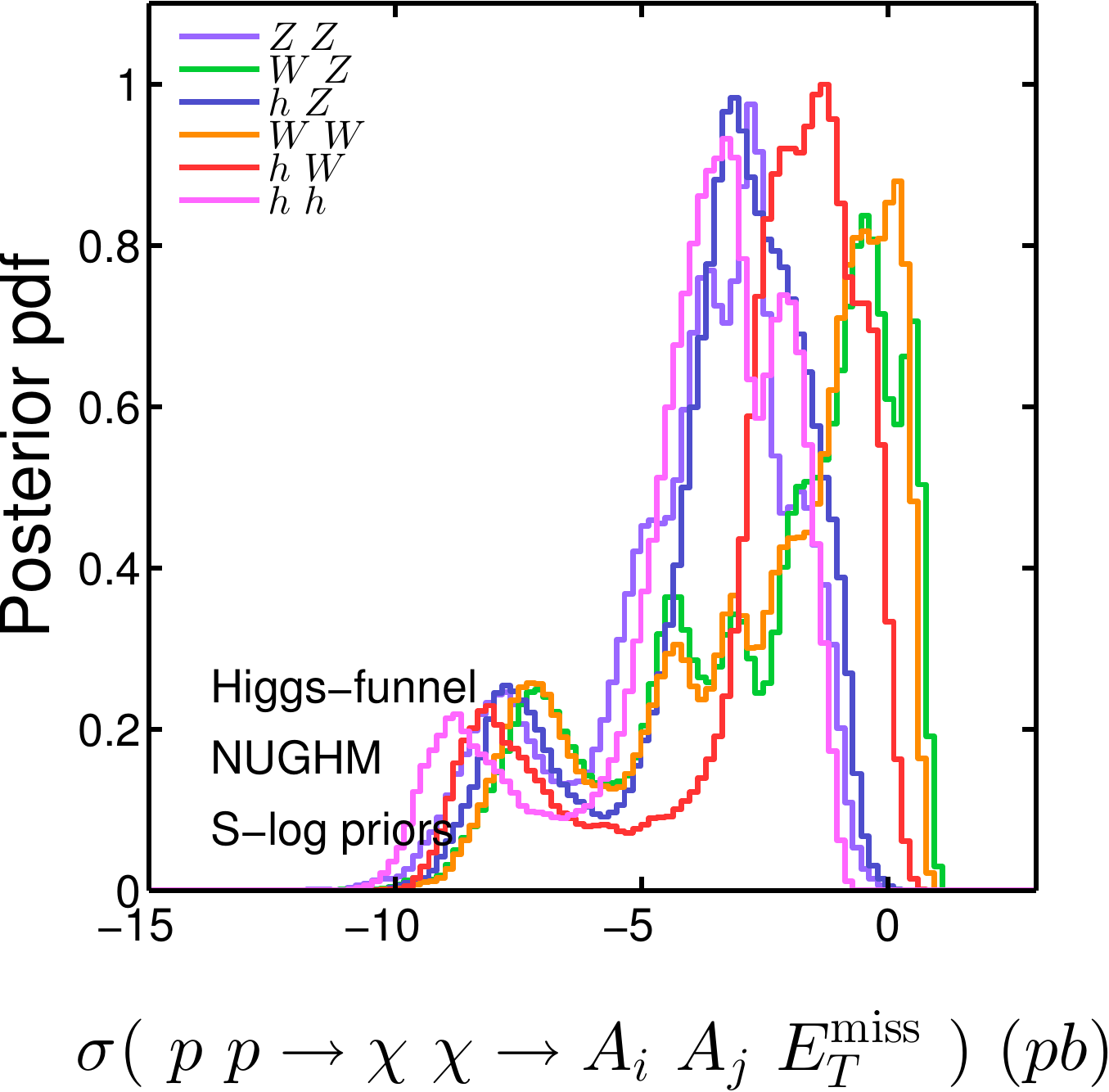}
\caption{1D posterior probability distribution of production cross section $\times$ branching ratios into different final states at $\sqrt{s}=14$ TeV for
  the Higgs-funnel region.}
\label{fig:low_M1_sigma2}
%\label{fig:}
\end{figure}

Let us consider now the phenomenology of this scenario at LHC. Fig.~\ref{fig:low_M1_sigma1} shows the probability of production of the most relevant pairs of electroweakinos. As usual, the production of $\chi_1^+\ \chi_1^-$ and $\chi_2^0\ \chi_1^\pm$ pairs is the most abundant one. Since in this case these states are not quasi-degenerate with $\chi_1^0$, they produce potentially-visible signals at LHC. After multiplying by the corresponding branching ratios (see Fig.~\ref{fig:BRNC} below),  we obtain the probability of electroweakino-mediated production of different final states, which is shown in Fig.~\ref{fig:low_M1_sigma2}. The most relevant production is $WZ$ and $WW$, which can yield a detectable signal in a substantial part of the parameter space. Actually, the present bounds on tri-lepton + missing energy \cite{ATLAS-CONF-2013-035, CMS-PAS-SUS-13-006} and same-sign di-lepton + missing energy \cite{ATLAS-CONF-2013-049,CMS-PAS-SUS-13-013} are likely to discard part of the models.

\begin{figure}[ht]
\centering 
\includegraphics[width=0.4\linewidth]{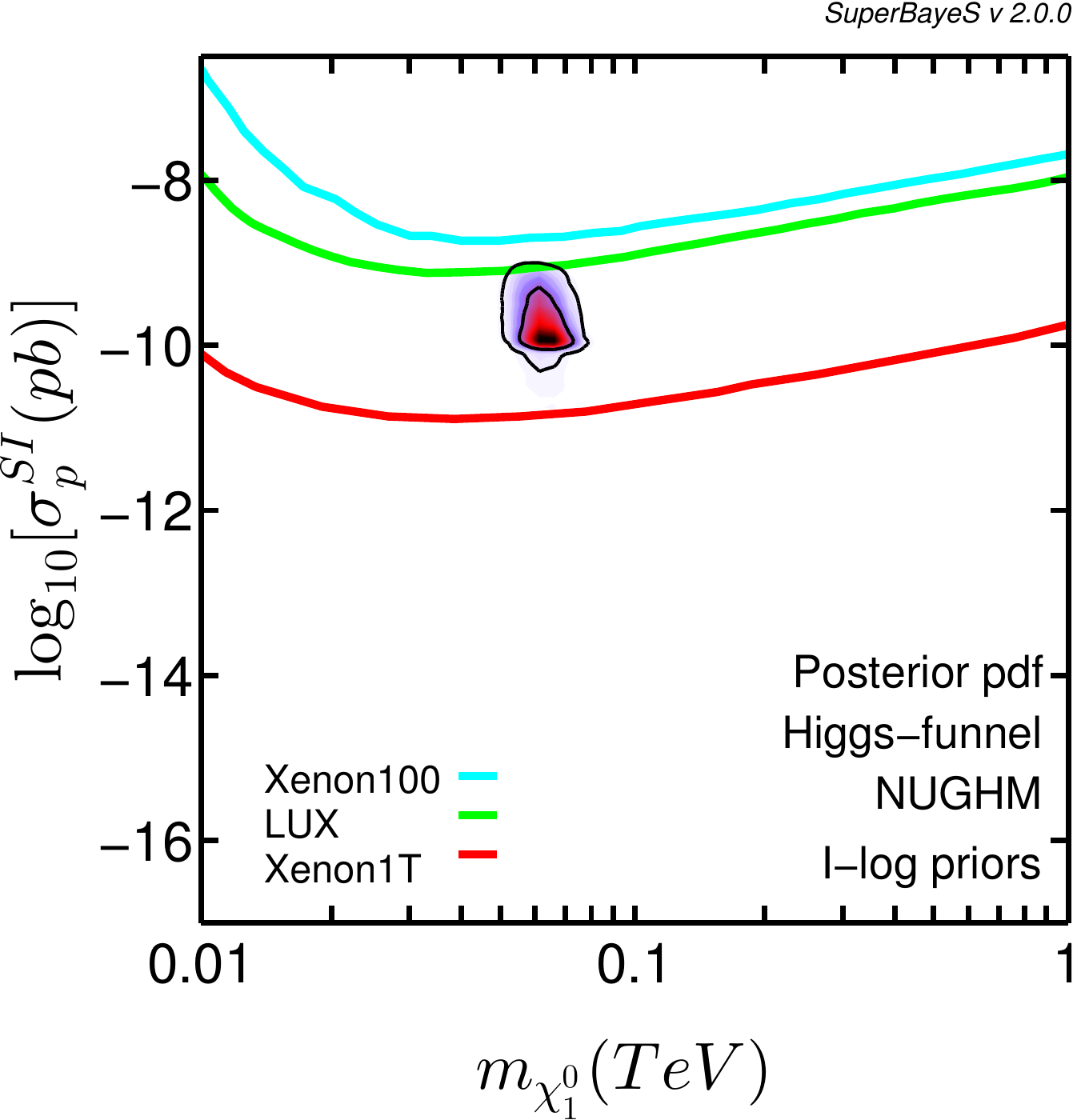}\hspace{1.0cm}
\includegraphics[width=0.4\linewidth]{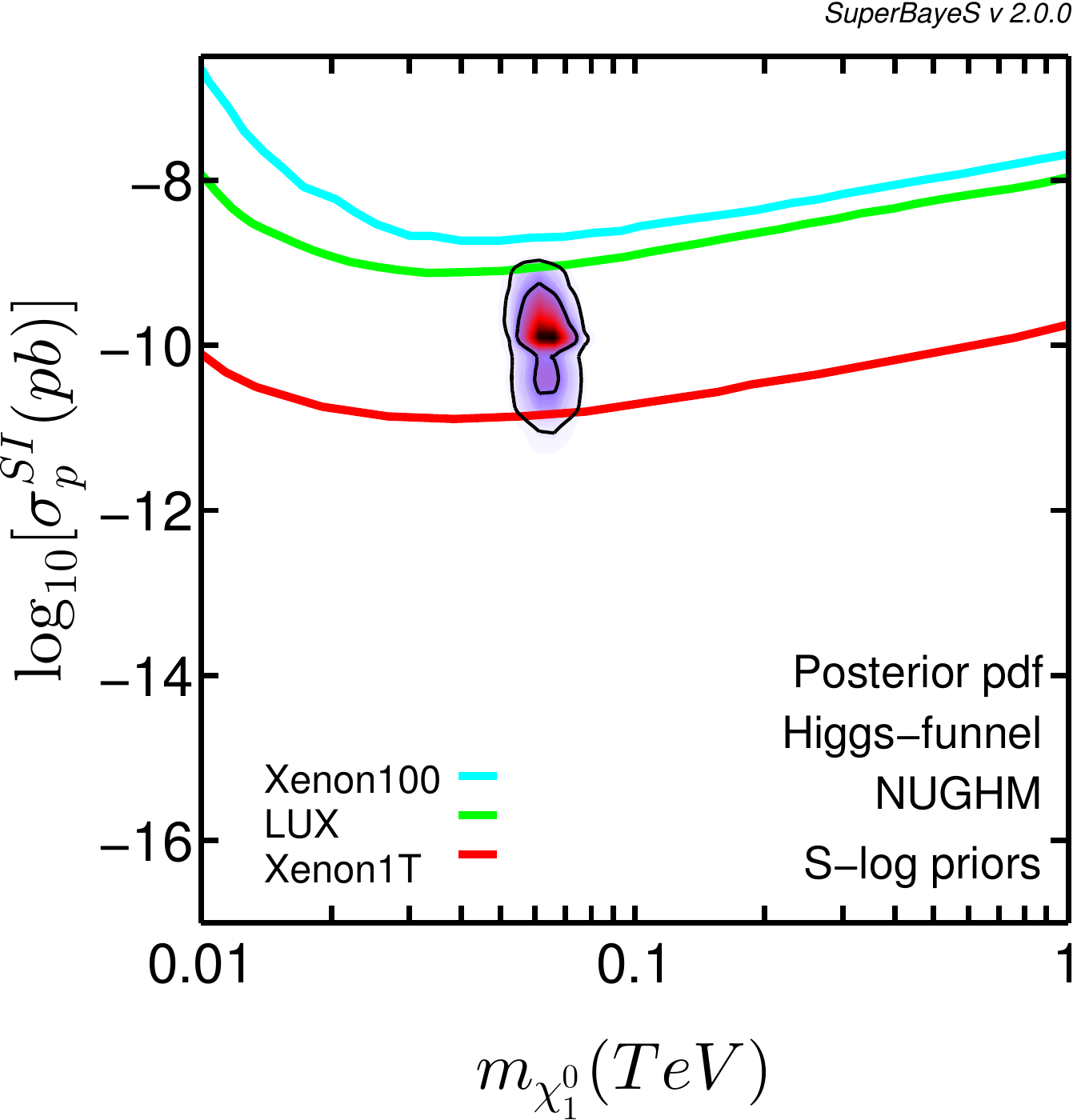}
\caption{2D marginalized posterior probability distribution in the $m_{\chi_1^0}-\sigma^{SI}$ plane for the Higgs-funnel region. The
  contours enclose respective 68\% and 95\% joint regions. The color code is as in Fig.~\ref{fig:single2D}. Present and future observational limits are shown.}
\label{fig:multCDM2D_lowM1}
\end{figure}

As for the general NUGHM, direct DM searches offer  a more complete way of testing the parameter space of this special scenario in the future. Again, the dominant diagram for direct detection occurs via Higgs-interchange in $t$-channel, where the relevant vertex is bino-Higgs-Higgsino, thanks to the non-vanishing Higgsino component of the lightest neutralino. Fig.~\ref{fig:multCDM2D_lowM1} shows the pdf in the $m_{\chi_1^0}-\sigma^{SI}$ plane, where $\sigma^{SI}$ is the Spin--Independent cross section for DM direct detection. We have included the present XENON100 and LUX limits, as well as the future XENON1T ones  \cite{XENON1T}, which are able to probe completely the scenario.

%%%%%%%%%%%%%%%%%%%%%%%%%%%%%%%%%%%%%%%%%%%%%%%%%%%%%%%%%%%%%%%%%%
\subsubsection{The A-funnel and the stau-co-annihilation regions}
%%%%%%%%%%%%%%%%%%%%%%%%%%%%%%%%%%%%%%%%%%%%%%%%%%%%%%%%%%%%%%%%%%

\begin{figure}[ht]
\centering 
\includegraphics[width=0.4\linewidth]{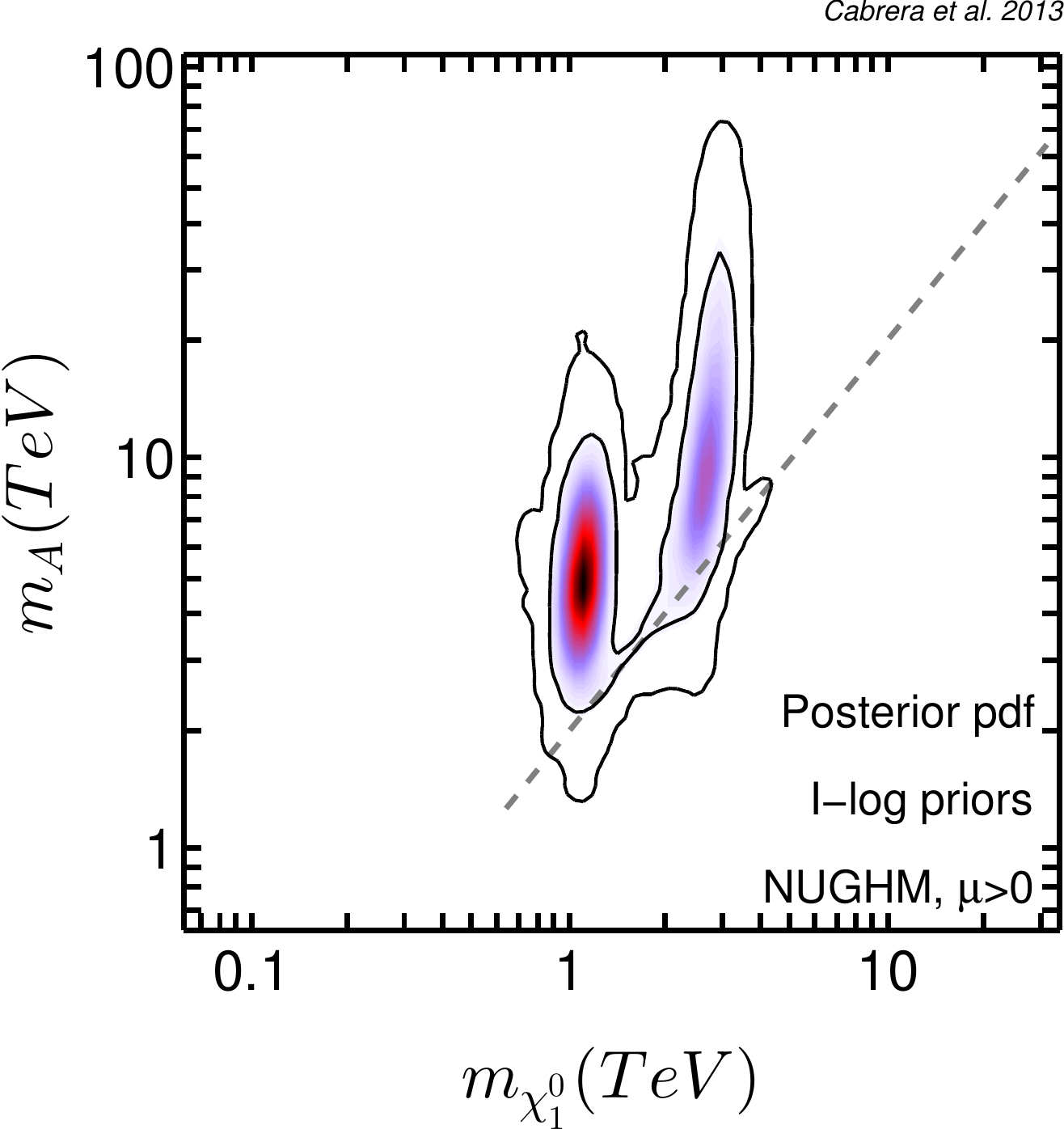}\hspace{1.0cm}
\includegraphics[width=0.4\linewidth]{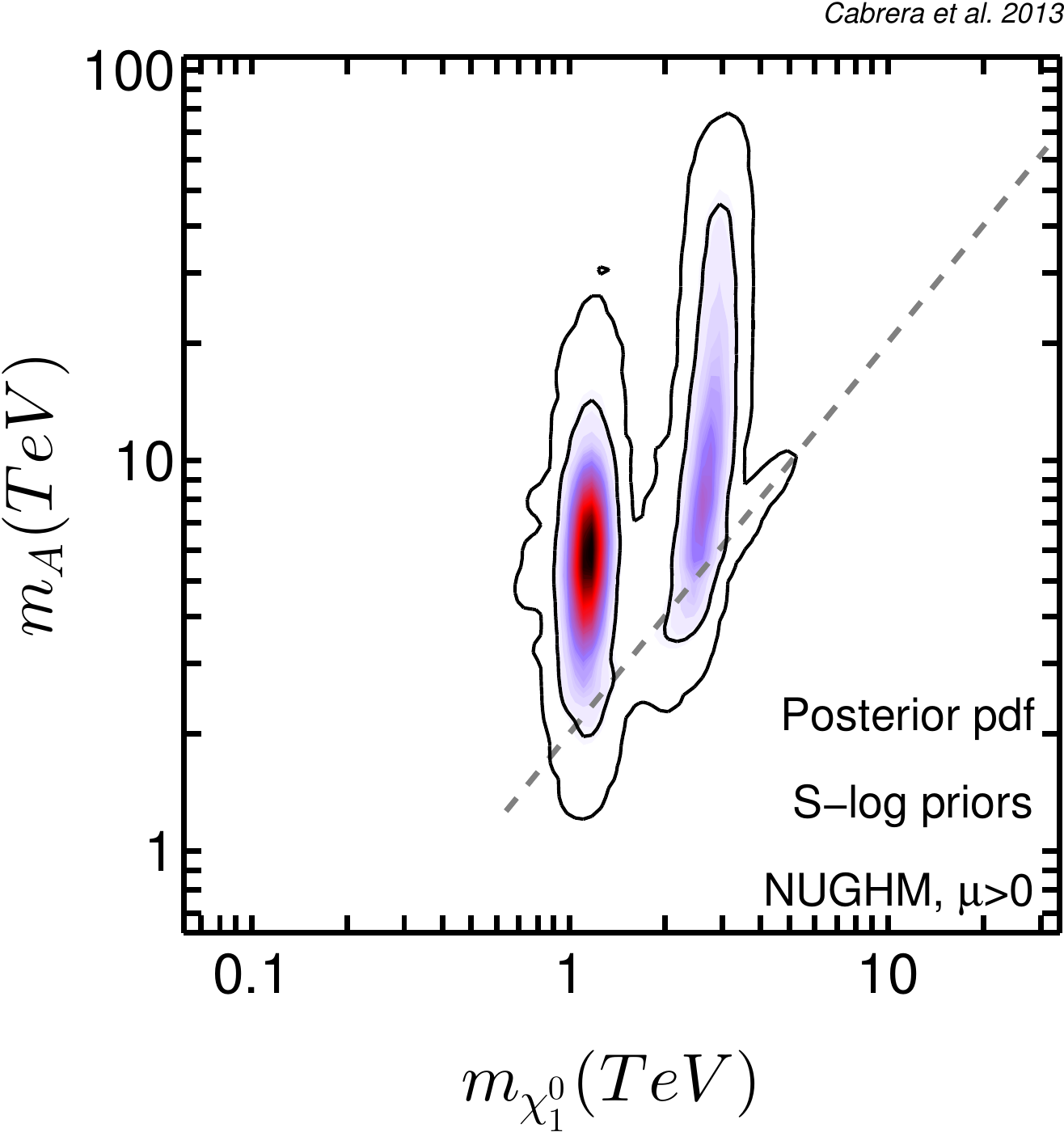}
\caption{2D marginalized posterior probability distribution for single-component CDM in the $m_{\chi_1^0}-m_A$ plane . 95\% and 99.9\% CL contours are shown. The $A-$funnel region shows up as a narrow band around the $m_{\chi_1^0}=m_A/2$ straight line.}
\label{fig:CDM2D_Afunnel}
\end{figure}

Let us finally consider the $A-$funnel region and the stau-co-annihilation regions.  The $A-$funnel annihilation mechanism requires $m_{\chi_1^0}$ to be close to $m_A/2$. This is not easy to achieve, even in a scenario of free gaugino masses, as the one at hand. The reason is that, for moderate-to-large $\tan\beta$, $m_A$ is close to $m_{H_d}$, and thus receives substantial RG contributions from the wino and bino masses. Hence, one needs to go to very large $\tan\beta$ ($\simgt 30$), so that the bottom and tau Yukawa couplings get sizeable and thus conveniently decrease $m_{H_d}$ along the running \cite{Roszkowski:2001sb}. As discussed in subsect.~3.2, such large values of $\tan\beta$  amount to an additional fine-tuning. Consequently the $A-$funnel region is strongly  disfavored statistically, although it is still there. To visualize it, we have to show the $99.9\%$ C.L. region in the NUGHM parameter space. Fig.~\ref{fig:CDM2D_Afunnel} displays this region in the $m_{\chi_1^0}$--$m_A$ plane. It shows up as a diagonal white band around the $m_{\chi_1^0}=m_A/2$ straight line, which passes across the two main isles, where the LSP is mostly Higgsino and wino respectively (discussed in detail in subsect.~4.1).

From Fig.~\ref{fig:CDM2D_Afunnel} it is clear that the electroweakino
sector of the $A-$funnel region is quite heavy and thus very difficult to
probe at the LHC. On the other hand, the phenomenology of this region in the
NUGHM is similar to the CMSSM case, which has been analyzed elsewhere
\cite{Roszkowski:2001sb, Ellis:2002iu,Roszkowski:2009sm}

Let us finally turn to the stau co-annihilation region. In this case
$m_{\chi_1^0}$ should be close to $m_{\tilde \tau}$ and not too heavy
\cite{Ellis:1998kh}. Certainly LHC puts strong lower bounds on the squarks, but the
limits are much weaker for sleptons, so a scenario of this type is still
viable. Note, in this sense, that even starting with sfermion-mass universality
at the high scale (as it happens for the NUGHM), the slepton masses can be
much lower than the squark ones since they do not receive the strong
contribution proportional to the gluino-mass-squared. Still, an important
fine-tuning is necessary to achieve the quasi-degeneracy of the stau and the
LSP. Actually, the region is so tiny that it remains almost invisible when the $99.9\%$ C.L. contours are displayed. This is why we do not show any additional figure, which would be hardly illustrative. Once more, the phenomenology of the co-annihilation region in the NUGHM is similar to the CMSSM case, and can be consulted in the standard references \cite{Roszkowski:2001sb, Ellis:2002iu}

\begin{figure}[ht]
\centering 
\includegraphics[width=0.415\linewidth]{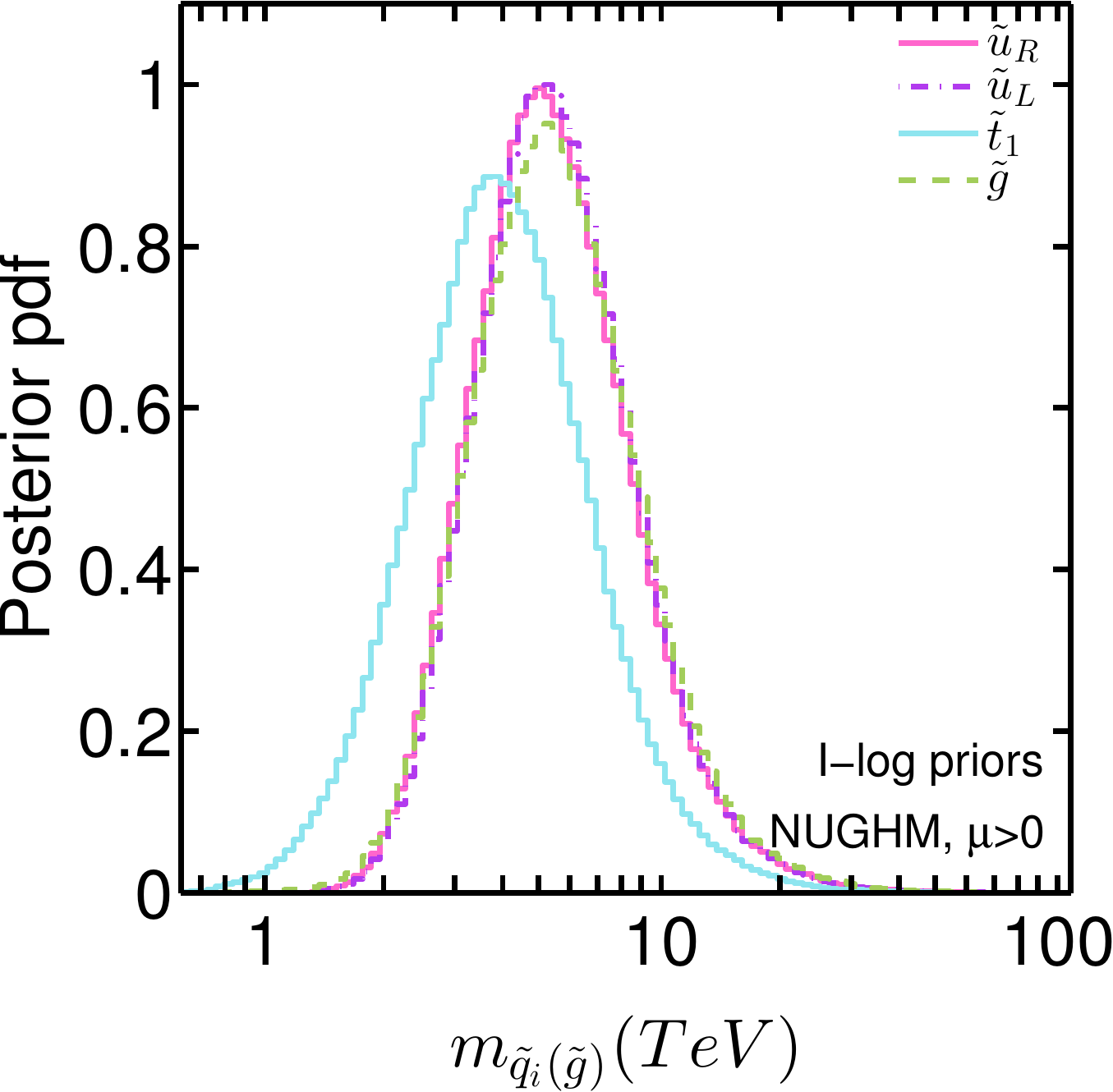}\hspace{1.0cm}
\includegraphics[width=0.4\linewidth]{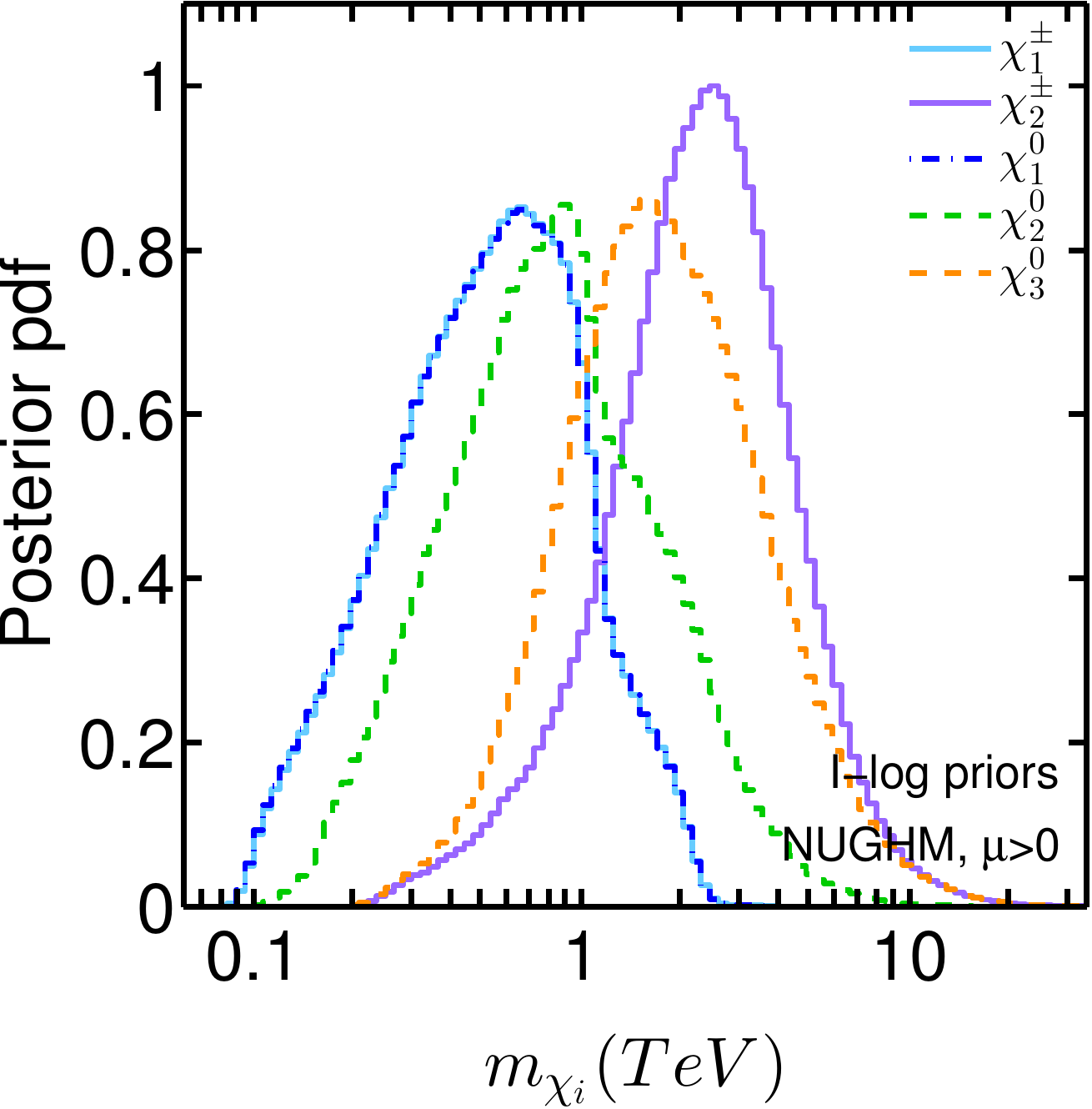}\\ 
\vspace{0.5cm} 
\includegraphics[width=0.415\linewidth]{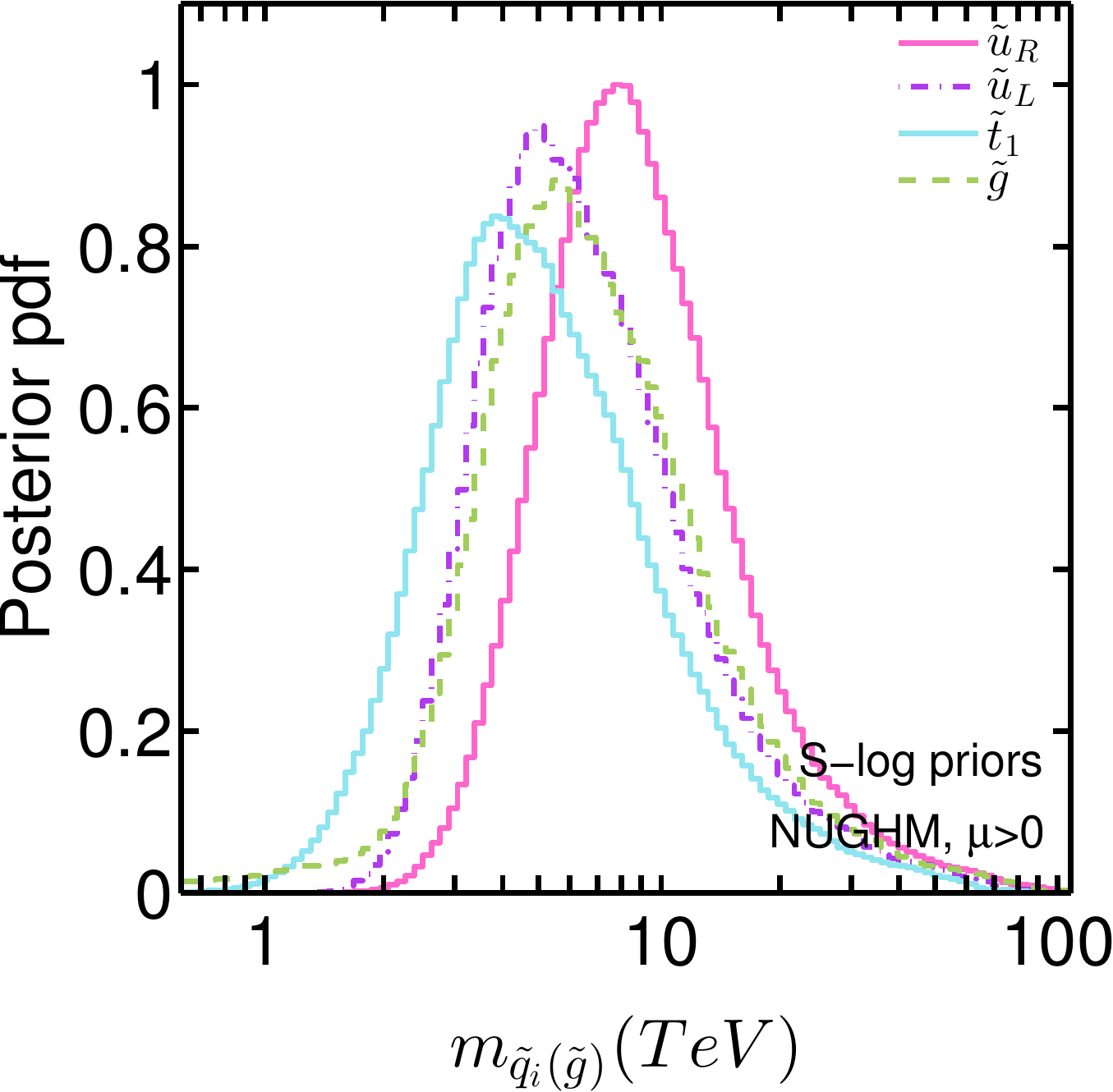}\hspace{1.0cm}
\includegraphics[width=0.4\linewidth]{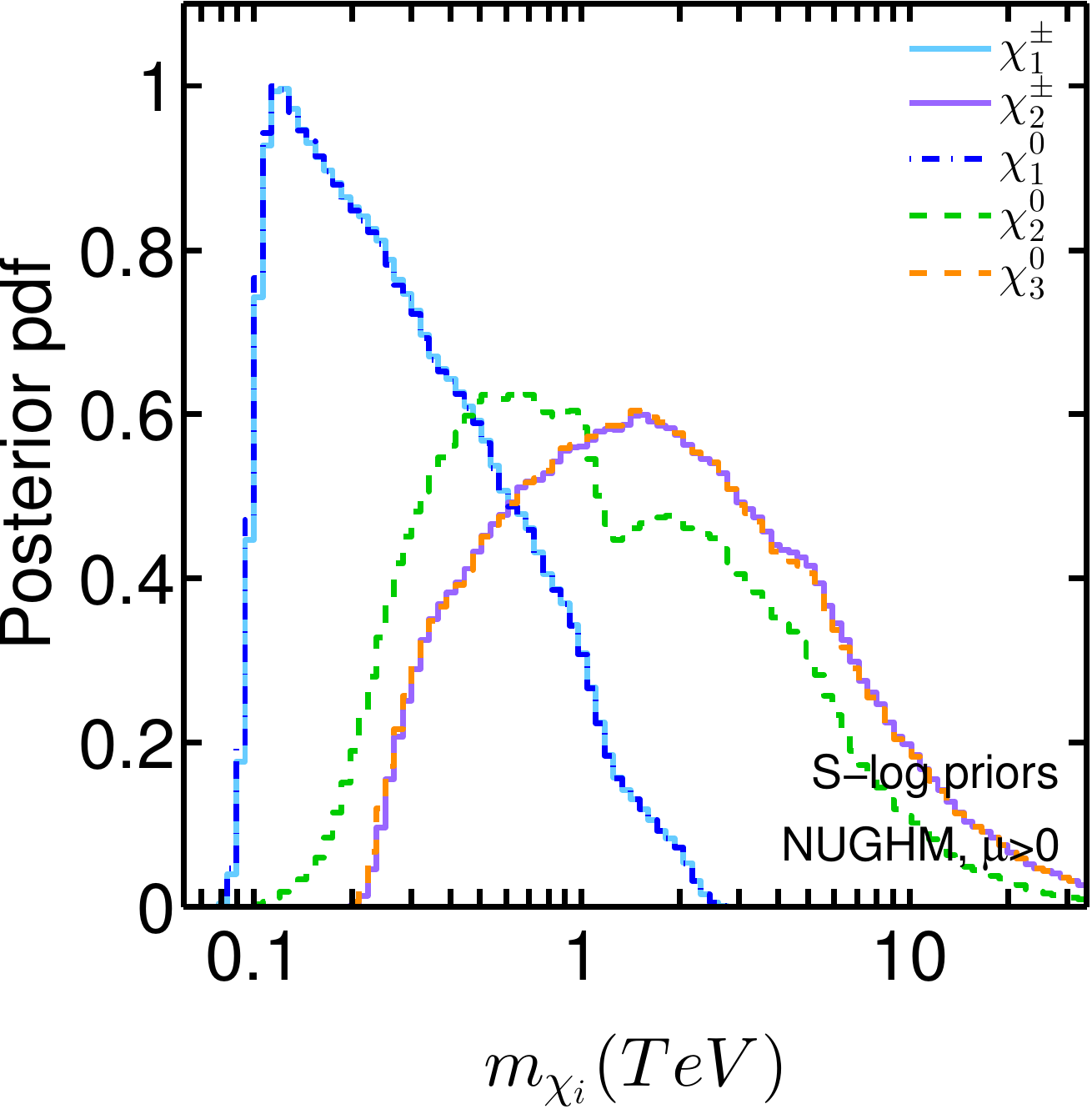}
\caption{1D marginalized posterior probability distribution of masses of supersymmetric particles for multi-component CDM. Upper (lower) plots correspond to I-log (S-log) priors.}
\label{fig:multi1D}
\end{figure}

%%%%%%%%%%%%%%%%%%%%%%%%%%%%%%%%%%%%%%%%%%%%%%%%%%%%%%%%%%%%%%%%%%
\subsection{Multi-Component CDM\label{sec:MCCDM}}
%%%%%%%%%%%%%%%%%%%%%%%%%%%%%%%%%%%%%%%%%%%%%%%%%%%%%%%%%%%%%%%%%%

Relaxing the requirement that all of the DM is made of LSPs implies that the
PLANCK measurement is just an upper bound on its abundance. As it is shown in
the Appendix of ref. \cite{Bertone:2010ww}, in this case, the correct
effective likelihood is given by the expression 
\be 
\label{eq:upperbound}
\like_\text{PLANCK}(\OhLSP) = \like_0 \int_{\OhLSP/\siW}^\infty
e^{-\frac{1}{2}(x-r_\star)^2} x^{-1}{\rm d}x , 
\ee 
where $ \like_0$ is an irrelevant normalization constant, $\OhLSP$ is the predicted relic density of the LSP as a
function of the NUGHM and SM parameters; and $r_\star \equiv
\muW/\siW$,  where $\muW$ and $\siW$ are the mean value and the standard deviation of the measure of $\Omega_{\rm DM}h^2$ by PLANCK.

When the LSP is not the only constituent of DM, the rate of events in a
direct-detection experiment is smaller, since it is proportional to the local
density of the LSP, $\rho_\chi$, which is now smaller than the total local DM
density, $\rho_\text{DM}$. The suppression is given by the factor $\xi \equiv
\rho_\chi / \rho_{\rm DM}$. Following ref. \cite{Bertone:2010rv}, we assume
that ratio of local LSP and total DM densities is equal to that for the cosmic
abundances, thus $\xi \equiv \rho_\chi / \rho_{\rm DM}= \Omega_\chi /
\Omega_{\rm DM}$. For $\Omega_{\rm DM}$ we adopt the central value of the PLANCK
determination see Table 2, while for $\rho_{\rm DM}$ we adopt, following
ref. \cite{Pato:2010zk}, the value $0.4\,$GeV\,cm$^{-3}$. This allows to
evaluate $\xi$ for each point in the NUGHM parameter space.

\begin{figure}[ht]
\centering 
\includegraphics[width=0.3\linewidth]{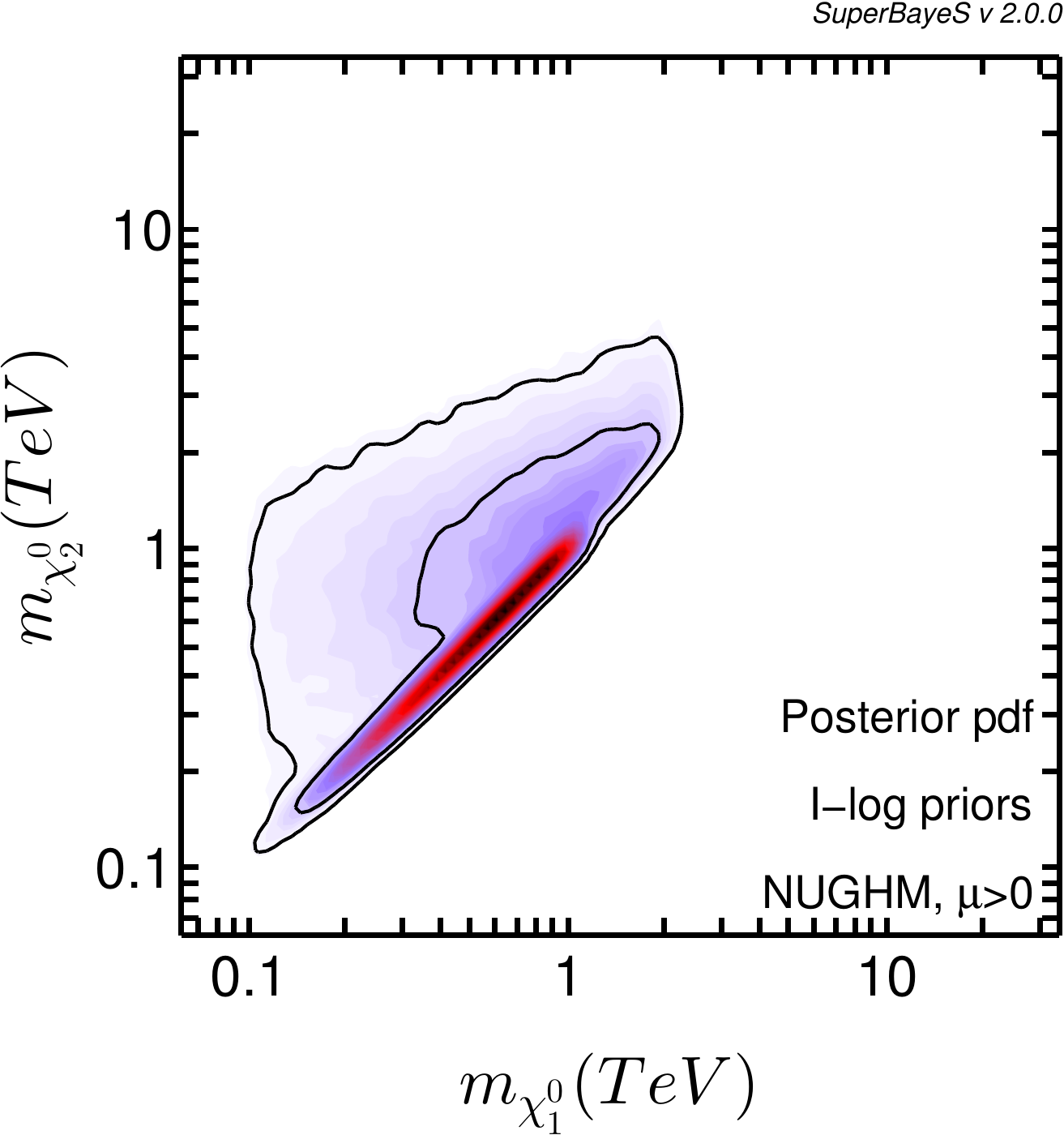}\hspace{0.5cm}
\includegraphics[width=0.3\linewidth]{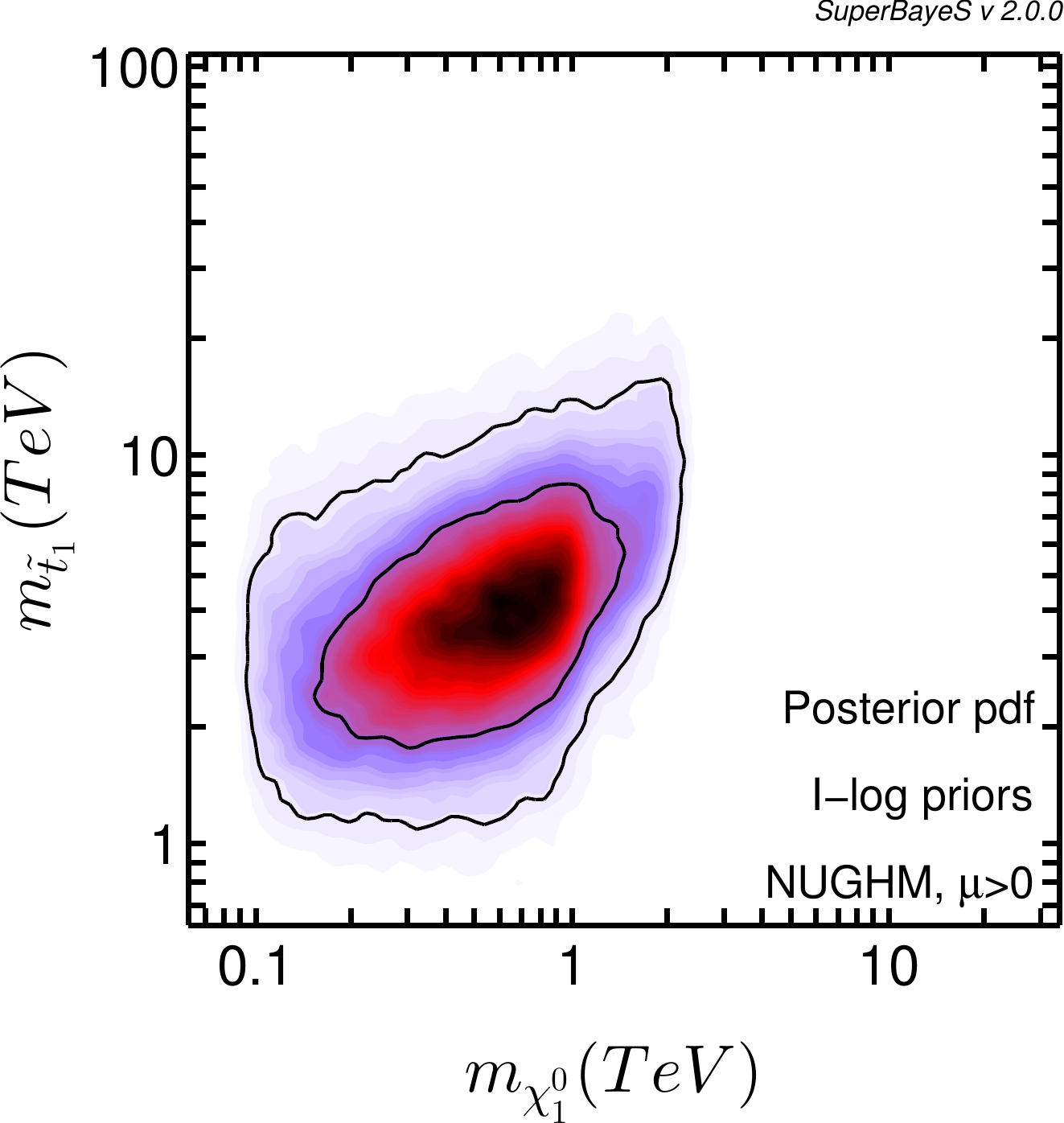}\hspace{0.5cm}
\includegraphics[width=0.3\linewidth]{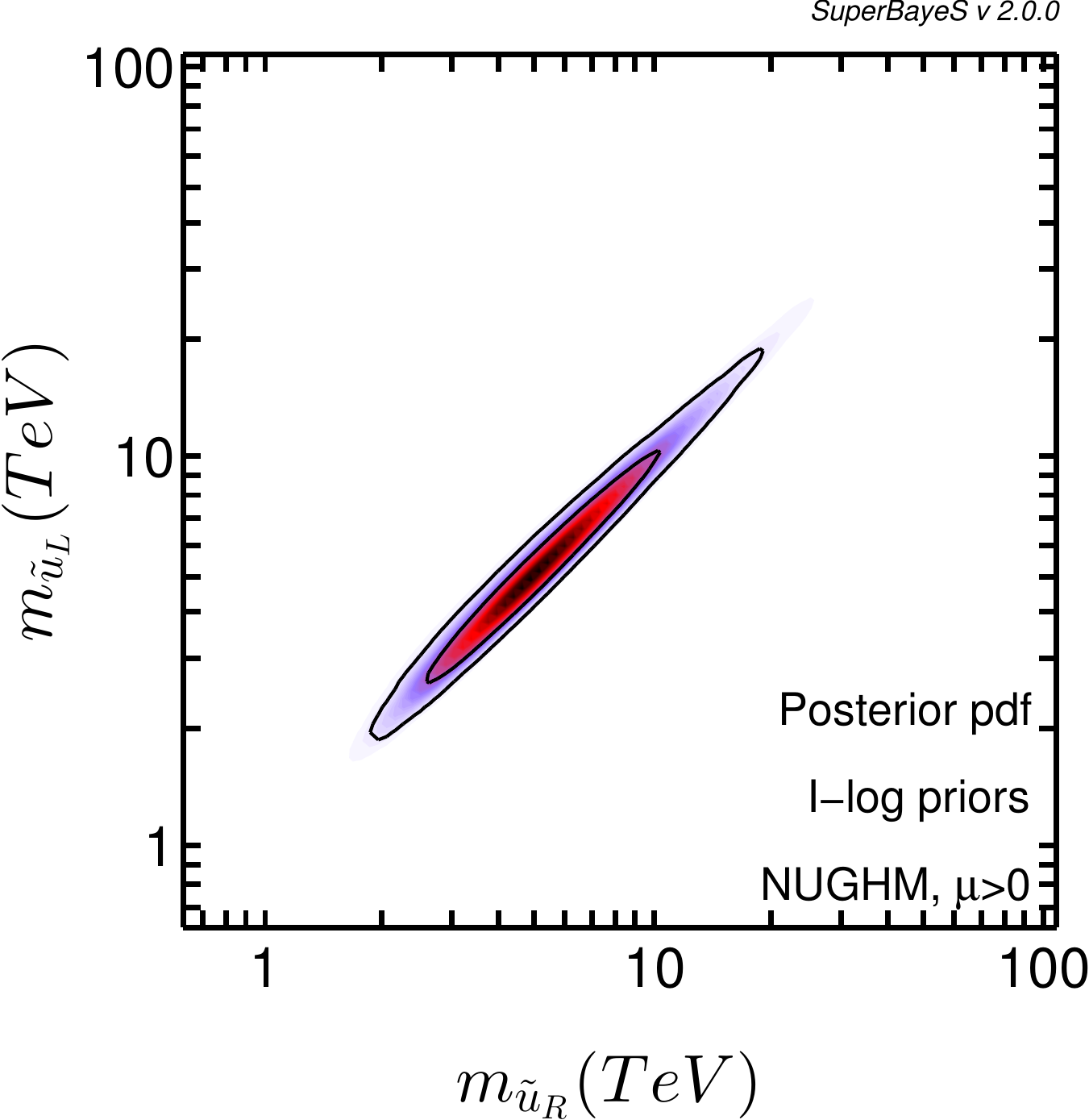}\vspace{0.5cm}
\includegraphics[width=0.3\linewidth]{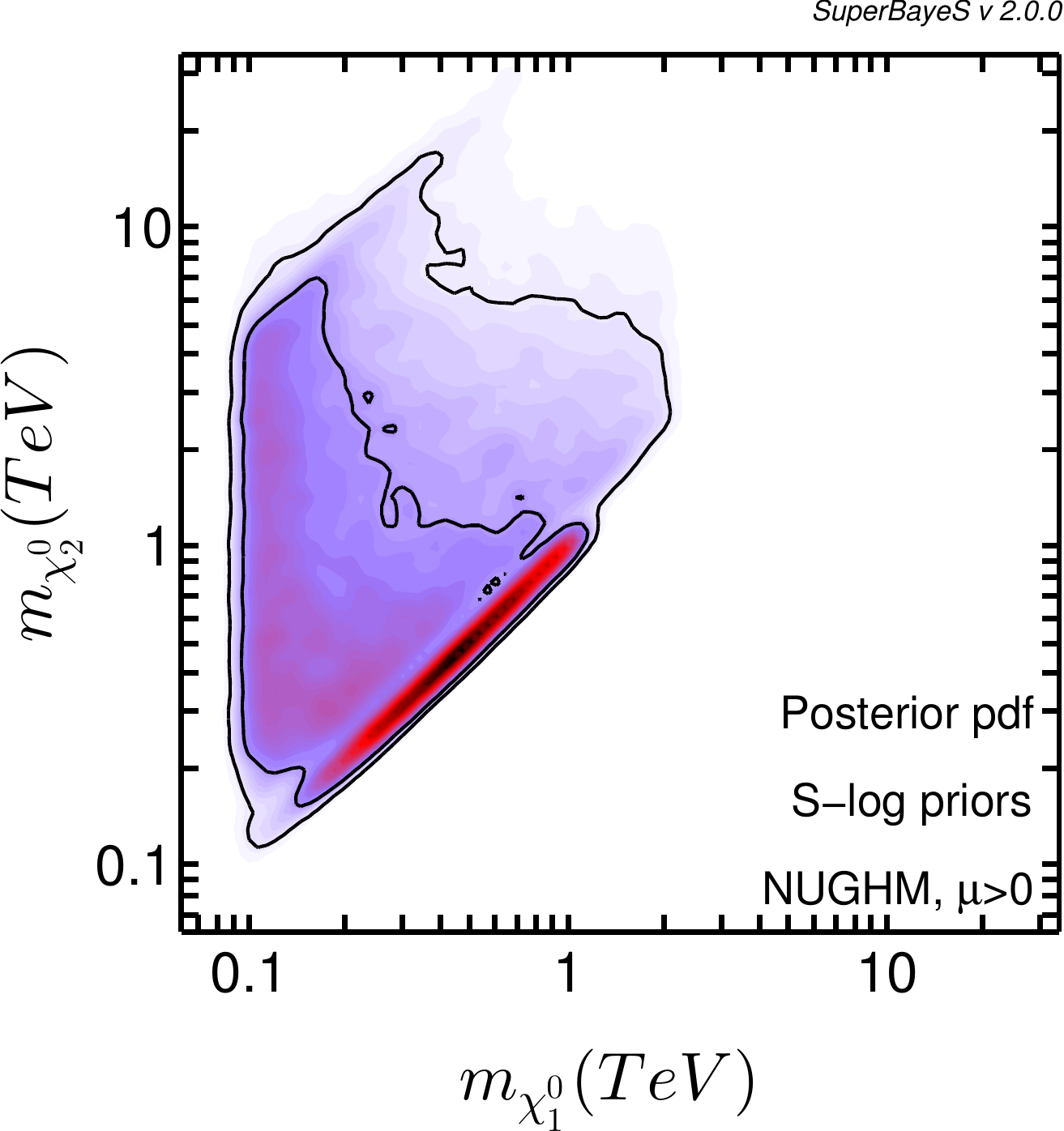}\hspace{0.5cm}
\includegraphics[width=0.3\linewidth]{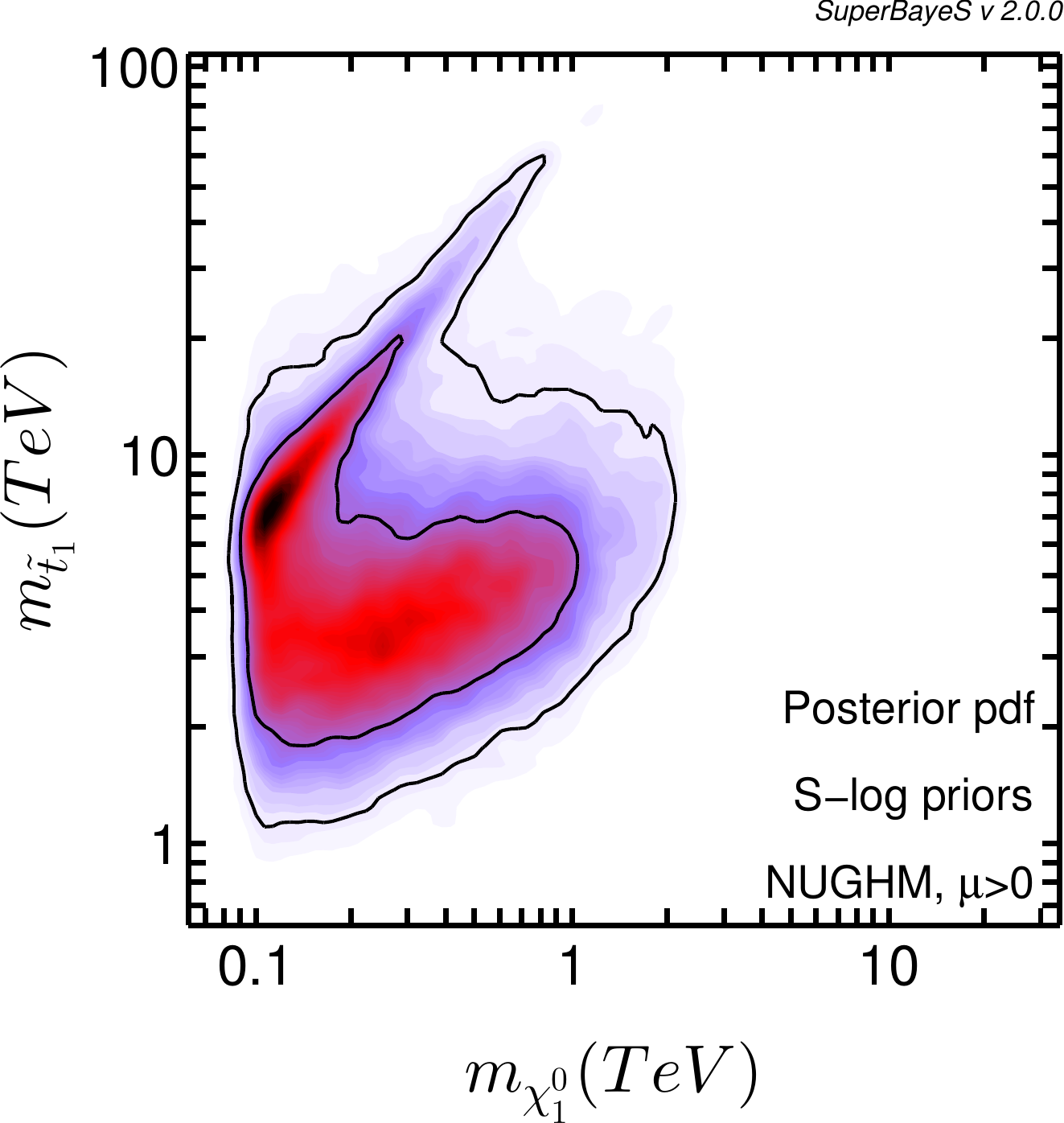}\hspace{0.5cm}
\includegraphics[width=0.3\linewidth]{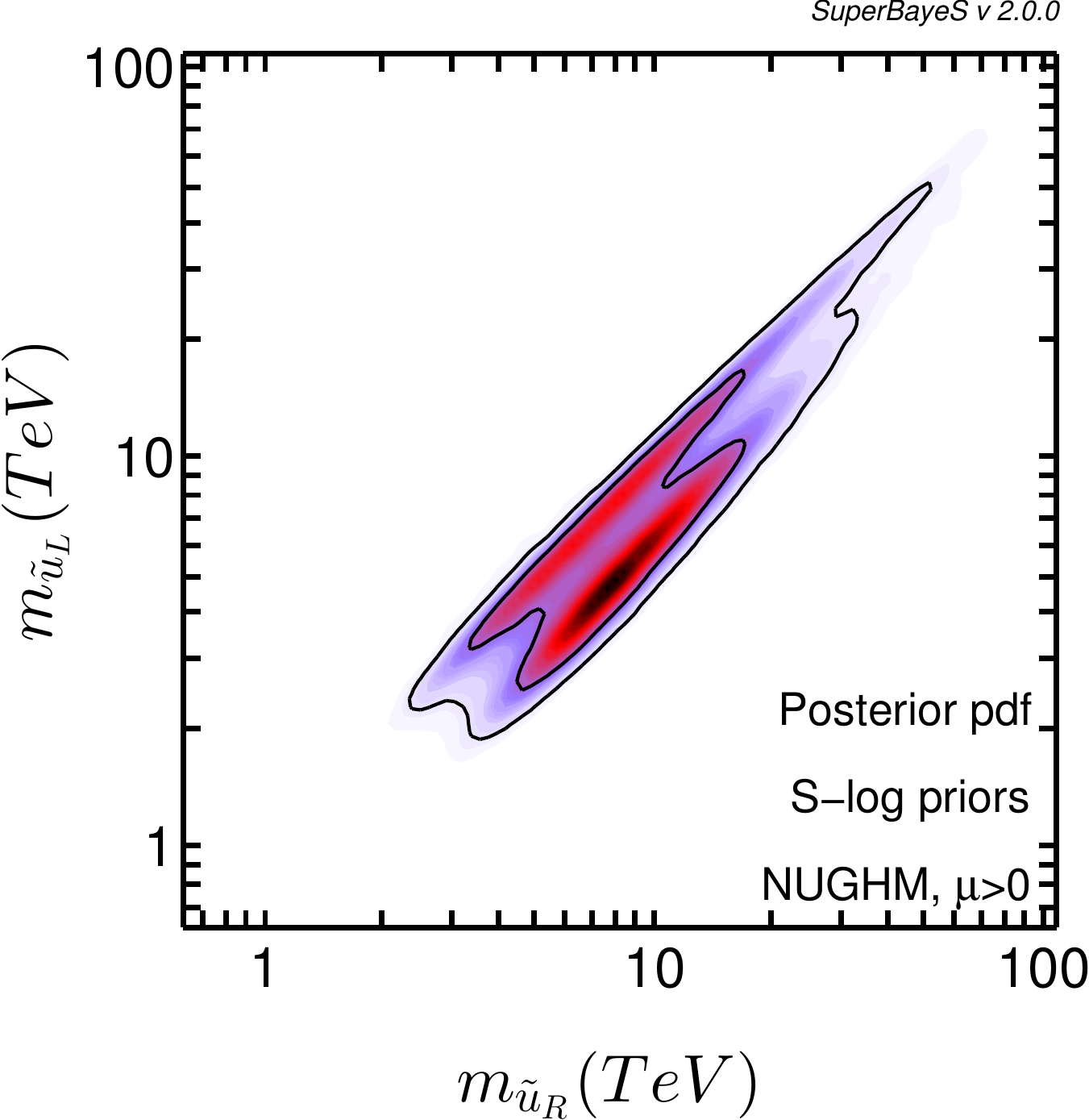}
\caption{2D marginalized posterior probability for multi-component CDM on different planes defined by couples of supersymmetric masses. Upper (lower) panels correspond to I-log (S-log) priors. The 68\% and 95\% CL contours are shown. The color code is as in Fig.~\ref{fig:single2D}.}
\label{fig:mult2D}
\end{figure}

Since we are relaxing the DM requirements, the preferred regions in the parameter space are wider\footnote{This effect is however almost invisible for the Higgs-funnel, A-funnel and stau co-annihilation regions analyzed in subsect.~4.2. The reason is that in these scenarios the primordial annihilation of DM is achieved thanks to a fine-tuning of the neutralino mass with half of the Higgs or $A$ masses (for the Higgs and $A-$funnel) and with the stau mass (for the co-annihilation region). Actually, this tuning is one of the main reasons for their low statistical weight. As a consequence, to achieve {\em less} supersymmetric DM than the actual abundance requires an even stronger fine-tuning and does not extend appreciably the corresponding allowed region of the parameter-space.}. Recall that for single-component CDM the two preferred
cases ocurred when $\chi_1^0$ was Higgsino-like and wino-like, and the
corresponding mass was essentially determined by the requirement of the DM
abundance to be $\sim 1$ TeV or $\sim 3$ TeV, respectively. Now, the
$\chi_1^0-$mass can be lighter, since this  translates into a more efficient
annihilation and thus a smaller supersymmetric DM abundance, which is an
acceptable possibility. Besides, the fact that the supersymmetric DM abundance can be smaller relaxes the XENON100 limits correspondingly. Thus in principle $\chi_1^0$ would not need to be a Higgsino or wino state with the same level of purity.

These facts are reflected in Fig. \ref{fig:multi1D}, which
shows the 1D posterior pdfs for squark and gluinos (left panels) and for charginos and neutralinos (right panels), in the same way as Fig.\ref{fig:single1D} did for the single-component CDM case.
Now there are not sharply selected ranges for the mass of the lightest neutralino. On the other hand, the fact that ${\chi}_1^0$ and
${\chi}_1^\pm$ continue to be typically almost degenerate, indicates that $\chi_1^0$ is still most often Higgsino- or wino-like. However, due to the fact that the displayed distributions are now broader, it is not as easy as before to determine with which probability $\chi_1^0$ is one or another. Nevertheless,  recalling that for the Higgsino-like case ${\chi}_1^0$ and ${\chi}_2^0$ are almost degenerate, we can get useful information from the pdf of $m_{\chi_2^0}$. In particular, for the S-log prior, the $m_{{\chi}_1^0}$ pdf has a
peak and edge at $\sim 100$ GeV, where the $m_{{\chi}_2^0}$ pdf is almost zero, implying that the peak
corresponds to wino-like $\chi_1^0$. Notice that the peak is essentially at the LEP lower bound on $m_{{\chi}_1^0}$, which means that for S-log priors, ${\chi}_1^0$ and ${\chi}_1^\pm$ prefer to be as light as they can. This is a typical consequence of the use of S-log priors, since they favor the smallness of any parameter, even if the others are large. 
On the other hand, for I-log priors, the $m_{{\chi}_1^0, \chi_1^\pm}$ and $m_{{\chi}_2^0}$ distributions are smoother an similar, implying that both possibilities, Higgsino- and wino-like $\chi_1^0$ are likely for a wide range of $m_{{\chi}_1^0}$ masses.

The left panels of Fig \ref{fig:multi1D} show the squarks and gluino posterior
pdfs. As for the single-component CDM case, for I-log priors the squark and gluino masses are close to each
other. On the other hand, for S-log priors there is a large splitting between
left and right squarks: $\tilde{u}_R$ is typically heavier than $\tilde{u}_L$, the
inverse case of single-component CDM. The reason is the following. As commented above, for S-log priors
the dominant region is now the one with wino-like
$\chi_1^0$, so $M_2$ is typically smaller than $M_1$. Actually $M_1$ can be quite large since, as discussed in sect. 3, for S-log priors large splittings between parameters are not specially penalized. Along the RG running both $m_{\tilde{u}_R}^2$ and $m_{\tilde{u}_L}^2$ grow with $M_1^2$,
but the factor of the $M_1^2$ term in the RGE of the former is 15 times larger than
the one of the latter, thus driving $\tilde{u}_R$ heavier than $\tilde{u}_L$.

\begin{figure}[ht]
\centering 
\includegraphics[width=0.4\linewidth]{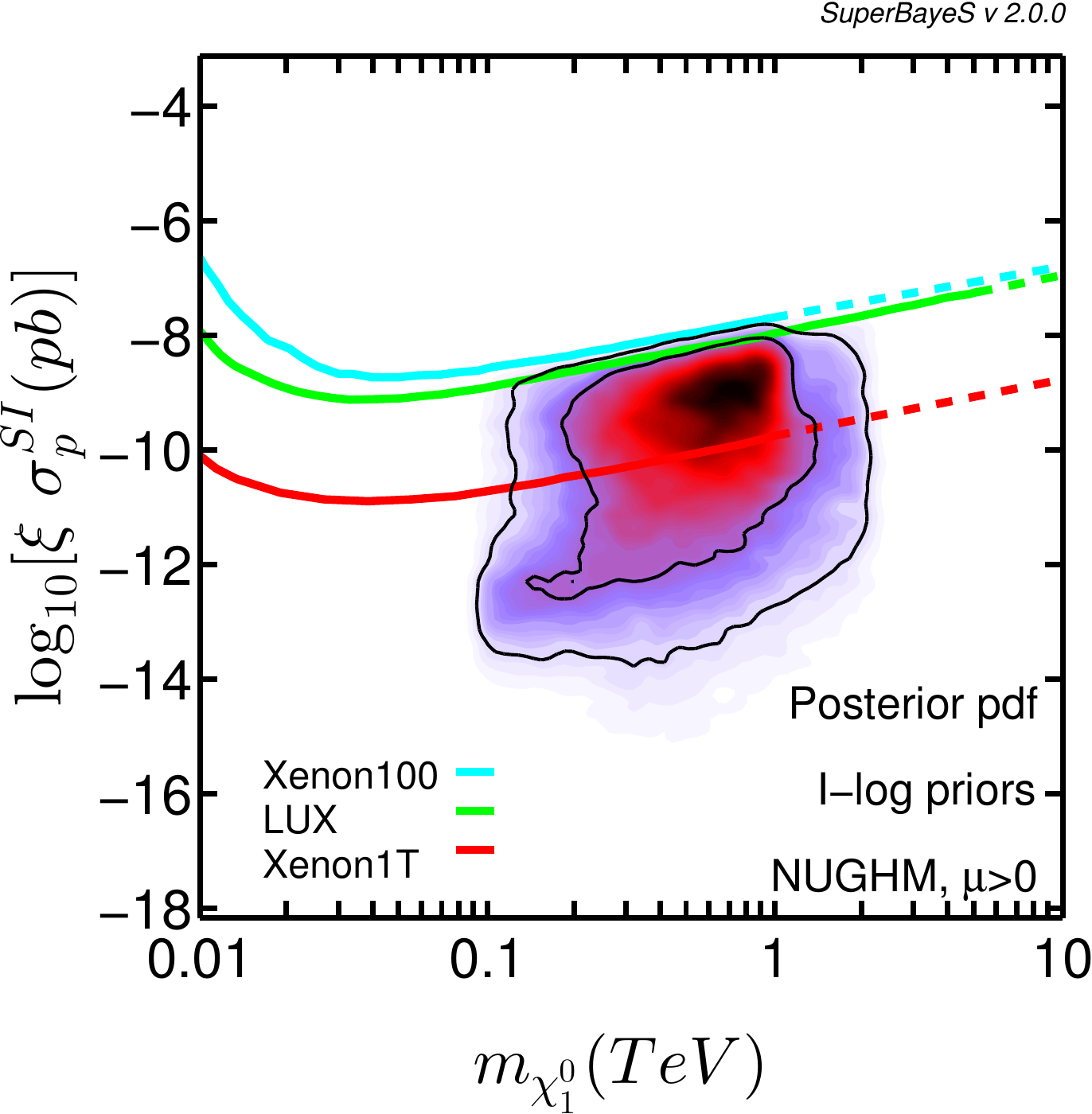}\hspace{1.0cm}
\includegraphics[width=0.4\linewidth]{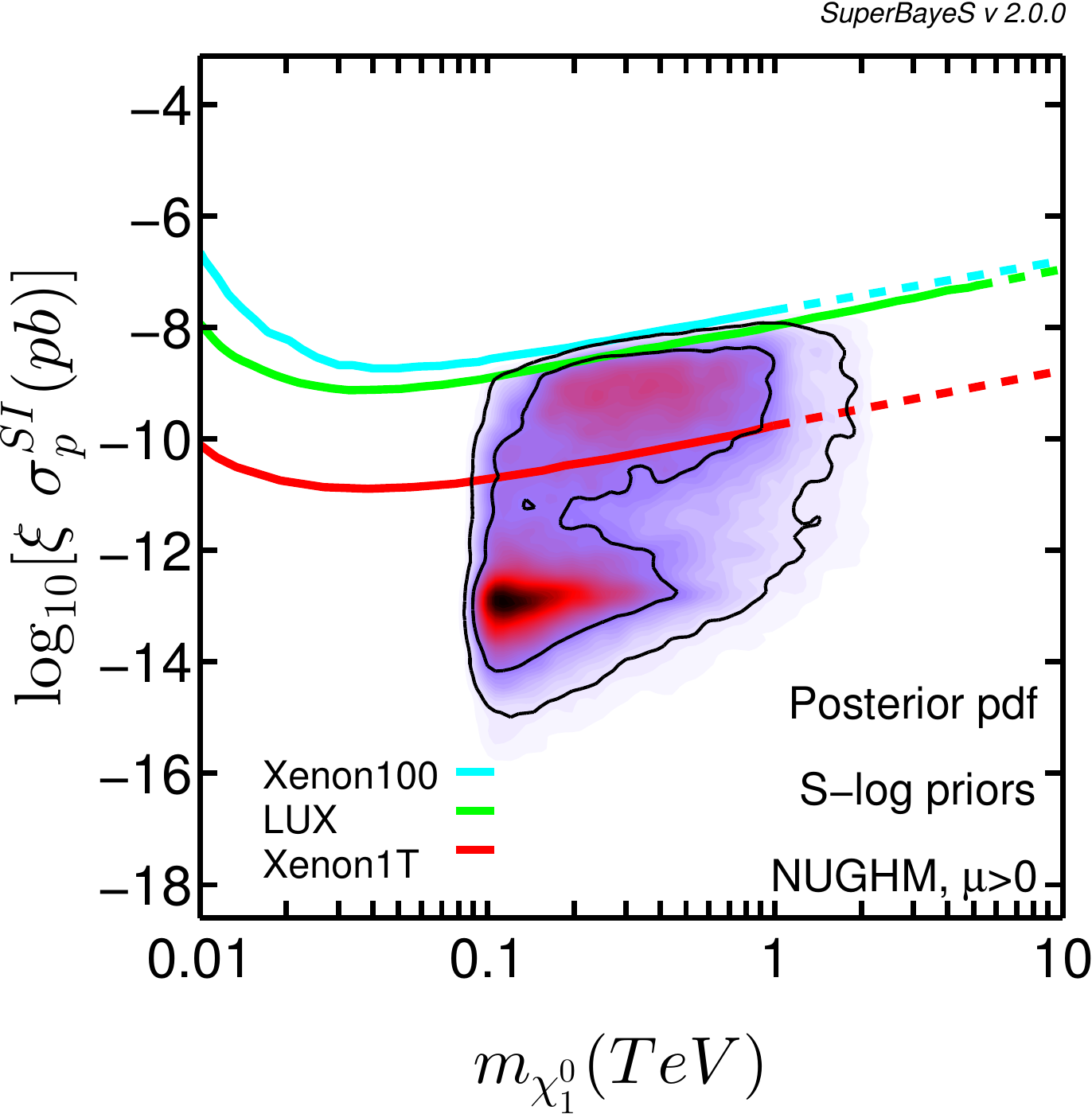}
\caption{2D marginalized posterior probability in the $m_{\chi_1^0}-\xi\sigma^{SI}$ plane for multi-component CDM. The left (right) panel corresponds to I-log (S-log) priors. The
  contours enclose respective 68\% and 95\% joint regions. Present and future observational limits are shown. The color code is as in Fig.~\ref{fig:single2D}.}
\label{fig:multCDM2D}
\end{figure}

Fig.~\ref{fig:mult2D} shows the 2D posterior pdfs of supersymmetric particles, similarly to Fig.~\ref{fig:single2D}. The region where ${\chi_1^0}$ is Higgsino(wino)-like can be identified in the left panels by the fact that the second lightest neutralino, ${\chi_2^0}$, is (is not) quasi-degenerate with it. As mentioned above, the region where ${\chi_1^0}$ is wino-like is much more important for S-log priors than for I-log priors. 
The central and right panels, show the 2D pdfs in the $m_{\chi_1^0}-m_{\tilde{t}_1}$  and  $m_{\tilde{u}_L}-m_{\tilde{u}_R}$ planes, where one can appreciate the above-discussed differences between S-log and I-log priors.
On the other hand, comparing Figs.~\ref{fig:mult2D} and \ref{fig:single2D}, it is clear that in the multi-component CDM case the neutralinos (and also charginos) are likely to be much lighter than in the previous single-component CDM one, while stops and squarks remain more or less as heavy as before.

Fig~\ref{fig:multCDM2D} shows the pdf in the $m_{\chi_1^0}-\xi\sigma^{SI}$ plane, in a similar way to Fig.~\ref{fig:singleCDM2D}. As for the other pdfs, the differences between S-log and I-log priors come from the abnormally important wino-like region with small $M_2$ for S-log priors.  This also makes a difference with respect to the prospects of detection in future experiments, like XENON1T. While for the Higgsino-like case the bulk of the probability in the parameter-space will be tested, for the wino-like case is exactly the opposite. This is precisely the dominant case for S-log priors.

Fortunately, for the multi-component CDM scenario the supersymmetric masses (particularly in the electroweakino sector) are typically smaller than in the single-component one, which makes LHC searches complementary to the DM ones. This is analyzed in further detail in the next subsection.

%%%%%%%%%%%%%%%%%%%%%%%%%%%%%%%%%%%%%%%%%%%%%%%%%%%%%%%%%%%%%%%%%%
\subsection{``Low-energy'' NUGHM}
%%%%%%%%%%%%%%%%%%%%%%%%%%%%%%%%%%%%%%%%%%%%%%%%%%%%%%%%%%%%%%%%%%

In order to discuss the phenomenology of the NUGHM at LHC, we perform here a modified analysis in which we assume that SUSY is in a region potentially accesible to LHC. This is equivalent to perform a zoom into the phenomenologically interesting region of the parameter space (the one which is in principle testable at LHC). In this way we can derive the most likely signatures of SUSY, provided the model is well described by a NUGHM. 

For this task we then impose the additional conditions listed in  table~\ref{tab:lowSusyCuts}. The requirement is not that those condidtions must be all realized at the same time, but that at least one of them should be satisfied. Note that the constraint on the electroweakino sector is just a bound on the $\chi_2^\pm$ mass. The reason for this will be explained soon. To avoid confusion with the general analysis of the NUGHM (presented in the previous subsections), we will call this scenario ``low-energy NUGHM" (NUGHM-low in the following figures).

\begin{table}[ht]
  \centering
  \setlength{\extrarowheight}{4pt}
  \begin{tabular}[h]{|c|c|}
    \hline
    Observable & Condition \\ \hline\hline
    $m_{\tilde{q}}$ & $\leq 3$ TeV \\
    $m_{\tilde{g}}$ & $\leq 3$ TeV \\
    $m_{\tilde{t}_1}$ & $\leq 1$ TeV \\ 
    $m_{\chi_2^\pm}$ & $\leq 800$ GeV \\ 
    \hline
  \end{tabular}
  \caption{Conditions imposed for `Low-energy' NUGHM. At least one of the conditions is required to be satisfied.}
  \label{tab:lowSusyCuts}
\end{table}

\begin{figure}[ht]
  \centering
  \includegraphics[width=0.415\linewidth]{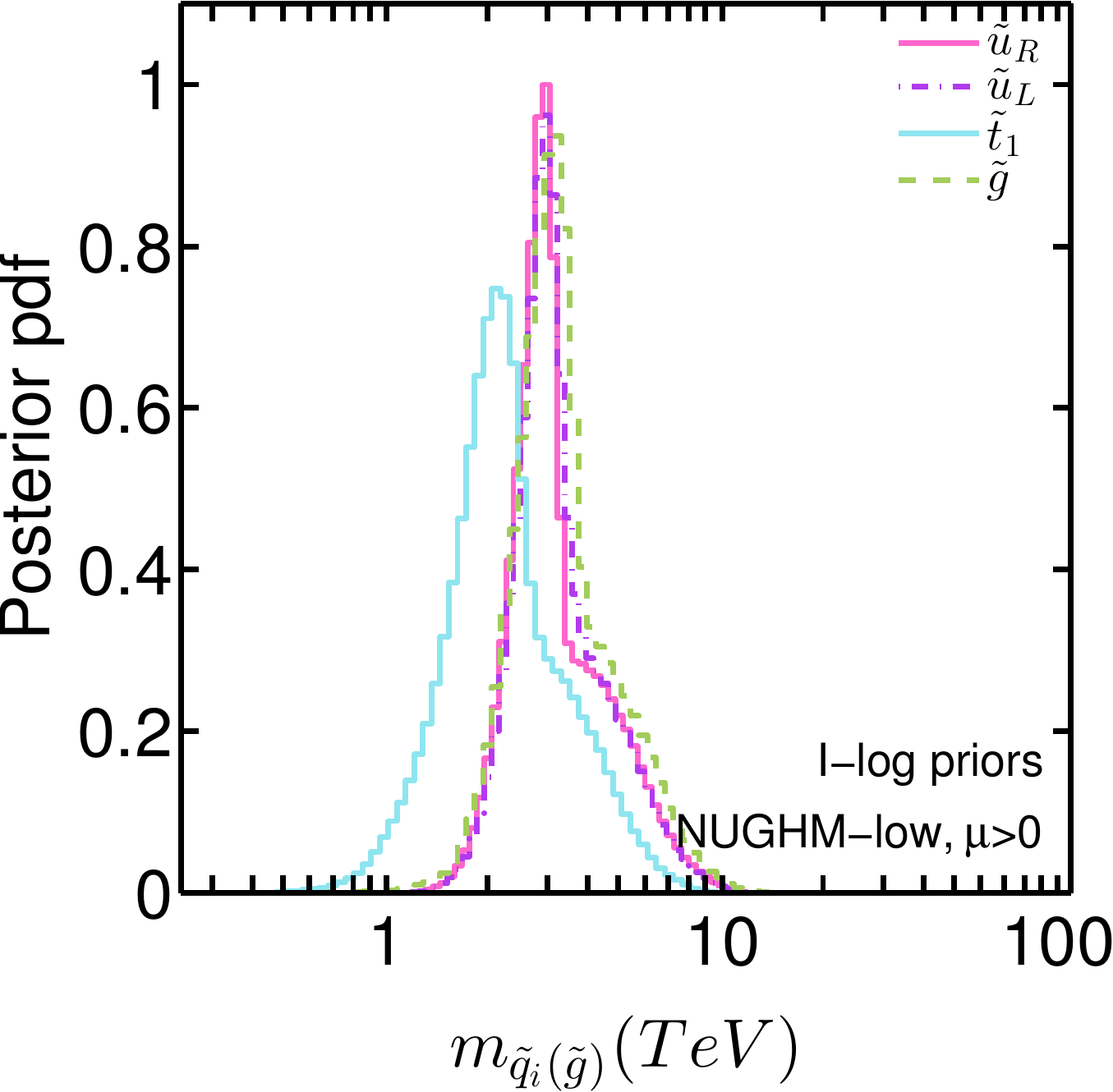}\hspace{1.0cm}
  \includegraphics[width=0.4\linewidth]{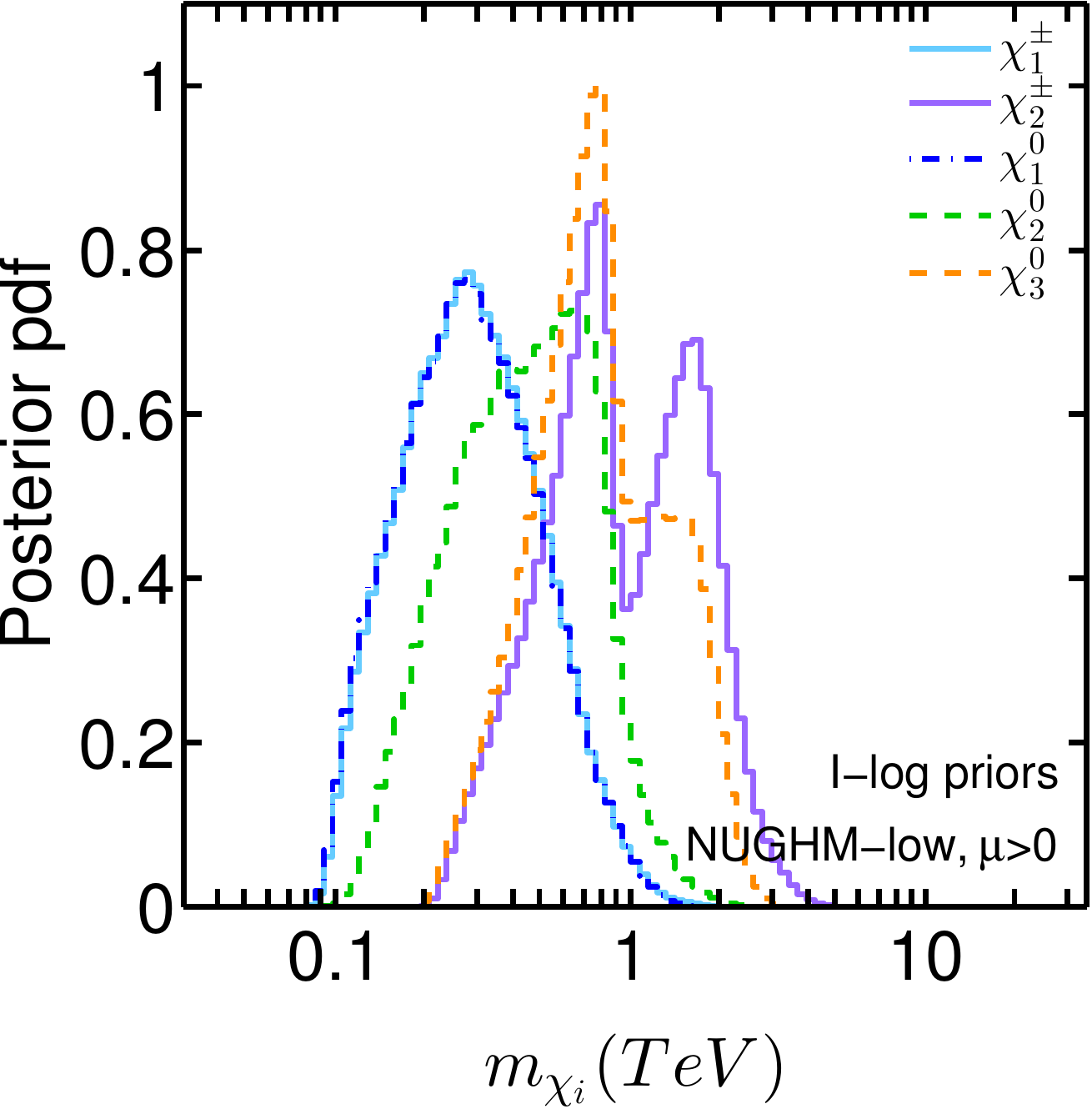}\\ \vspace{0.5cm}
  \includegraphics[width=0.415\linewidth]{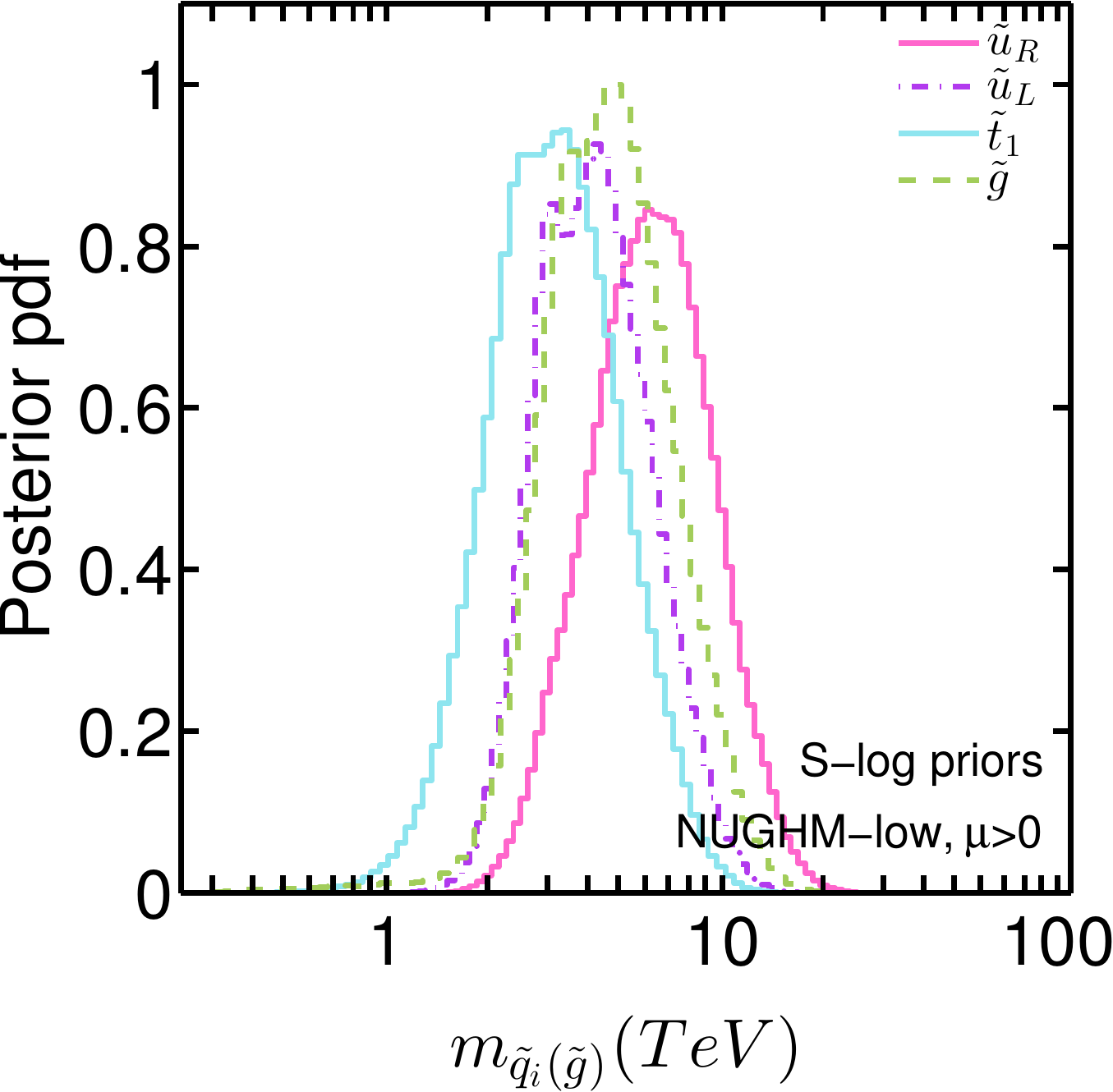}\hspace{1.0cm}
  \includegraphics[width=0.4\linewidth]{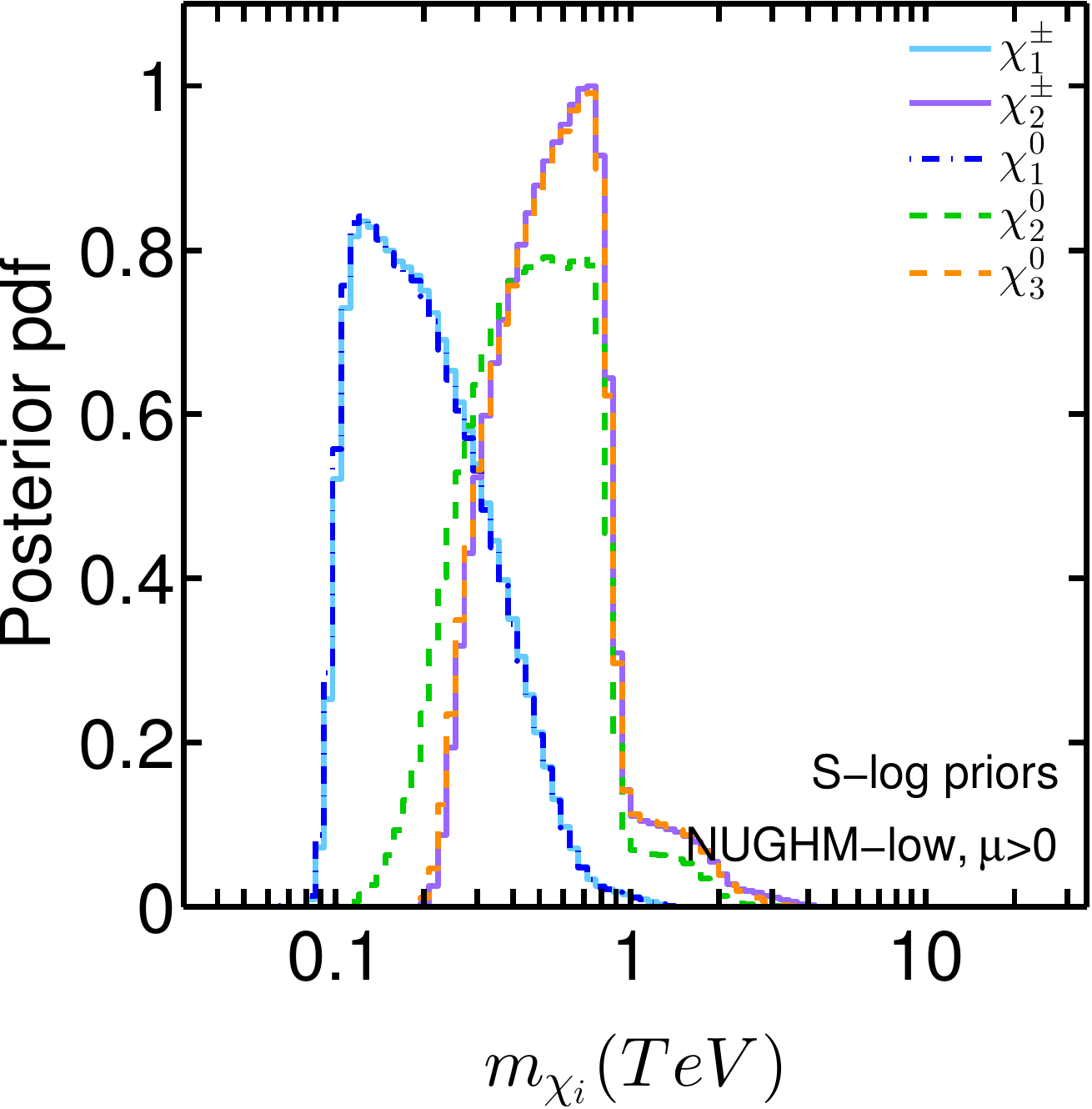}
  \caption{As Fig.\ref{fig:single1D} but for ``Low energy" NUGHM.}
  \label{fig:lowSusy1D}
\end{figure}

\begin{figure}[ht]
  \centering
  \includegraphics[width=0.3\linewidth]{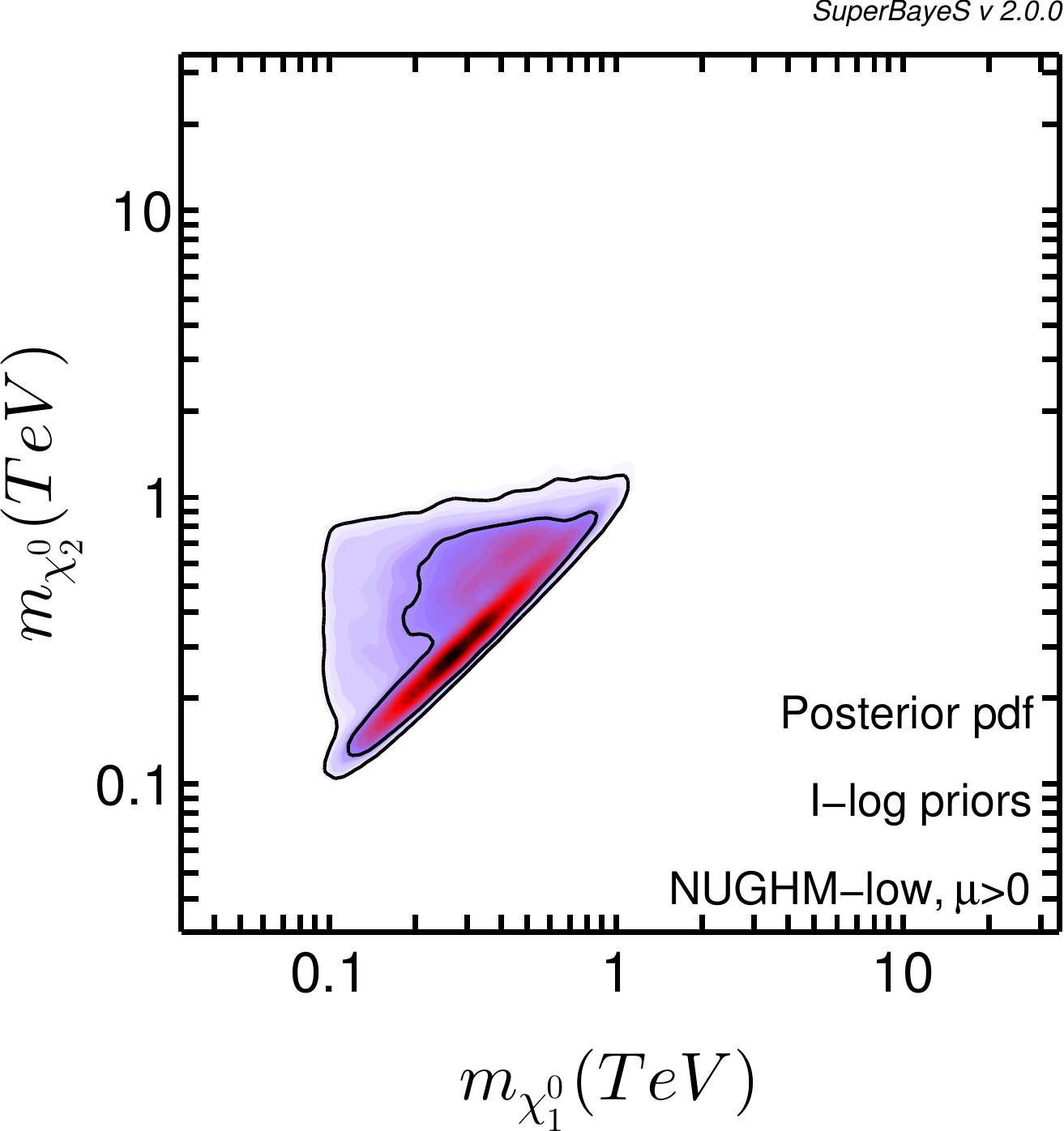}\hspace{0.5cm}
  \includegraphics[width=0.3\linewidth]{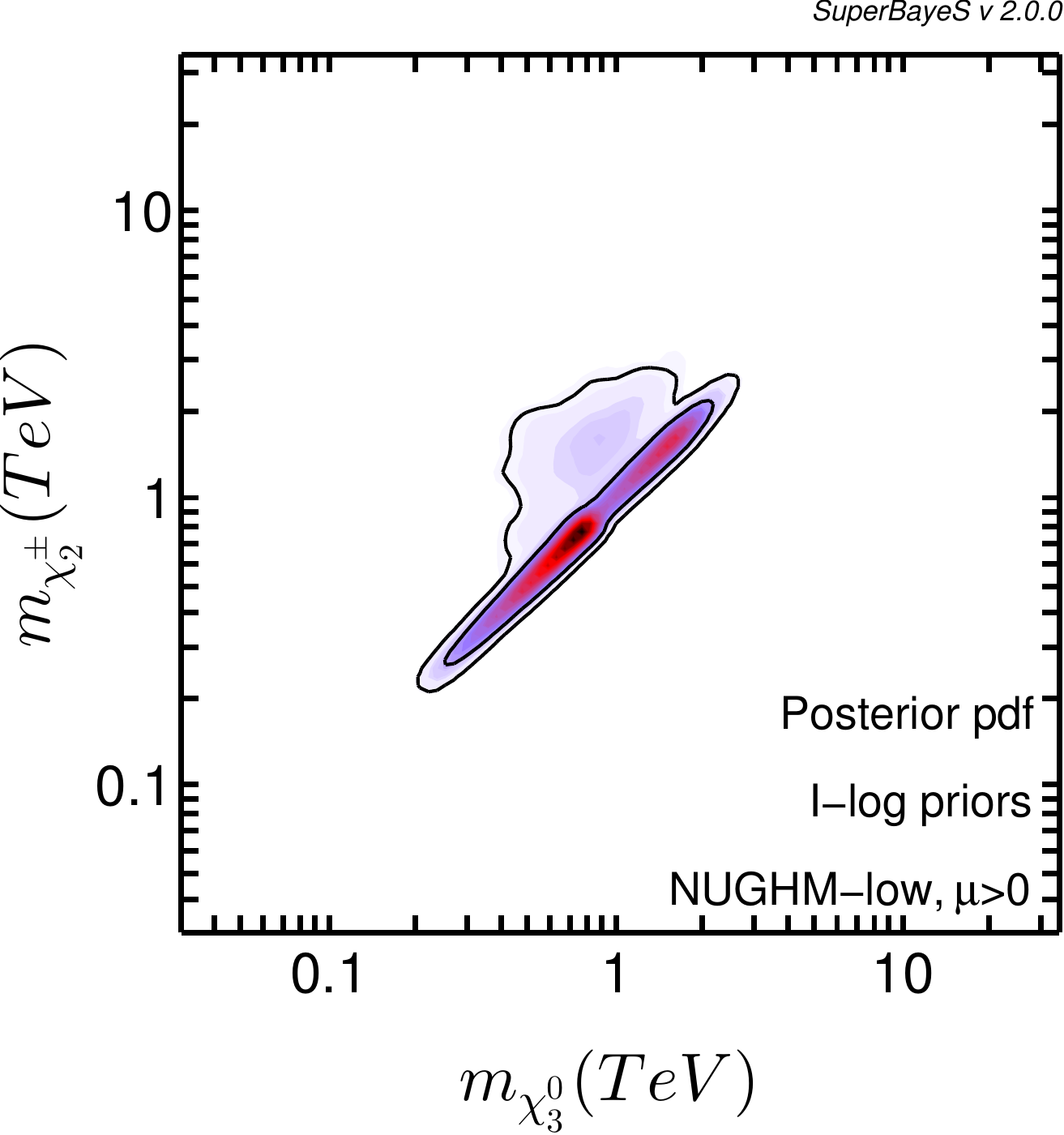}\hspace{0.5cm}
  \includegraphics[width=0.3\linewidth]{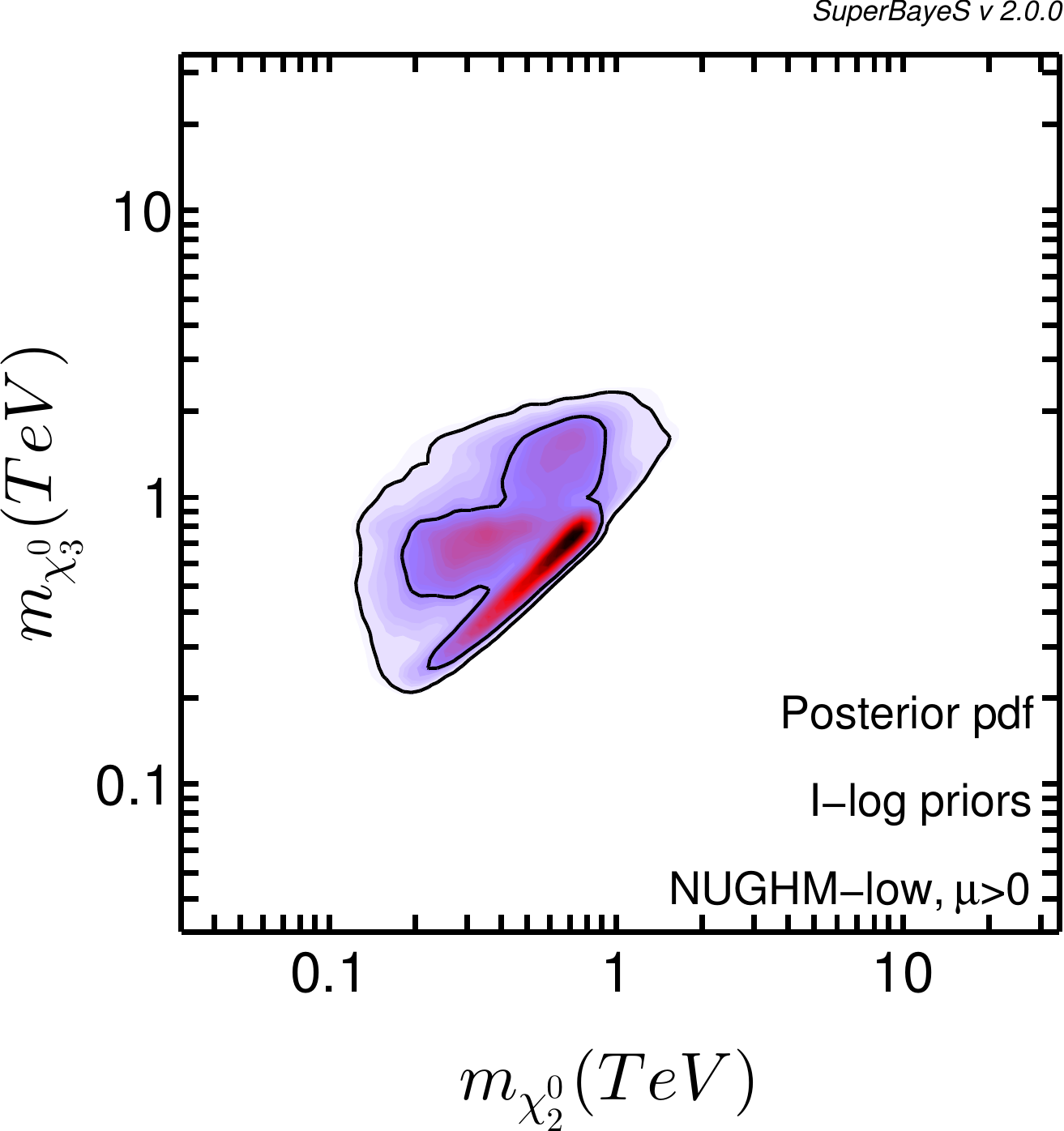}\vspace{0.5cm}
  \includegraphics[width=0.3\linewidth]{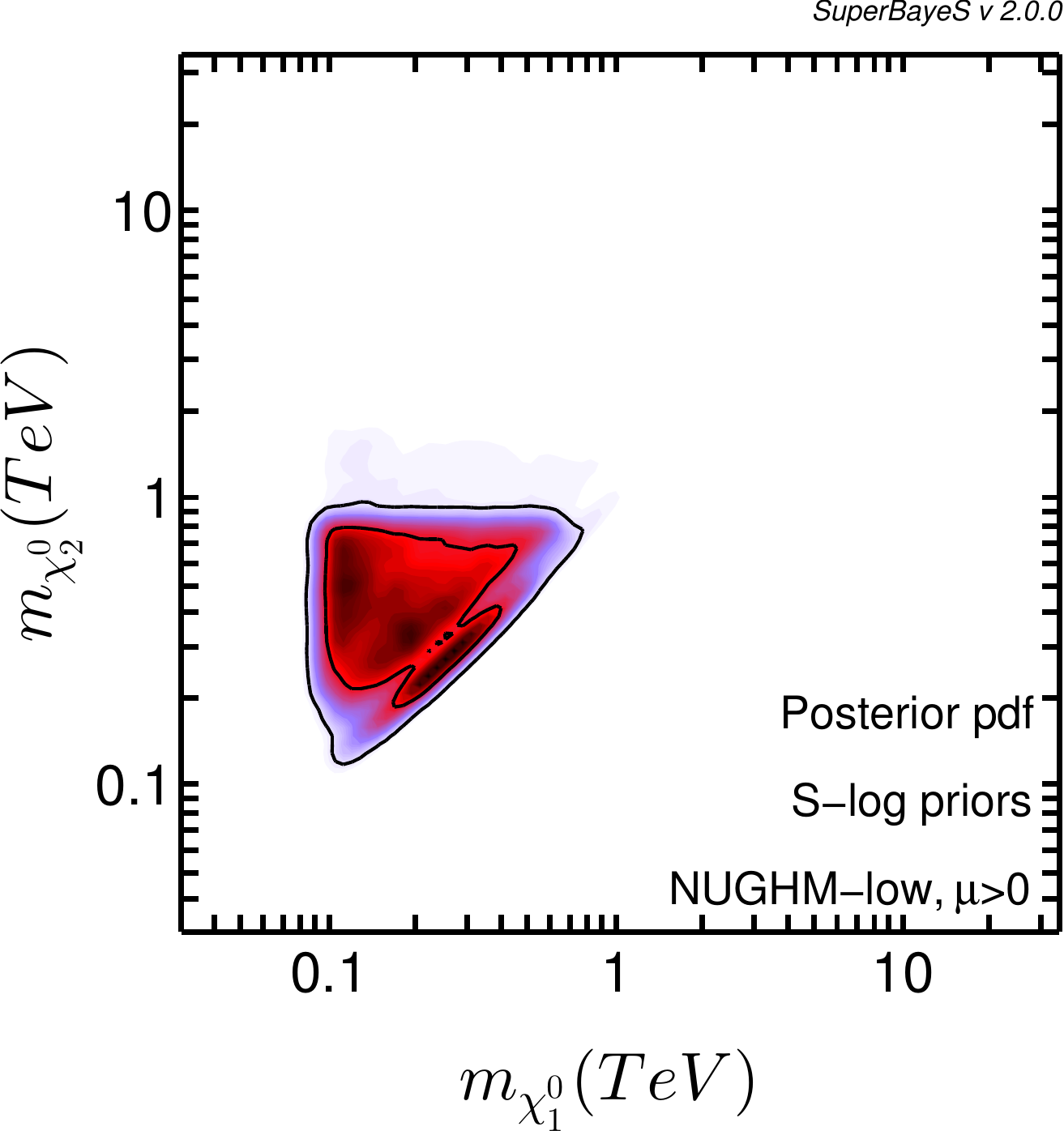}\hspace{0.5cm}
  \includegraphics[width=0.3\linewidth]{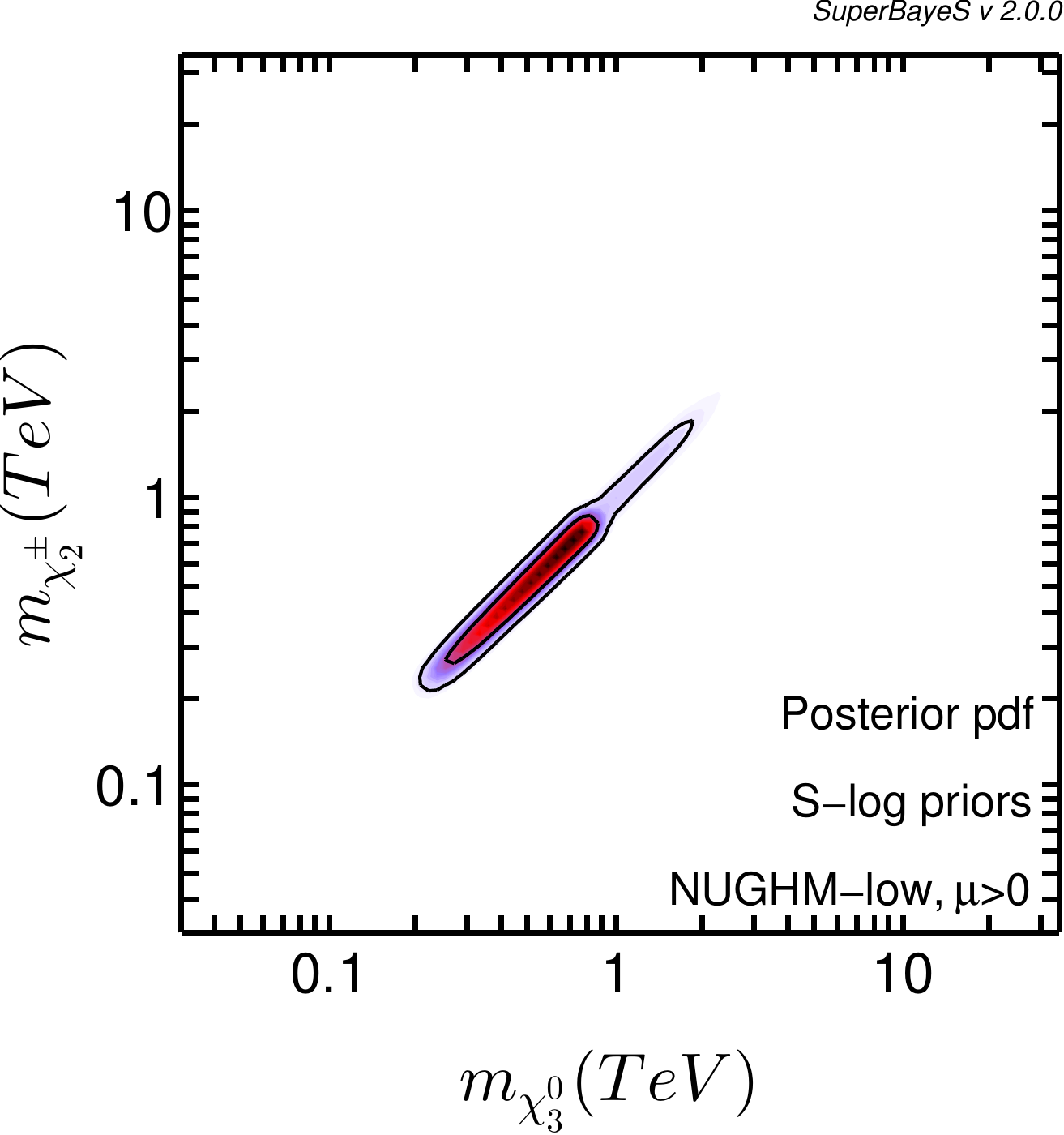}\hspace{0.5cm}
  \includegraphics[width=0.3\linewidth]{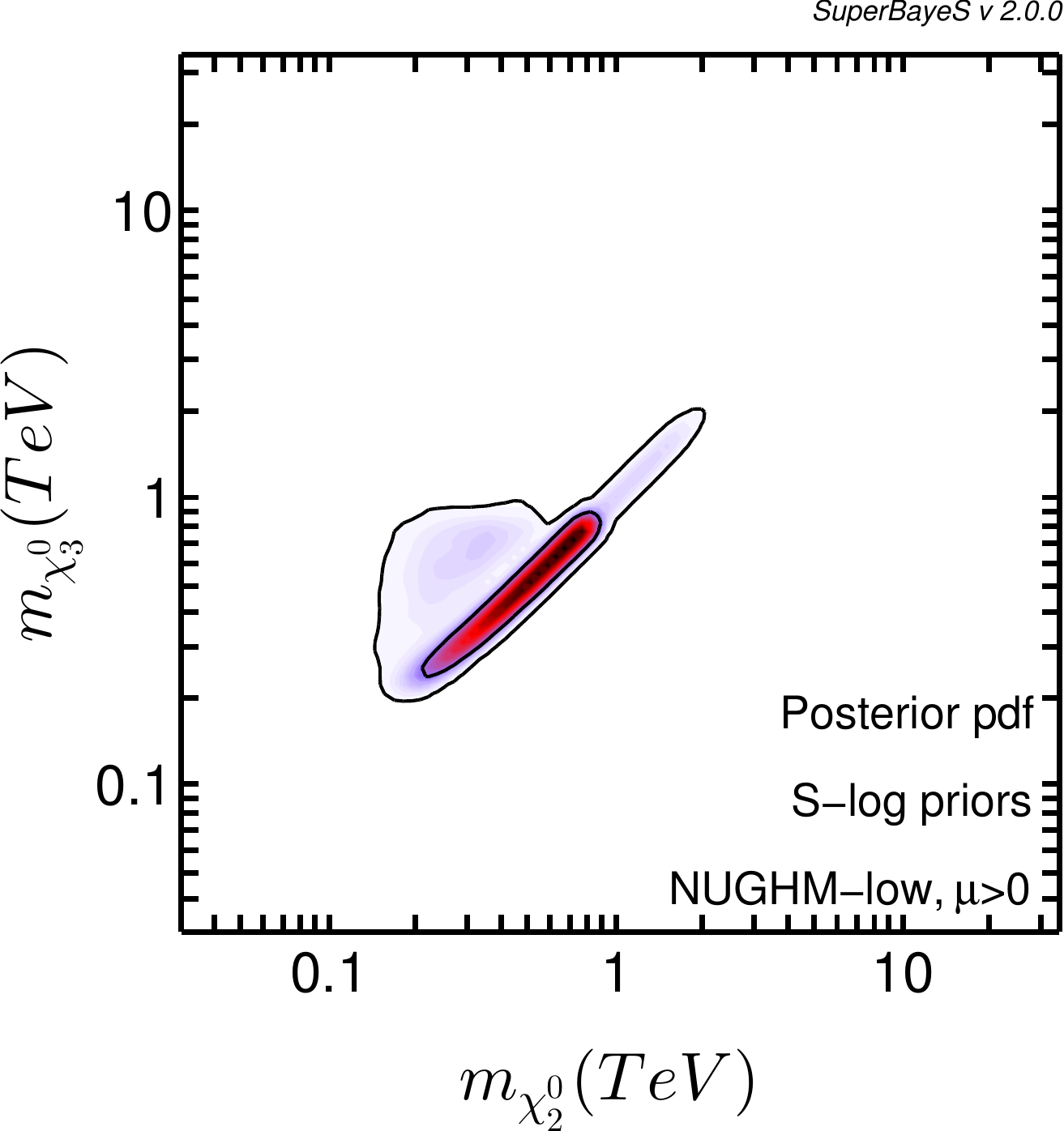}
  \caption{2D marginalized posterior probability for ``Low Energy" NUGHM on different planes defined by couples of supersymmetric masses. Upper (lower) panels correspond to I-log (S-log) priors. The 68\% and 95\% CL contours are shown. The color code is as in Fig.~\ref{fig:single2D}.}
  \label{fig:lowSusy2DNC}
\end{figure}

In addition, we will only require that the LSP abundance is equal or less than the observed DM abundance, thus allowing for a multi-component composition of DM, which is the most conservative assumption.

Figs.~\ref{fig:lowSusy1D} and \ref{fig:lowSusy2DNC} illustrate the spectrum of low-energy NUGHM. 
Concerning the strongly-interacting sector, it is quite similar to the spectrum before imposing the cuts of table~\ref{tab:lowSusyCuts}, compare to Fig.\ref{fig:multi1D}.

\begin{figure}[ht]
\centering 
\includegraphics[width=0.4\linewidth]{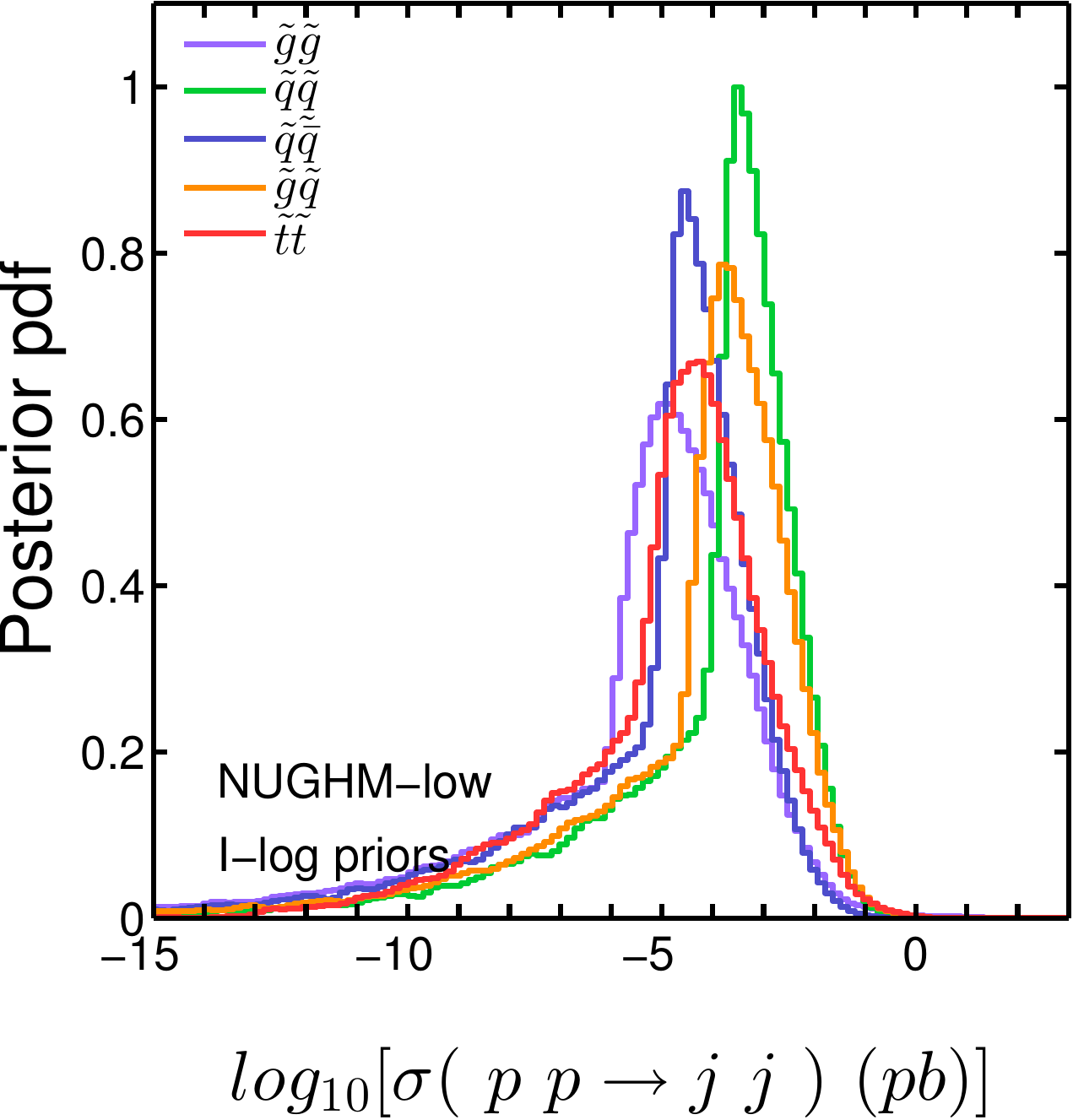}\hspace{1.0cm}
\includegraphics[width=0.4\linewidth]{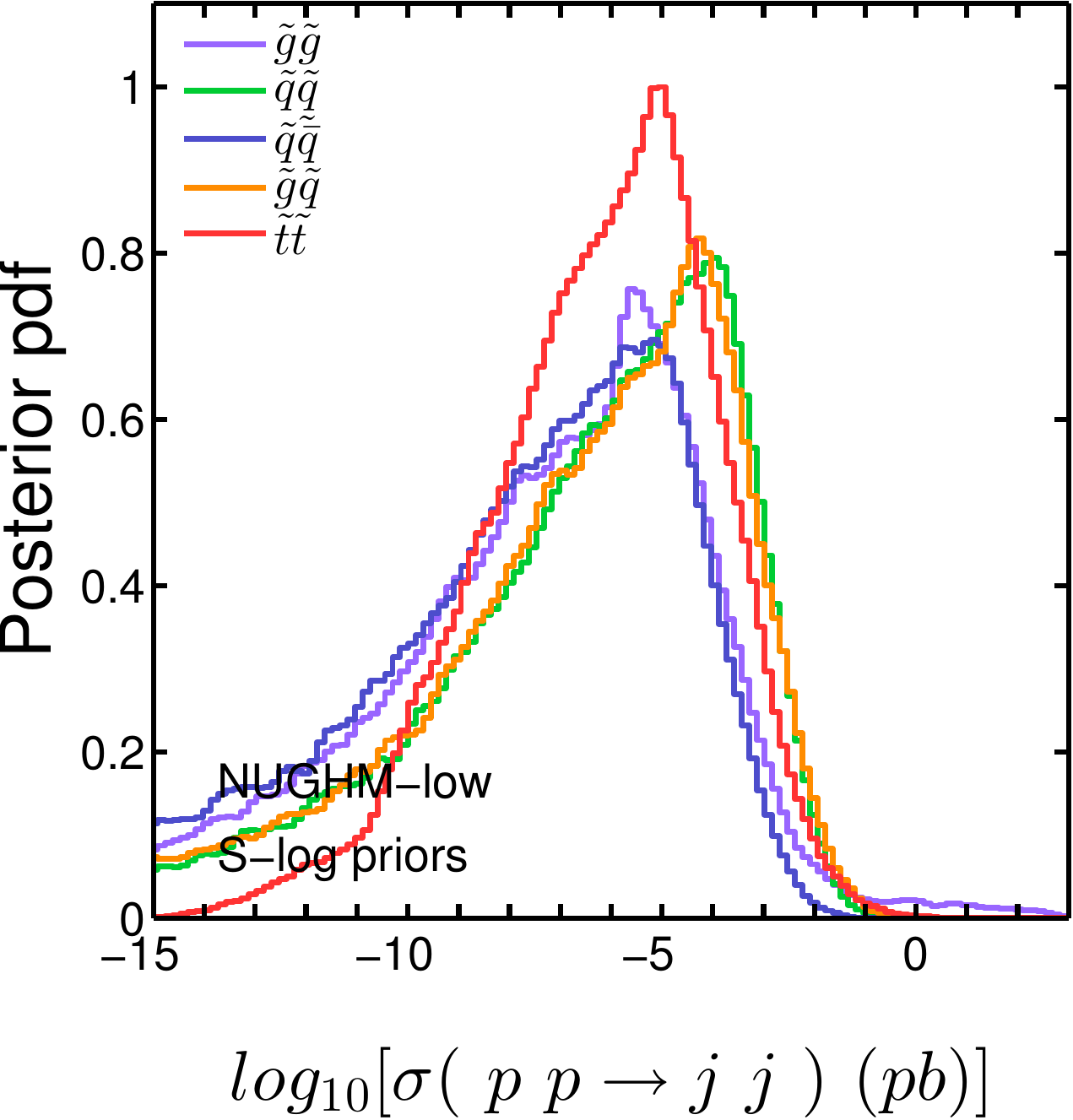}
\caption{1D posterior probability distribution of production cross-section of different pairs of colored supersymmetric particles pairs at $\sqrt{s}=14$ TeV for ``Log energy'' NUGHM. The left (right) panel corresponds to I-log (S-log) priors.}
\label{fig:sigma_lowSusy_color}
\end{figure}

Concerning the electroweakino sector, there are important differences, especially for the heavy chargino and neutralino states, which are crucial for the detectability at LHC. Fig.~{\ref{fig:lowSusy1D} show that $\chi_1^0$ and $\chi_1^\pm$ are (almost) always quasi-degenerate, meaning that the LSP is typically Higgsino or wino, exactly as it happened for the general NUGHM studied before. This is the reason why there is no condition on $\chi_1^\pm$ in table~\ref{tab:lowSusyCuts}, as it is difficult to detect an electroweakino that is quasi-degenerate with the LSP. Consequently, there are two different regions in the parameter space: either $\chi_1^0$ and $\chi_1^\pm$ are mostly winos (and quasi-degenerate), or $\chi_1^0$, $\chi_1^\pm$ {\em and} $\chi_2^0$ are mostly Higgsinos (and quasi-degenerate). 
In the former case the most relevant neutralino for LHC phenomenology is the lightest Higgsino-like one, which can be $\chi_2^0$ or $\chi_3^0$, being always quasi-degenerate with  $\chi_2^\pm$. In the latter case $\chi_2^0$ is mostly Higgsino and quasi-degenerate with $\chi_1^0$, so it  can easily escape LHC detection. Then the most relevant neutralino for phenomenology is the one which is mostly wino (again, the bino-like one is difficult to produce). It can be $\chi_3^0$ or $\chi_4^0$, but in either case is quasi-degenerate with $\chi_2^\pm$. In summary, in all cases the most relevant chargino is $\chi_2^\pm$ and the most relevant neutralino is quasi-degenerate with it; thus the constraint on $\chi_2^\pm$ given in table~\ref{tab:lowSusyCuts}.

According to the previous paragraph, there are four possible types of chargino and neutralino spectrum:

\begin{enumerate}
\item $\chi_1^0$, $\chi_2^0$ and $\chi_1^\pm$ are Higgsinos; $\chi_3^0$ and $\chi_2^\pm$ are winos; $\chi_4^0$ is bino.
\item $\chi_1^0$, $\chi_2^0$ and $\chi_1^\pm$ are Higgsinos; $\chi_3^0$ is bino; $\chi_2^\pm$ and $\chi_4^0$ are winos.
\item $\chi_1^0$ and $\chi_1^\pm$ are winos; $\chi_2^0$, $\chi_3^0$ and $\chi_2^\pm$ are Higgsinos; $\chi_4^0$ is bino
\item $\chi_1^0$ and $\chi_1^\pm$ are winos; $\chi_2^0$ is bino; $\chi_3^0$, $\chi_4^0$ and $\chi_2^\pm$ are Higgsinos.
\end{enumerate}

\begin{figure}[ht]
\centering 
\includegraphics[width=0.4\linewidth]{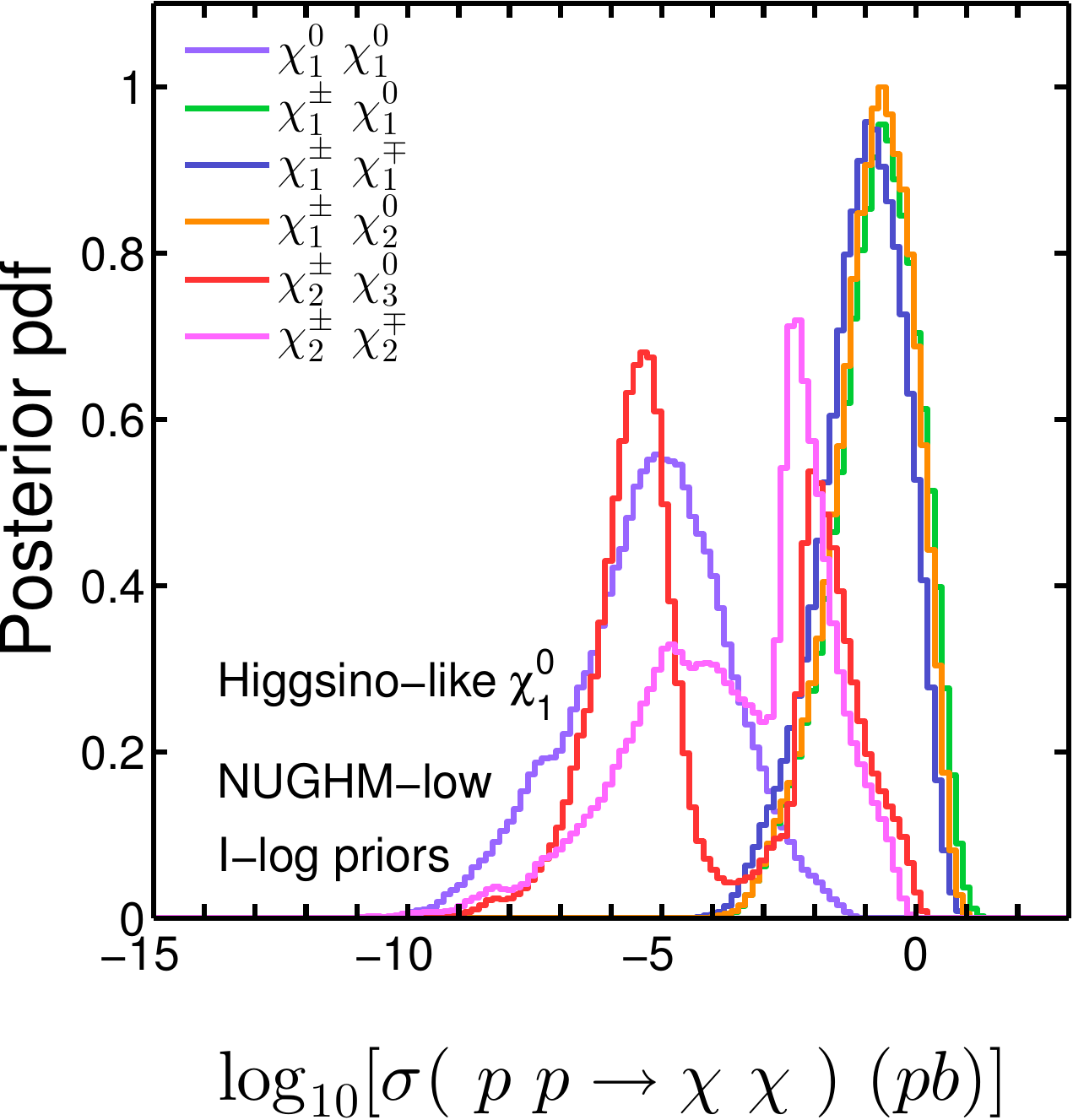}\hspace{0.5cm} 
\includegraphics[width=0.4\linewidth]{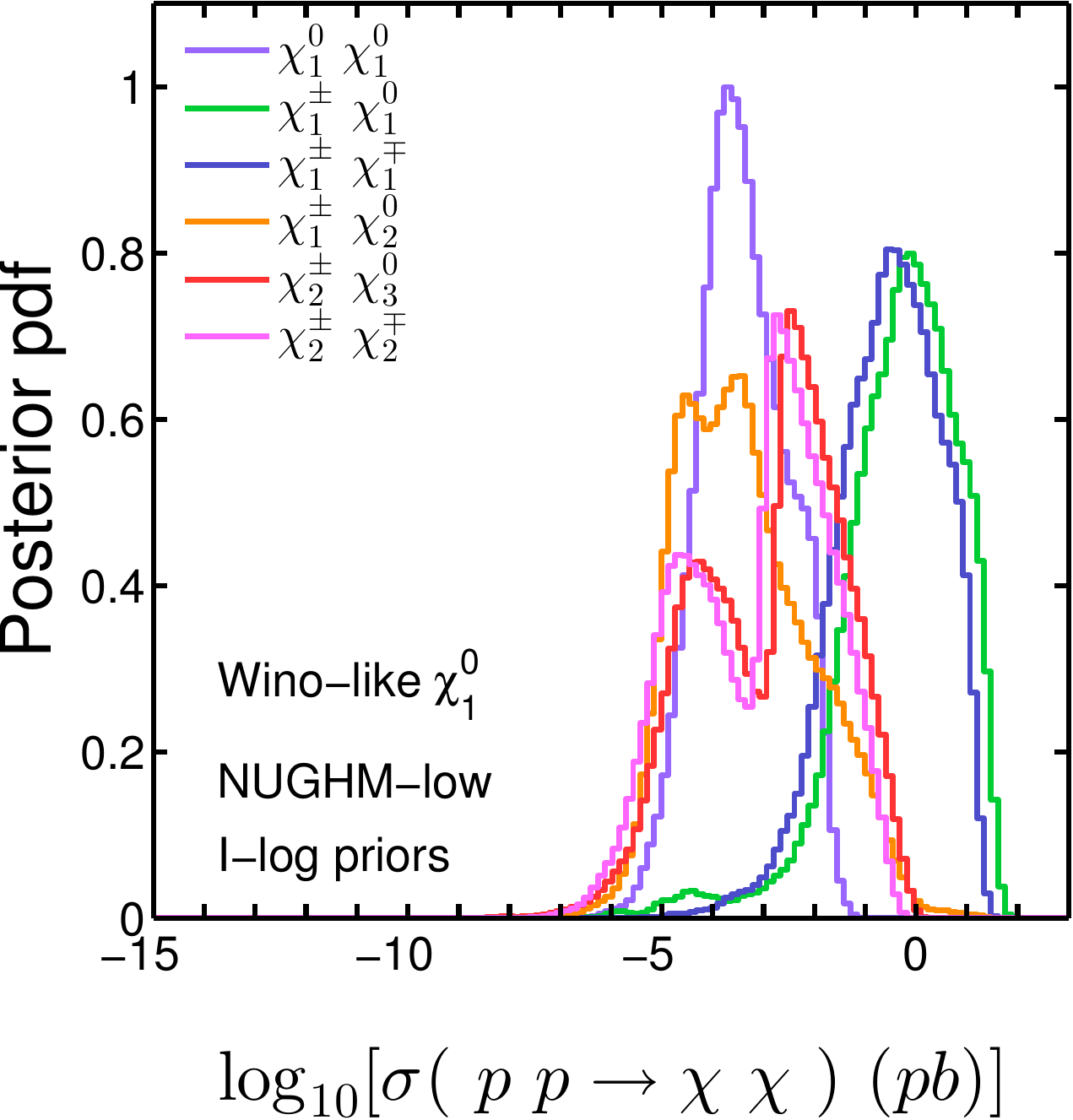}\\ 
\vspace{0.6cm}
\includegraphics[width=0.4\linewidth]{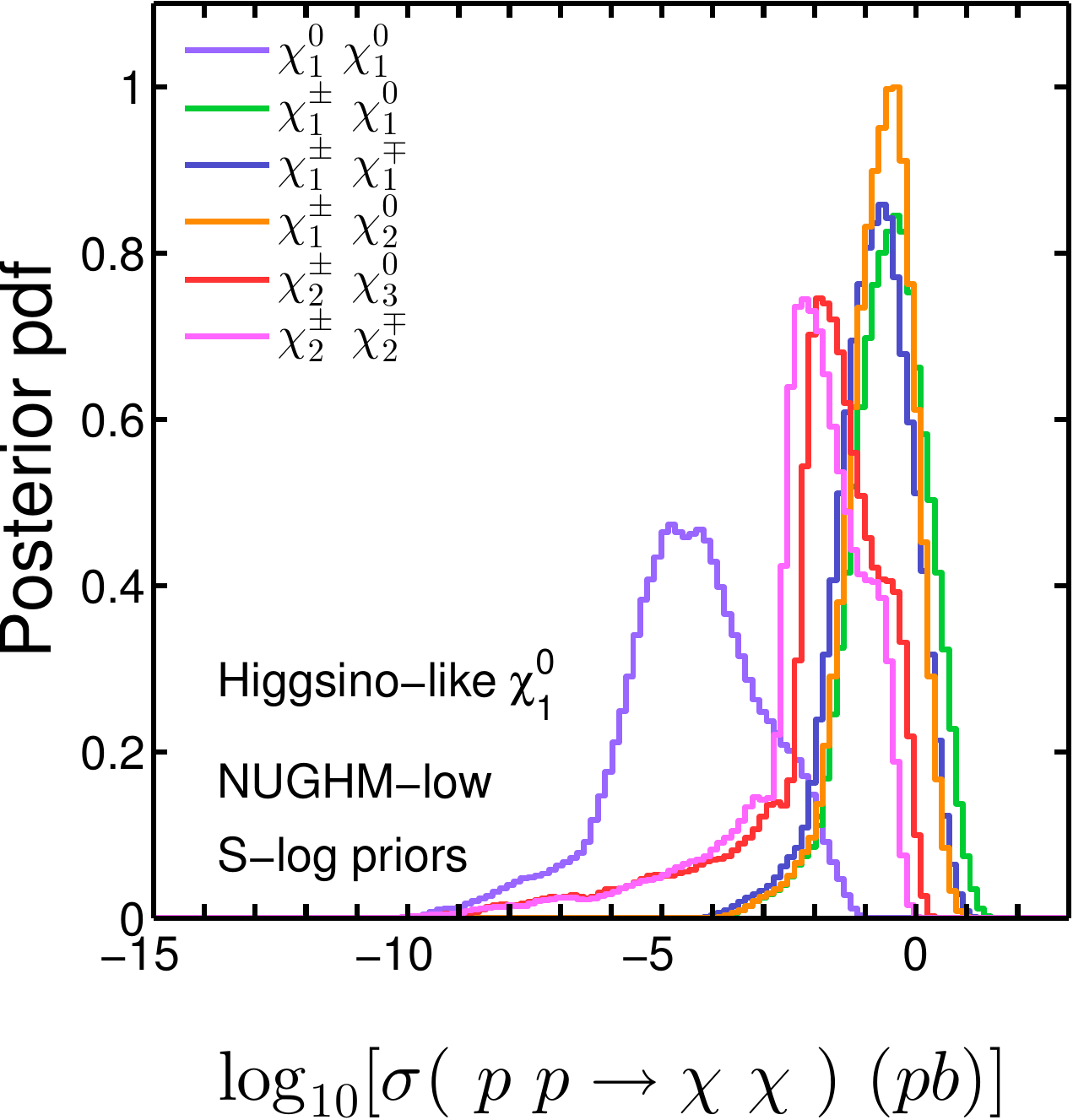}\hspace{0.5cm} 
\includegraphics[width=0.4\linewidth]{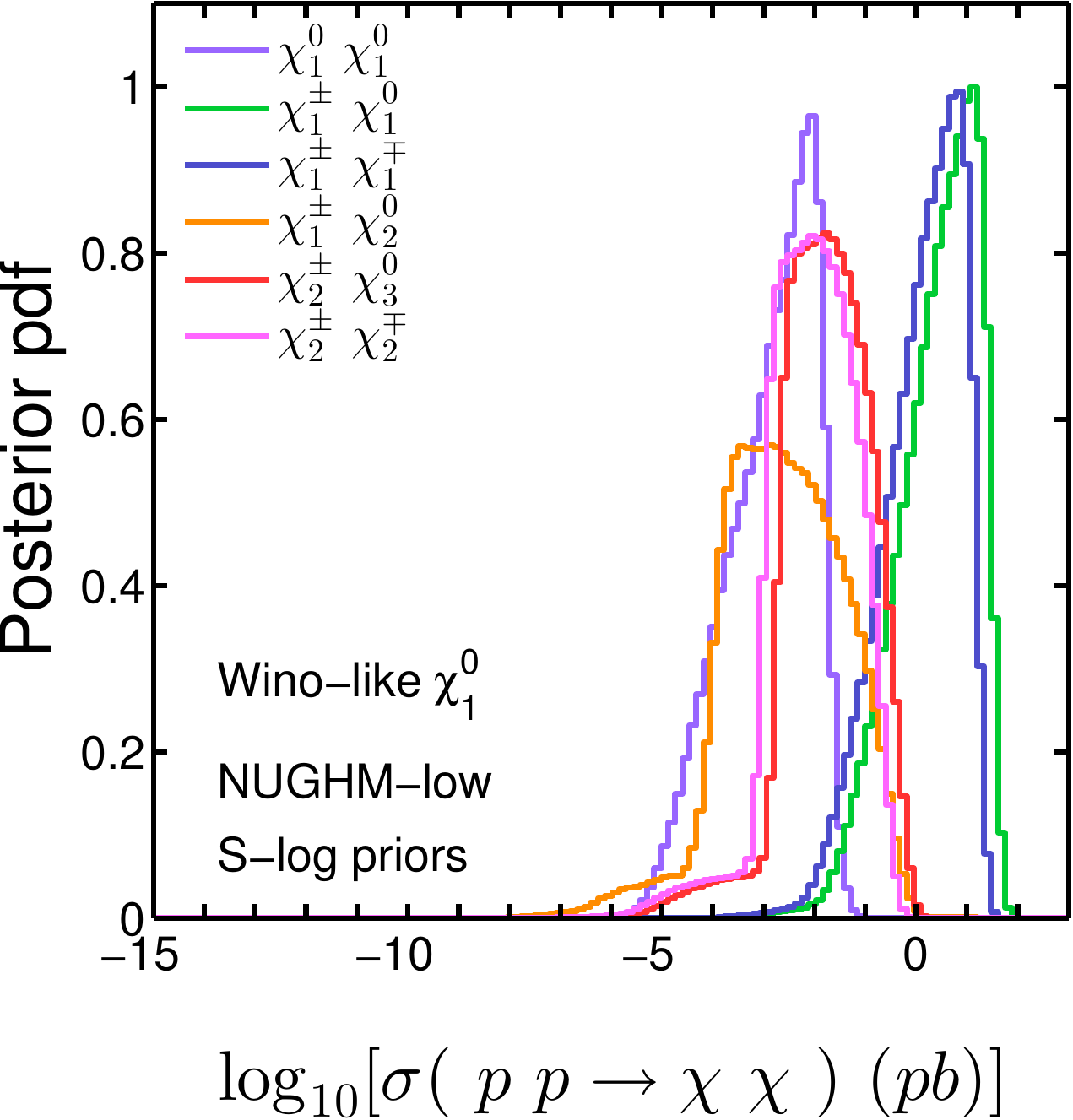}
\caption{1D posterior probability distribution of production cross-section of different electroweakino pairs at $\sqrt{s}=14$ TeV for ``Log energy'' NUGHM. Upper (lower) panels correspond to I-log (S-log) priors.}
\label{fig:sigma_lowSusy}
\end{figure}

The plots of Fig. \ref{fig:lowSusy2DNC} are useful to see the relative statistical weight of these possibilities using I-log priors (upper panels) or S-log priors (lower panels).
Namely, from the left panels we see the probability that $\chi_2^0$ is quasi-degenerate with  $\chi_1^0$, indicating the Higgsino-like nature of both (and $\chi_1^\pm$). For S-log priors the Higgsino-like character of the light electroweakinos is less abundant than the wino-like case, while for I-log priors both characters are more or less equally probable. The other plots give information about the character of the heavier charginos and neutralinos. The right panels show the probability that $\chi_2^0$ and $\chi_3^0$ are quasi-degenerate, indicating that $\chi_1^0$ is wino (which, as mentioned, is the typical case for S-log priors) and that $\chi_2^0$ and $\chi_3^0$ are mainly Higgsinos (and thus $\chi_4^0$ should be bino). Clearly, this is the usual situation for S-log priors, corresponding to spectrum 3 in the above list. For I-log priors the possibility that $\chi_2^0$ and $\chi_3^0$ are {\em not} quasi-degenerate becomes also relevant, as the butterfly-like plot shows. The regions outside the degeneracy correspond to spectra 1 and 2 in the above list (upper `wing" of the butterfly), and to 
spectrum 4 (lower wing of the butterfly), as can be easily deduced taking into account the constraints from table~\ref{tab:lowSusyCuts}. Finally, the central panels show that the probability that $\chi_3^0$ and $\chi_2^\pm$ are {\em not} quasi-degenerate is small (negligible for S-log priors), indicating that the heaviest neutralino is typically bino. This is completely different to what happens in the extensively studied CMSSM scenario, where the bino-like neutralino is always lighter than the wino-like one, due to the unification of gaugino masses at $M_X$ and the different RG equation for each gaugino. We see here that, once one allows for non-universality of gauginos (and there is nothing against this possibility), that instance becomes the less likely one. In summary, for S-log priors the typical spectrum is of type 3, while for I-log it is 1 or 3, though spectra 2 and 4 are also possible.

\begin{figure}[ht]
\centering 
\includegraphics[width=0.4\linewidth]{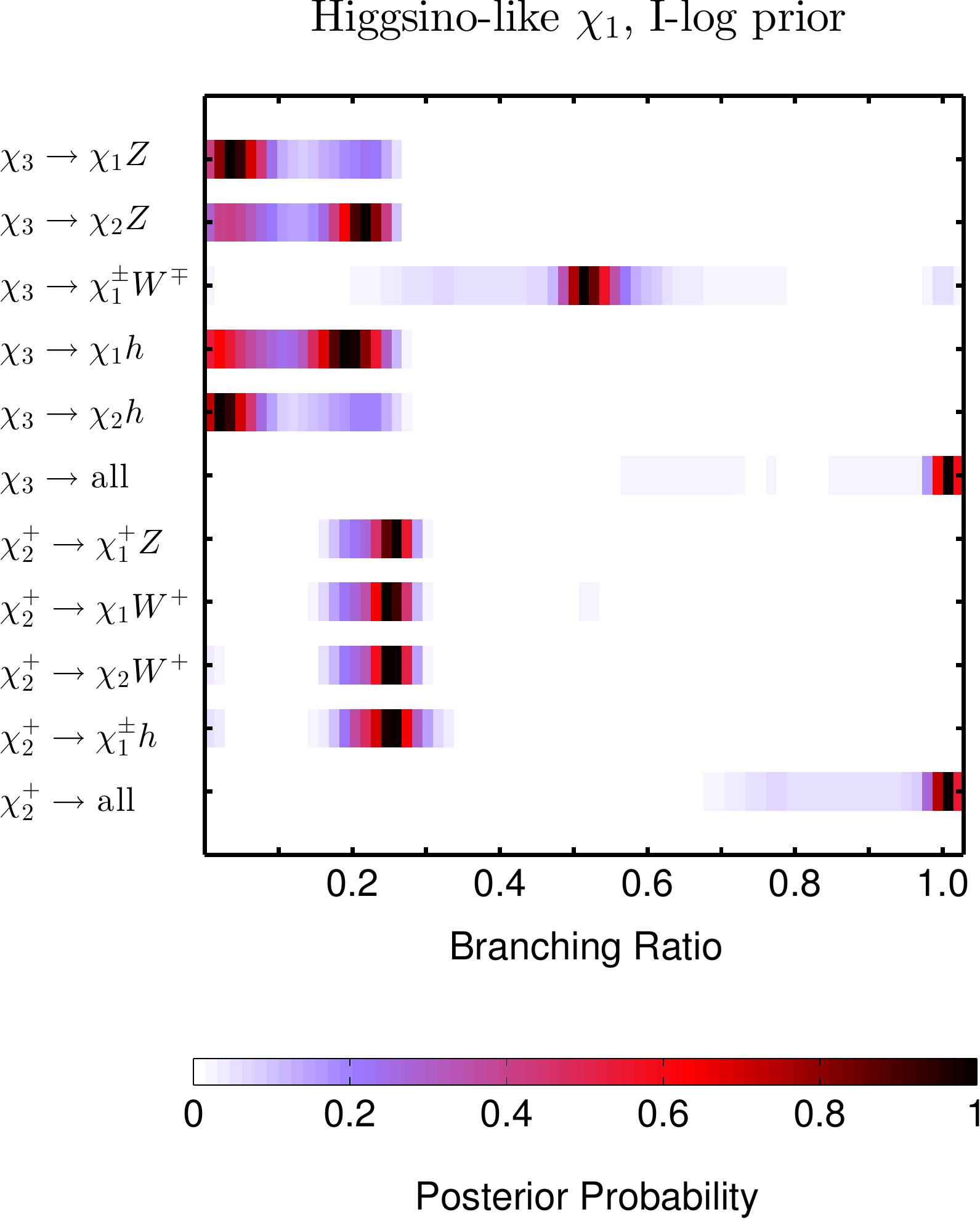}\hspace{0.5cm} 
\includegraphics[width=0.4\linewidth]{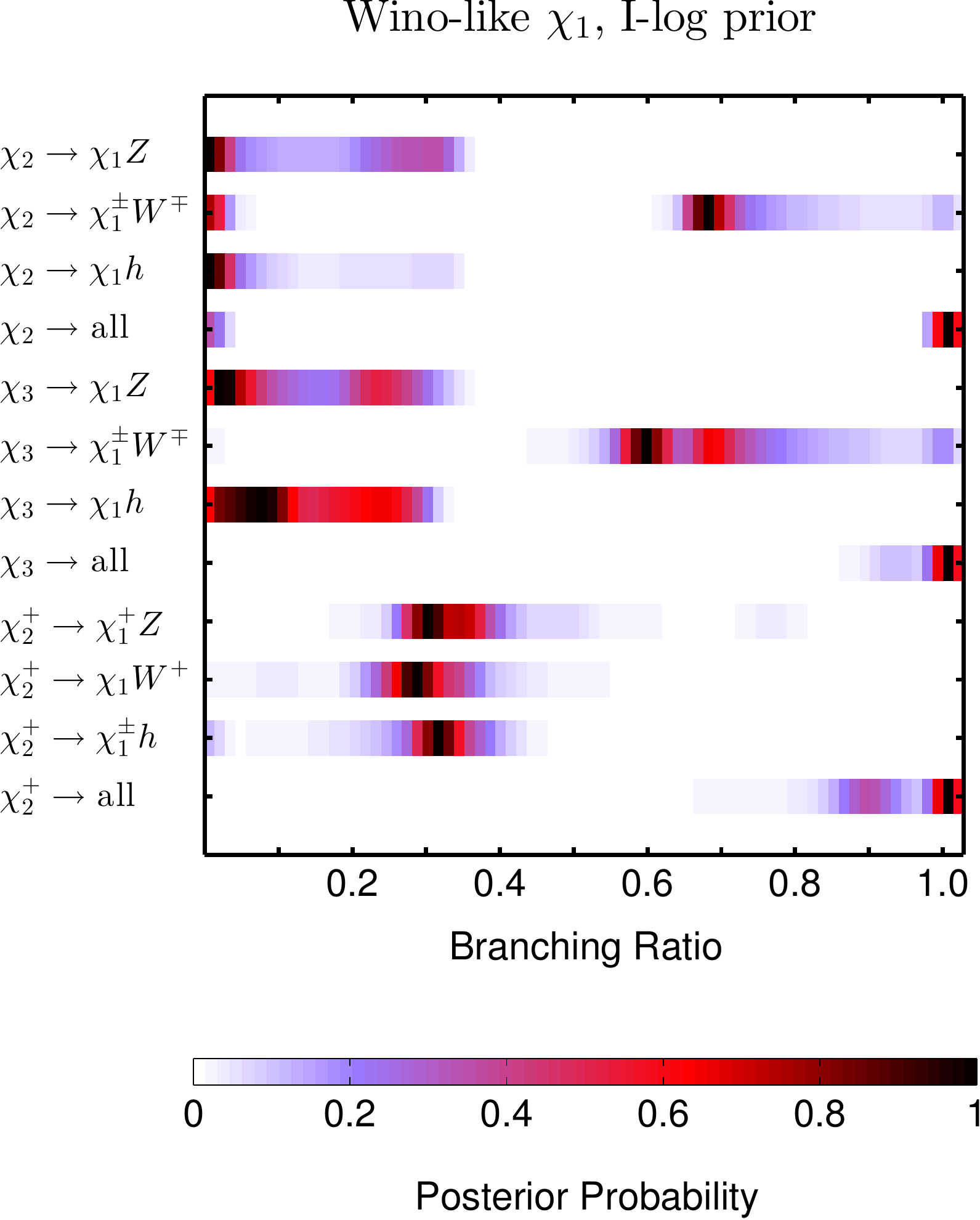}
\caption{Posterior pdf of the branching ratios of neutralinos and charginos
 with mass smaller than 800 GeV. The label ``$\chi_i \rightarrow$ all'' means the sum of
 the BRs of the processes showed in the plot.}
\label{fig:BRNC}
\end{figure}

Let us stress that the previous kinds of electroweakino spectrum are, to a large extent, determined by DM constraints. The observed DM abundance 
requires an efficient annihilation mechanism for the supersymmetric LSP. If the LSP is mainly Higgsino or wino, the anihilation rate is naturally much higher than for the bino case and, besides, it is reinforced by co-annihilation processes since in the Higgsino or wino cases the LSP is quasi-degenerate with other states. In the CMSSM this possibility can only be realized if the LSP is Higgsino-like (not wino-like).

Let us now consider the most likely supersymmetric signals at LHC. Concerning the colored sector, there exists a region of relatively light squarks, gluinos and stops (see Fig.~\ref{fig:lowSusy1D}), which will be testable at LHC, using standard techniques. This is illustrated in Fig.~\ref{fig:sigma_lowSusy_color}, which shows the probability of production of different pairs of colored supersymmetric particles ($\tilde q \tilde q$, $\tilde g \tilde g$, $\tilde q \tilde g$, $\tilde q \bar{\tilde q}$ and $\tilde t \bar{\tilde t}$) [vertical axis] with an specified cross section at 14 TeV center of mass energy  (horizontal axis) for S-log and I-log priors which have been computed at NLO with PROSPINO \cite{prospino}. There is an appreciable portion of the probability distribution which is potentially detectable at LHC. The part which is not corresponds to the region of the low-energy NUGHM parameter-space, that is potentially testable through the electroweakino sector, which we discuss next.

\begin{figure}[ht]
\centering 
\includegraphics[width=0.45\linewidth]{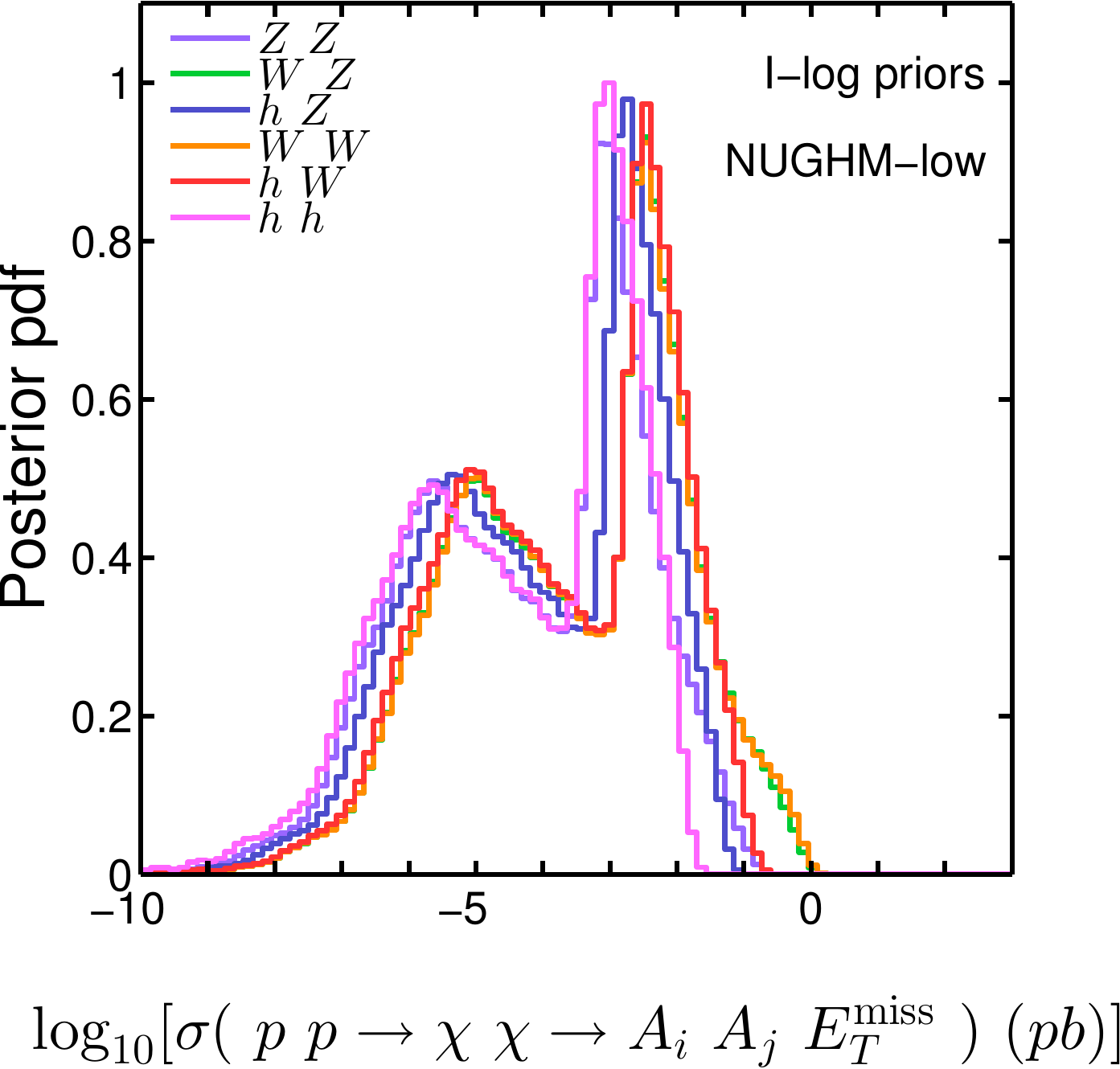}\hspace{0.5cm} 
\includegraphics[width=0.45\linewidth]{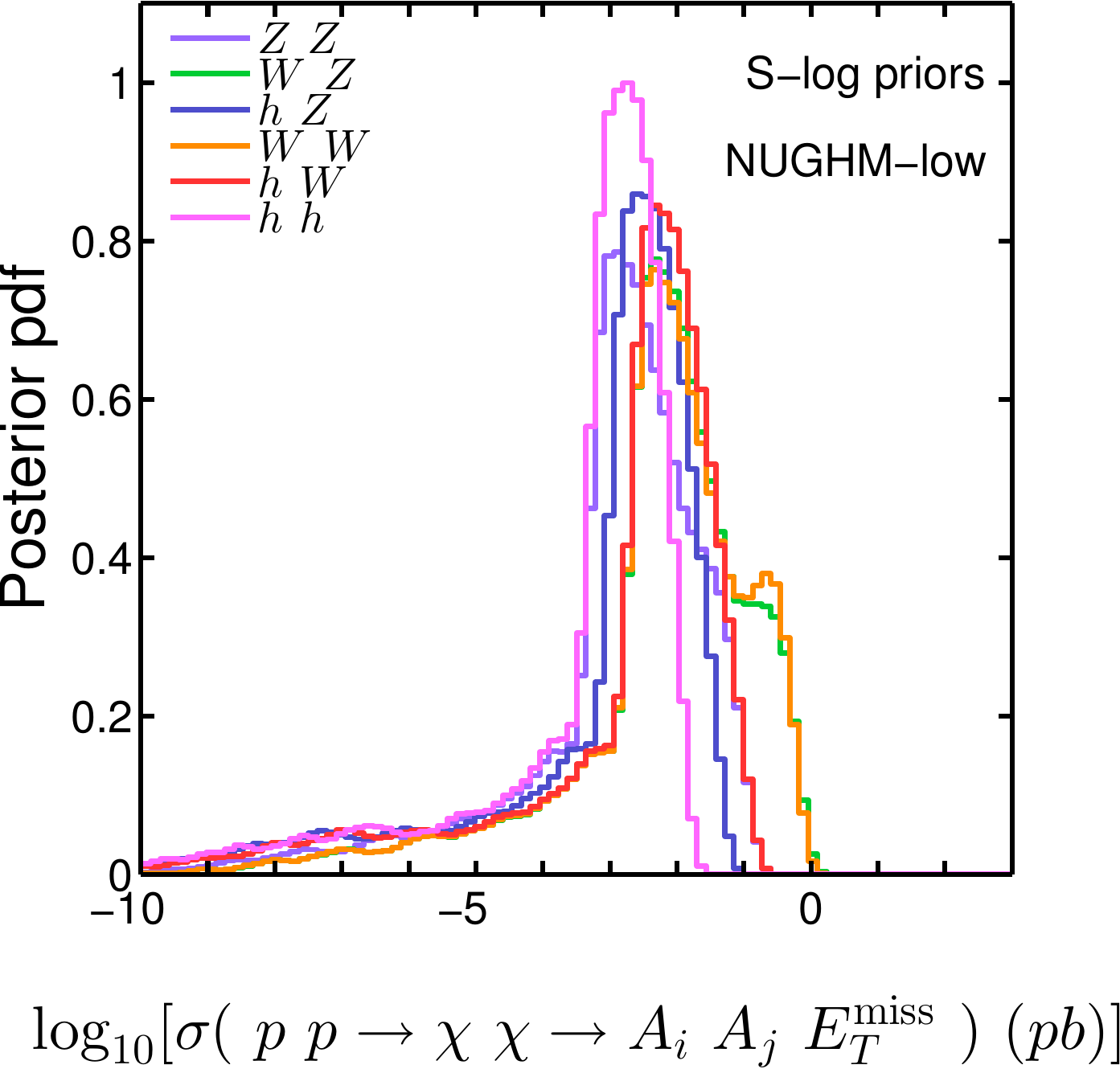}
\caption{1D posterior probability distribution of production cross section $\times$ branching ratios into different final states at $\sqrt{s}=14$ TeV for ``Low energy'' NUGHM. The left (right) panel corresponds to I-log (S-log) priors.}
\label{fig:sigmaBR_lowSusy}
\end{figure}

Indeed the electroweakino sector is also relevant for detection, and more intricate than the colored sector due to the different possibilities of spectrum and composition. Fig.~\ref{fig:sigma_lowSusy} shows the probability of production of different pairs of electroweakinos (vertical axis) with an specified cross section at 14 TeV center of mass energy (horizontal axis) for S-log and I-log priors as computed with prospino \cite{prospino}. Some of the pairs can be quite copiously produced; however this does not mean they are detectable at LHC. Note e.g. that the somewhat ``standard" $\chi_2^0\ \chi_1^\pm$ production is typically quite high, but the $\chi_1^\pm$ chargino is normally quasi-degenerate with $\chi_1^0$, thus leading to observable particles (e.g. leptons) with very small $p_T$, plus missing energy. Hence the usual analysis for the detection of this pair \cite{ATLAS-CONF-2013-035, CMS-PAS-SUS-13-006,Cabrera:2012gf} is not useful here. The $\chi_2^0$ neutralino is also quasi-degenerate with $\chi_1^0$ in the case where both are Higgsino-like.  On the other hand, from Fig.~\ref{fig:sigma_lowSusy} we see that there exists an non-negligible probability of production of heavier states of neutralinos and charginos, with appreciable cross-sections, able to give detectable signals. This is therefore an additional way to detect NUGHM at LHC, besides the production of squarks and gluinos. 

In order to determine the most probable final states coming from the decay products of the heavy electroweakino states, one has to multiply the previous production cross sections by the corresponding branching ratios. Fig.~\ref{fig:BRNC} illustrates, for I-log priors, the probability  in the NUGHM parameter space of having different values for the branching ratios of the various decay chanels of $\chi_2^0$ (irrelevant if the LSP is Higgsino),  $\chi_3^0$ and $\chi_2^\pm$ as computed with SUSY-HIT \cite{shit}. It is worth-noticing that the combination of the production cross sections (Fig.~\ref{fig:sigma_lowSusy}) and the branching ratios (Fig.~\ref{fig:BRNC}) somehow define the simplified models that describe the most likely LHC phenomenology of the NUGHM.

\begin{figure}[ht]
\centering 
\includegraphics[width=0.4\linewidth]{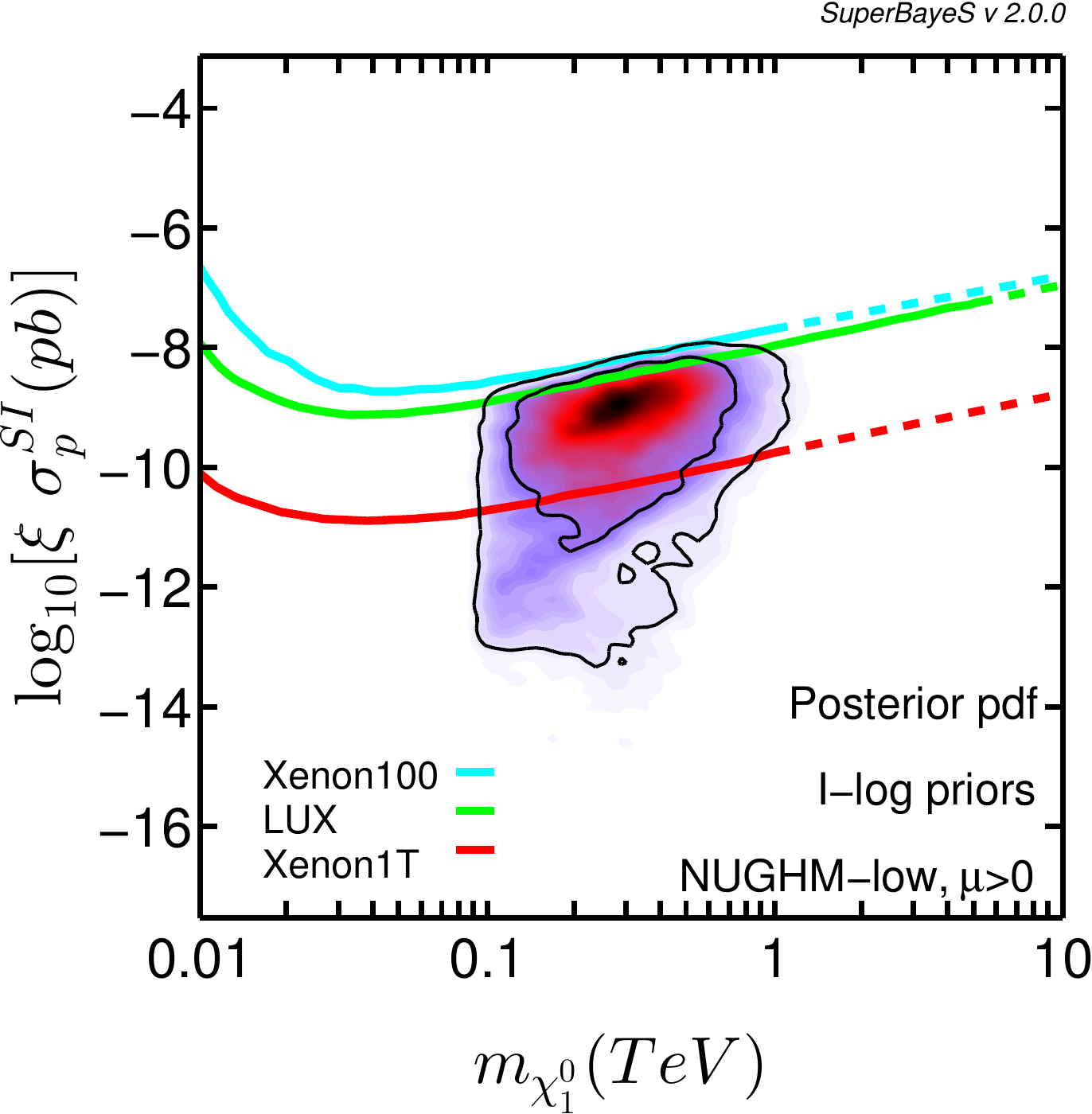}\hspace{1.0cm}
\includegraphics[width=0.4\linewidth]{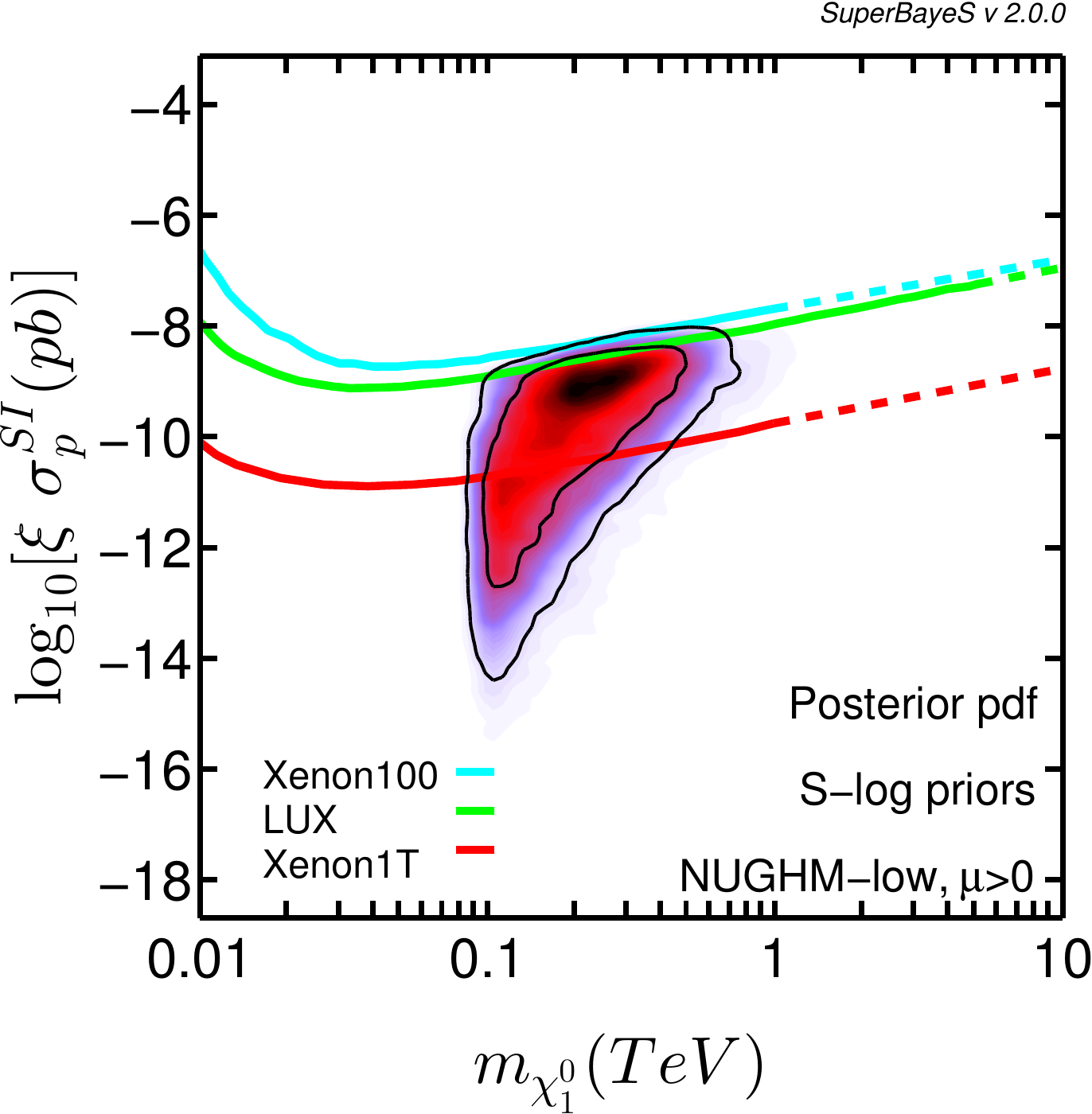}
\caption{As Fig.\ref{fig:singleCDM2D} but for ``Low Energy" NUGHM.}
\label{fig:lowCDM2D}
\end{figure}

Combining the production cross sections with the branching ratios, one finally obtains the probability of electroweakino-mediated production of different final states, which is shown in Fig.~\ref{fig:sigmaBR_lowSusy}. The most promising final states are $WZ$, $WW$, and also $ZZ$ and $hW$. Note that the $WW$ production does not require the $W$s to have opposite signs, which increases the signal/background signal. On the other hand the $WZ$ final state can be analyzed in the standard way (studying 3-lepton production), although the interpretation is not the usual $\chi_2^0\ \chi_1^\pm$ production. Notice that the supersymmetric $WZ$ and $WW$ productions are predicted to be of the same size in all cases, a distinctive feature of the scenario. 

{Note that the portion of the probability distribution in Fig.~\ref{fig:sigmaBR_lowSusy} which is not testable at LHC corresponds to the region that is potentially testable through the production and detection of colored supersymmetric particles. This also allows to see that both types of searches --colored supersymmetric particles or electroweakinos-- are more or less equally favorable for I-log priors, while for S-log priors electroweakino searches represent the most favorable strategy to test the NUGHM at LHC.

In summary, the most distinctive signals of the low-energy NUGHM are, besides the standard production of colored supersymmetric particles, the production of (heavy) electroweakinos, giving $WZ$ and same-sign $WW$ final states plus missing energy.

The previously discussed potential signals at LHC are complementary to the DM detection prospects. Fig.~\ref{fig:lowCDM2D} shows the pdf in the $m_{\chi_1^0}-\sigma^{SI}$ plane, where $\sigma^{SI}$ is the Spin--Independent cross section for DM direct detection. The present XENON100 and LUX limits, as well as the future XENON1T ones, are shown as well. The region of large $m_{\chi_1^0}$ (say $\simgt 500$ GeV) is very difficult to test at LHC, however it will be almost completely tested by XENON1T. For lower $m_{\chi_1^0}$ the region testable by XENON1T decreases, but this is precisely the region where the LHC signals discussed above become potentially visible. Still, inside that region, the LHC discovery potential depends on the mass of the $\chi_2^\pm$ chargino. In summary, some regions of the ``Low Energy" NUGHM are more easily testable in DM-detection experiments than at LHC. Other regions are more easily testable at LHC. A substantial part can be tested in both ways, which is extremely interesting; and there remains a part which is difficult to test in any of them.

%%%%%%%%%%%%%%%%%%%%%%%%%%%%%%%%%%%%%%%%%%%%%%%%%%%%%%%%%%%%%%%%%%
\section{Conclusions}
%%%%%%%%%%%%%%%%%%%%%%%%%%%%%%%%%%%%%%%%%%%%%%%%%%%%%%%%%%%%%%%%%%

In this paper we have considered a quite generic MSSM model, namely one with non-universal gaugino masses and Higgs masses (NUGHM). Although not completely general, this scenario is well-motivated by a number of theoretical and phenomenological facts and goes far beyond the CMSSM and NUHM in complexity and phenomenological richness. 
 
From the theoretical point of view, the universality of gaugino masses is not supported by any robust theoretical fact or phenomenological requirement, in contrast with sfermion-mass universality, which is, at least partially, endorsed by flavour violation constraints. From the phenomenological side, the main merit of the NUGHM is that it essentially includes all the possibilities for DM candidates within the MSSM, since the neutralino spectrum and composition (as well as the chargino ones, which are relevant for co-annihilation processes) are as free as they can be in the general MSSM framework; and they are not correlated to the gluino mass, which is severely constrained by LHC. On the other hand, even if the sfermion masses are heavy --as the experimental Higgs mass, the present bounds on squarks and flavor issues may suggest-- there are reasons to expect fermionic supersymmetric states to be around the TeV range or even lighter. Namely, this is required not only for DM issues but also to  keep the succesful gauge unification that occurs in the MSSM. In addition, the presence of light charginos and neutralinos is probably the most robust consequence of ``Natural SUSY" scenarios, i.e. those with as-small-as-possible fine-tuning. Consequently, the production of charginos and neutralinos ("electroweakinos") at the LHC may be one of the best motivated avenues to detect SUSY at the LHC. In this sense, the NUGHM captures the rich phenomenology associated to these states in the general MSSM. Hence, one of the advantages of the NUGHM is that, with eight initially independent parameters (seven plus the $\mu-$parameter, which is solved in terms of $M_Z$), it catches most of the phenomenological subtleties of the MSSM without the enormous complexity of the general MSSM parameter-space (or the so-called pMSSM, with 21 free parameters). 

We have performed a Bayesian analysis to obtain a map of relative probability of the different regions of the NUGHM parameter-space, i.e. a forecast for the NUGHM. For this analysis, we have been very careful in the treatment of the naturalness issue. Actually, a penalization for the fine-tuned regions appears automatically from the Bayesian analysis itself (without any ad-hoc assumptions) once $M_Z^{\rm exp}$ is treated on the same footing as the rest of the experimental information. This procedure has the additional advantage that the results are independent of the ranges chosen for the parameters, indeed we do not need to impose that they are smaller than any ${\cal O}(10)$ TeV limit (as usual in Bayesian analyses). Concerning the choice of the prior for the parameters, we have used an improved version of the logarithmic prior, that incorporates the fact that all the soft parameters have a common origin. However, we have performed the whole analysis using the standard logarithmic prior as well, in order to visualize the prior-dependence of the results. For the likelihood piece of the study we have used all the present experimental observables (except $g-2$) plus the most recent constraints on DM direct detection.

If one requires that the neutralino makes all of the DM in the Universe (single-component CDM case), the results of the analysis show two preferred regions around $m_{\chi_1^0}= 1\ {\rm TeV}$ and $3\ {\rm TeV}$, which correspond to the (almost) pure Higgsino and wino cases. The masses of squarks and gluino are typically above 3 TeV; and the stops' above 1 TeV. All this makes quite challenging to detect such scenario at LHC. On the other hand, it is remarkable that the $95\%$ c.l. regions, though almost `untouched' by the recent LUX results on direct DM detection, will be almost fully tested by the future XENON1T and similar experiments. Thus DM detection becomes a favorite fashion to detect a supersymmetric single-component DM scenario.
Beside the previous regions, there are other viable regions in the parameter-space corresponding to the Higgs funnel annihilation (not viable in the CMSSM), $A-$funnel and stau-co-annihilation. All of them have little statistical weight compared to the pure Higgsino and wino case. Anyway we have addressed their typical features and associated phenomenology (especially for the Higgs-funnel, which is a bonus chance in the NUGHM). Again, direct DM detection becomes a more efficient way to test these somewhat marginal regions.

When the DM constraints are relaxed, just requiring that the supersymmetric DM abundance is equal {\em or less} than the observed one (multi-component CDM case), the prospects for probing this scenario at the LHC become more promising. The masses of the lightest neutralinos and charginos become typically between 100 GeV and 1 TeV. However, this is not enough for LHC detection, at least using the production of $\chi_1^+\ \chi_1^-$ and $\chi_2^0\ \chi_1^\pm$ pairs. The reason is that those states are typically quasi-degenerate with
$\chi_1^0$, and do not produce final  states with high $p_T$ in their decays. On the other hand, the production and decay of heavier neutralino and chargino states can give detectable signals, whenever they are light enough. We have performed a zoom into the region which is potentially detectable at LHC, finding that the most typical detectable signals of NUGHM at LHC are precisely the production of heavy charginos and neutralinos, which decay giving either WZ or (same sign) WW final states, plus missing energy; together with the ``traditional" signals associated to the production of gluinos and squarks. Finally, once more, even though the actual DM is not assumed to be mostly supersymmetric, DM detection remains a complementary, and even more powerful, way to probe the scenario.

\em Acknowledgements.}  M.E.C thanks Paul de Jong and Francesca Calore for
useful comments an discussions. We also thank Marco Cirelli and St\'ephane
Lavignac for bringing our attention to the Sommerfeld-enhancement effect for
wino annihilation, and to Andrzej Hryczuk for lending us a numerical routine
and giving extremely useful advice to check the Sommerfeld-enhancement
effect. We also thank Dan Hooper for useful advise. R. RdA, is supported by
the Ram\'on y Cajal program of the Spanish MICINN and also thanks the support
of the Invisibles European ITN project (FP7-PEOPLE-2011-ITN,
PITN-GA-2011-289442-INVISIBLES). GB acknowledges the support of the European
Research Council through the ERC starting grant {\it WIMPs Kairos}.  This work
has been partially supported by the MICINN, Spain, under contract
FPA2010-17747; Consolider-Ingenio CPAN CSD2007-00042, as well as MULTIDARK
CSD2009- 00064. We thank as well the Comunidad de Madrid through Proyecto
HEPHACOS S2009/ESP-1473 and the European Commission under contract
PITN-GA-2009-237920.  The authors acknowledge as well the support of the
Spanish MINECO \emph{Centro de Excelencia Severo Ochoa} Programme under grant
SEV-2012-0249.  The use of IFT-UAM High Performance Computing Service is
gratefully acknowledged.
 
\bibliography{references} \bibliographystyle{IEEEtran}

\end{document}